\documentstyle[prd,aps,preprint,tighten,epsfig]{revtex}

\begin{document}

\draft

\title{Running Neutrino Masses, Leptonic Mixing Angles and
CP-Violating Phases: From $M_{\rm Z}$ to $\Lambda_{\rm GUT}$}
\author{{\bf Jianwei Mei}}
\address{CCAST (World Laboratory), P.O. Box 8730, Beijing 100080, China \\
and Institute of High Energy Physics, Chinese Academy of Sciences, \\
P.O. Box 918 (4), Beijing 100049, China
\footnote{Mailing address} \\
({\it Electronic address: jwmei@mail.ihep.ac.cn})} \maketitle

\begin{abstract}
We derive renormalization group equations for neutrino masses,
leptonic mixing angles and CP-violating phases running at energies
above the seesaw threshold, both in the Standard Model and in the
Minimal Supersymmetric Standard Model extended with three
right-handed neutrinos. With these equations, we carry out a
systematic study of the radiative correction that may arise to
neutrino parameters, via their renormalization group evolution
from the electroweak scale ($M_{\rm Z}$) to the scale of Grand
Unified theories ($\Lambda_{\rm GUT}$). We study in detail three
typically interesting neutrino mass patterns: normal hierarchy,
near degeneracy and inverted hierarchy. Magnitudes of possible
corrections in each case are carefully investigated. We also
emphasize the significance of CP-violating phases in controlling
the evolution behavior of all neutrino parameters.
\end{abstract}


\newpage

\section{Introduction}

\noindent Experimental information on neutrino masses and mixings
has opened up a new playground for efforts\cite{yukawaReview} of a
better understanding of the Yukawa couplings in the Standard Model
(SM) and its extensions. In the lepton sector, it is conceptually
natural to use the elegant seesaw mechanism
\cite{seesaw1,seesaw2,seesaw3} to give masses to the SM neutrinos.
In the most simple version of this mechanism, the dimension 5
neutrino mass term is the low energy relic of some more
fundamental theories with very heavy right-handed neutrinos. So in
this framework, we often need to relate physics at vastly
different energy scales.

The only way to compare the high energy theoretical prediction and
the low energy experimental observation is to use renormalization
group equations (RGEs). It has been found that corrections arising
from renormalization group (RG) evolution can be very significant
for leptonic mixing angles and neutrino mass splitting, especially
in the case of nearly degenerate left-handed neutrinos. So in
principle, the RG correction should not be neglected in the
discussion of models suggested at high energy scales. With RGEs
derived in Refs.\cite{wetterich,9306333,0108005}, early
discussions\cite{9508247,9808251,9904395,9810471,0011217,0001310,0011263,0006229,0203222}
of the RG running effect are mainly concerned with neutrino mass
(or Yukawa coupling) matrices. The evolution of neutrino masses,
leptonic mixing angles and possible CP-violating phases is studied
by diagonalizing the relevant Yukawa matrices at different energy
scales, and the behaviors are discussed in a numerical or
semi-analytical way. However, authors of
Refs.\cite{9910231,9910420,9911481,0104131} have also emphasized
the significance of RGEs for individual neutrino parameters. Such
equations not only make it possible to predict the evolution
behavior of each parameter\cite{0301234,0302181,0312167,0406103},
but also can help people appreciate interesting features such as
the existence of (pseudo-) fixed points in the evolution of mixing
angles and CP-violating phases
\cite{0011263,0203222,9910231,0110249,0306243}. The derivation of
these equations {\it below} the seesaw threshold has been done in
Refs.\cite{9910231,9910420,0305273,0312167}. And based on such
equations, a comprehensive study of the RG evolution of neutrino
parameters from the electroweak scale to the seesaw threshold has
been carried out in Ref.\cite{0305273}.

However, RG corrections above the seesaw threshold sometimes are
as important as or even more significant than those below the
threshold\cite{9508247,0006229,0203233,0206078,0404081}. Since the
physics responsible for neutrino mass generation is more likely to
exist at the scale of Grand Unified theories, a systematic study
of the RG correction above the seesaw threshold should be
necessary. And this is one of our main concerns in this work. In
much the same spirit as of Ref.\cite{0305273}, we derive RGEs for
individual neutrino parameters running above the seesaw threshold,
under the condition that eigenvalues of the Yukawa coupling matrix
that connects left- and right-handed neutrinos are hierarchical.
The contribution from the largest eigenvalue of this matrix is
explicitly shown. By setting this contribution to zero, we can
regain RGEs obtained earlier in Refs.\cite{0305273,0312167}, which
are valid at energies below the seesaw threshold.

The second purpose of this work is to carry out a systematic study
of the radiative correction that may arise via the RG evolution in
the full energy range from the electroweak scale ($M_{\rm Z}$) to
the scale of Grand Unified theories ($\Lambda_{\rm GUT}$). To
demonstrate main features of possible corrections, we study in
detail three typically allowed neutrino mass patterns: normal
hierarchy, near degeneracy and inverted hierarchy.

The paper is organized as follows. We write down {\it full}
one-loop RGEs for individual neutrino masses, leptonic mixing
angles and CP violating phases in Section II, with a brief
discussion. Then in Section III, we carry out a systematic study
of the correction that may arise during the RG evolution from
$M_{\rm Z}$ to $\Lambda_{\rm GUT}$, in theories with the three
neutrino mass patterns mentioned above. Section IV is devoted to a
Summary.

\section{Analytical formulae for neutrino parameters running above the
seesaw threshold}

\noindent Extended with three right-handed neutrinos, the
Lagrangian giving mass to leptons in the SM is
\begin{equation}
-{\cal L}_l=\overline{E_{L}}HY_{l}l_{R}^{}+\overline{E_{L}}
H_{}^{c}Y_\nu\nu_{R}^{}+\frac{1}{2}\overline{\nu
_{R}^{c}}M_{R}\nu_{R}+h.c.;  \label{effSM}
\end{equation}
and in the MSSM is
\begin{equation}
-{\cal L}_l=\overline{E_{L}}H_{1}Y_{l}l_{R}^{}+\overline{E_{L}
}H_{2}Y_\nu\nu_{R}^{}+\frac{1}{2}\overline{\nu
_{R}^{c}}M_{R}\nu_{R}+h.c.,  \label{effMSSM}
\end{equation}
where, $E_{L}$, $l_{R}^{}$ and $\nu_{R}^{}$ denote
$SU(2)_{L}$-doublets, right-handed charged leptons and
right-handed neutrinos, respectively. Both in Eqs.(\ref{effSM})
and (\ref{effMSSM}), the scale of $M_{R}$ is expected to be
extremely high, since there is not a protective symmetry. Around
this energy scale, mass eigenstates of $M_{R}$ are successively
integrated out at their respective masses
$\left(M_{1}<M_{2}<M_{3}\right) $, giving rise to a series of
effective theories at different energy scales\cite{0203233}. Then
at energies below the lightest right-handed neutrino mass, we
obtain the dimension 5 effective mass term for left-handed
neutrinos
\begin{equation}
-{\cal L}_\nu =-\frac{1}{2}\overline{E_{L}}\Phi \cdot \kappa \cdot
\Phi^{T}E_{L}^{c}+h.c.,  \label{effmass}
\end{equation}
where $\kappa$ is the effective Yukawa coupling matrix, and $\Phi$
is $H_{}^{c}$ in the SM but is $H_{2}$ in the MSSM. Since $\kappa$
is calculated from $Y_\nu$ and $M_R$ by decoupling right-handed
neutrinos at successive energy scales, step by step, the relation
between $\kappa$ and $Y_\nu$, $M_R$ is complicated. Only in the
most simplified procedure when all right-handed neutrinos are
decoupled at a common scale, can we obtain (at that chosen scale)
a simple equation because of the tree-level matching
condition\cite{0203233}:
\begin{equation} \kappa =Y_\nu M_{R}^{-1}Y_\nu^{T} ~.  \label{kappa}
\end{equation}
Then, when the Higgs field acquires a non-zero vacuum expectation
value $\langle \Phi \rangle =\left(v~~0\right)^T$ during the
electroweak symmetry breaking, Eq.(\ref{effmass}) yields an
effective mass matrix $M_\nu=v^{2}\kappa $ for left-handed
neutrinos. In the SM, $ v\simeq 174$ GeV; and in the MSSM,
$v\simeq 174\sin \beta $ GeV.

Since right-handed neutrinos are to be decoupled at their
respective thresholds, it will be too complicated to derive RGEs
for neutrino parameters between these thresholds. So we shall be
less ambitious than solving the whole problem, but shall simplify
it by (a) decoupling all right-handed neutrinos at a {\it common}
scale, which we take to be $M_{3}$, and (b) limiting our
derivation only to the case when eigenvalues of $Y_\nu$ are
\textit{hierarchical}. Here, assumption (a) can be justified when
the RG evolution through right-handed neutrino thresholds is not
too dramatic (As has been demonstrated in
Refs.\cite{0006229,0203233}, this is not always the case). And
assumption (b) is also well motivated since in a large class of
high energy models considered in literature (such as those based
on $U(1)$ symmetry or those base on $SO(10)$ Grand Unified
Theories), there is often a certain similarity or even
identification between $Y_\nu$\ and the up-quark Yukawa coupling
matrix. Such a similarity of matrices should lead to some likeness
between their eigenvalues.

One-loop RGEs for $\kappa $ running from $M_{\rm Z}$\ to $M_1$ and
those for $Y_\nu$ and $M_{R}$ running through right-handed
neutrino thresholds to $\Lambda_{\rm GUT}$ have been given in
Ref.\cite{0203233}. For readers' convenience, we have collected a
part of them together with those for $Y_{l}$ in Appendix A. To
discuss the RG evolution of neutrino parameters in the full energy
range from $M_{\rm Z}$ to $\Lambda_{\rm GUT}$, it is convenient to
make use of $\kappa $ also at energies above the seesaw
threshold\cite{0406103,0404081}. In this energy range, we find
from Eqs.(\ref{kappa}), (\ref{runYnu}) and (\ref{cR})
\begin{equation}
16\pi^2\frac{\rm d \it \kappa}{\rm d \it t} =\alpha_\kappa\kappa
+N_{\kappa }\kappa +\kappa N_\kappa^{T}, \label{RGEorigin}
\end{equation}
where $t=\ln\mu$ with $\mu$ being the energy scale, and details of
$\alpha_\kappa$ and $N_\kappa$ are given in Eq.(\ref{nKappa}) in
the Appendix.

For neutrino masses, leptonic mixing angles and CP-violating
phases, $\kappa$ is diagonalized by a unitary matrix $U_\kappa$:
\begin{equation}
\kappa =U_\kappa\kappa^{\prime}U_\kappa^{T};\ \ \ \kappa^{\prime
}={\rm diag}\left\{k_{1},k_{2},k_{3}\right\} , \label{mns1a}
\end{equation}
where $k_i$ (for $i=1,2,3$) at $M_{\rm Z}$ are proportional to
left-handed neutrino masses : $k_{i}\left(M_{\rm Z}\right) \equiv
m_{i}\left(M_{\rm Z}\right) /v^{2}$. At energy scales below
$M_{1}$, $Y_{l}$ is always diagonal during the RG evolution, if it
is diagonal at the beginning. In such a basis, the leptonic mixing
matrix is $U_{\rm MNS}\equiv U_\kappa$. However, $Y_{l}$ can not
be kept diagonal above the seesaw threshold when Eq.(\ref{runYl})
is used for its RG evolution. In this case, there is the
contribution to $U_{\rm MNS}$ from diagonalizing
$H_{l}=Y_{l}Y_{l}^\dagger$:
\begin{eqnarray}
&& H_{l} =U_{l}H_{l}^{\prime}U_{l}^\dagger;\ \ \ H_{l}^{\prime
}={\rm diag}\left\{y_{e}^{2},y_{\mu}^{2},y_{\tau}^{2}\right\}, \label{HL} \\
&& \Longrightarrow U_{\rm MNS} = U_{l}^\dagger U_\kappa.
\label{MNS}
\end{eqnarray}
For the MNS matrix, a convenient parametrization can be found in
Ref.\cite{0307359}
\begin{eqnarray}
U_{\rm MNS}&=& \left (\matrix{1 & & \cr & c_{y} & s_{y} \cr & -s_{y} & c_{y}}\right )%
\left (\matrix{c_{z} & & s_{z} \cr & e^{-i\delta} & \cr -s_{z} & & c_{z}}\right )%
\left (\matrix{c_{x} & s_{x} & \cr -s_{x} & c_{x} & \cr & & 1}\right )%
\left (\matrix{e^{i\rho } & & \cr & e^{i\sigma} & \cr & & 1}\right) \nonumber \\
&=&\left(\matrix{c_x c_z & c_z s_x & s_z \cr%
-c_y s_x e^{-i\delta} - c_x s_y s_z & c_x c_y e^{-i\delta} -s_x s_y s_z & c_z s_y \cr%
s_x s_y e^{-i\delta} -c_x c_y s_z & -c_x s_y e^{-i\delta} -c_y s_x s_z & c_y c_z}\right)%
\left (\matrix{e^{i\rho } & & \cr & e^{i\sigma} & \cr & & 1}\right) ~,%
\label{mns1b}
\end{eqnarray}
where $c_{x}\equiv \cos \theta_{x},s_{x}\equiv \sin \theta_{x}$
and so on. \footnote{Comparing to another parametrization used in
the literature\cite{PDG04}, apart from the notation of mixing
angles $\theta_{12} \Leftrightarrow \theta_x, ~\theta_{23}
\Leftrightarrow \theta_y$ and $\theta_{13} \Leftrightarrow
\theta_z$, the main difference resides only in the Majorana
phases: $\alpha_1 \Leftrightarrow 2(\rho- \delta), ~\alpha_2
\Leftrightarrow 2(\sigma -\delta)$, while the Dirac phase $\delta$
is the same.}The merit of this parametrization is that the Dirac
phase $\delta$ does not appear in the neutrinoless double beta
decay, while Majorana phases $\rho$ and $\sigma$ do not contribute
to the leptonic CP violation in neutrino oscillations. Thus two
different types of phases can be separately studied in different
types of experiments.

At energies below the seesaw threshold, RGEs for the running of
left-handed neutrino masses $\left(m_1, m_2, m_3\right) $,
leptonic mixing angles $\left(\theta_{x}, \theta_{y},
\theta_{z}\right) $ and CP-violating phases $\left(\delta ,\rho
,\sigma \right) $ have been derived in
Refs.\cite{9910231,9910420,0305273,0312167}. In order to obtain
the same kind of formulae at energies above the seesaw threshold,
we also need a parametrization of $Y_\nu$. We find that the
derivation is most straightforward if $Y_\nu$ is parameterized in
the diagonal basis of $\kappa$ by $\left(H_\nu\equiv Y_\nu^{}
Y_\nu^\dagger\right) $:
\begin{equation}
H_\nu =U_\nu H_\nu^{\prime}U_\nu^\dagger;\ \ \ H_\nu^{\prime}
=y_\nu^{2}\cdot {\rm diag}\left\{r_{1}^{2},r_{2}^{2},1\right\}\
,\label{Hnu}
\end{equation}
where (with $c_{1}=\cos \theta_{1}$,$\ s_{1}=\sin \theta_{1}$ and
so on)
\begin{equation}
U_\nu= \left (\matrix{e^{i\phi_{1}} & & \cr & e^{i\phi_{2}} & \cr
& & 1}\right ) \left (\matrix{1 & & \cr & c_{1} & s_{1} \cr &
-s_{1} & c_{1}}\right ) \left (\matrix{c_{2} & & s_{2} \cr &
e^{-i\delta_\nu} & \cr -s_{2} & & c_{2}}\right ) \left
(\matrix{c_{3} & s_{3} & \cr -s_{3} & c_{3} & \cr & & 1}\right ) .
\label{Unu}
\end{equation}
As mentioned above, we shall concentrate on cases in which
eigenvalues of $Y_\nu$ are hierarchical, i.e.
$r_{1}^{2}<<r_{2}^{2}<<1$. So the contribution of $r_{1}^{2}$ and
$r_{2}^{2}$ to the running of neutrino parameters can always be
neglected. This is equivalent to taking $r_{1}^{2}=r_{2}^{2}=0$
when deriving RGEs. From Eqs.(\ref{Hnu}) and (\ref{Unu}), it is
obvious that only $\theta_{1},\theta_{2},\phi _{1}$ and $\phi_{2}$
in $U_\nu$ contribute to the evolution of parameters $m_1, m_2,
m_3, \theta_{x}, \theta_{y}, \theta_{z}, \delta ,\rho$ and
$\sigma$.

Before writing down all the analytical formulae, we remark that RG
corrections to mixing angles and CP-violating phases come from
three separable sources. To clarify this point, we need
Eq.(\ref{nKappa}): $N_\kappa =C_\kappa^{l}H_{l} +C_\kappa^\nu
H_\nu$. In Eq.(\ref{RGEorigin}), only $N_{\kappa }$ contributes to
the evolution of mixing angles and CP-violating phases. So each of
the two terms in $N_\kappa$ is a source of RG corrections. Also,
there is a contribution from diagonalizing $Y_l$, and this is the
third one.

\begin{itemize}

\item The contribution from $C_\kappa^l H_l$ in $N_\kappa$ is
proportional to $C_\kappa^l y_{\tau}^{2}$ (we have omitted the
contributions from $y_e^2$ and $y_\mu^2$ for obvious reasons).
This contribution is \textit{exactly the same as} that governs the
evolution of neutrino parameters at energies below the seesaw
threshold. Analytical formulae of this contribution have been
derived and extensively discussed in Refs.\cite{0305273,0312167}.

\item The contribution from $C_\kappa^\nu H_\nu$ in $N_\kappa$ is
proportional to $C_\kappa^\nu y_\nu^{2}$. While $y_{\tau}$ can be
of ${\cal O}(1)$ only in the MSSM when $\tan \beta $ is large, it
is quite natural for $y_\nu$ to be of the same magnitude as the
top Yukawa coupling. Furthermore, just like the contribution from
$C_\kappa^l H_l$, the contribution from $C_\kappa^\nu H_\nu$ can
also be resonantly enhanced when eigenvalues of $\kappa $ are
nearly degenerate. This is because both contributions contain such
enhancing factors as $\zeta_{ij}^{-1}$ (for $i<j; ~~i,j=1,2,3$).
Note that
\begin{equation}
\zeta_{ij}\equiv \frac{k_{i}-k_{j}}{k_{i}+k_{j}} ~;~~~ i,j=1,2,3
~. \label{ratio}
\end{equation}

\item The third contribution comes from diagonalizing $Y_{l}$ and
is proportional to $C_l^\nu y_{\nu }^{2}$. Different from that of
$C_\kappa^\nu H_\nu$, there are no enhancing factors in this
contribution other than functions of mixing angles, such as
$s_{z}^{-1} \equiv (\sin \theta_z )^{-1}$ or $s_{x}^{-1} \equiv
(\sin \theta_x)^{-1}$ etc., which appear mostly in RGEs of
CP-violating phases and are important only when $s_{z}$ (or
$s_{x}$ etc.) $<<{\cal O}(1)$.

\end{itemize}

Now following the same procedure as described in
Refs.\cite{9910420,0305273} but including contributions from
$Y_\nu$ and $Y_{l}$ (yet with the above explained
simplifications), we obtain the following \textit{full} one-loop
RGEs for left-handed neutrino masses $(m_1,m_2,m_3)$, leptonic
mixing angles $(\theta_x, \theta_y, \theta_z)$ and CP-violating
phases $(\delta, \rho, \sigma)$ running at energies above the
seesaw threshold.

\footnote{Throughout this work, for what ever $F: ~\dot{F}\equiv
16\pi^2 \displaystyle\frac{d F}{d t}; ~t={\rm ln} \mu$, with $\mu$
being the energy scale.}For the running of left-handed neutrino
masses (at $M_{\rm Z}^{}: ~m_{i} \equiv v^2 k_{i} ~;~ i=1,2,3$):
\begin{eqnarray}
\dot{k}_{i} &=&a_{i}k_{i} ~,~~~ i=1,2,3 ~;  \nonumber \\
a_{1} &\equiv
&\alpha_\kappa+2C_\kappa^{l}y_{\tau}^{2}\left(s_{x}^{2}s_{y}^{2}-2c_{\delta
}^{}c_{x}c_{y}s_{x}s_{y}s_{z}+c_{x}^{2}c_{y}^{2}s_{z}^{2}\right)
+2C_\kappa^\nu y_\nu^{2}s_{2}^{2} ~;  \nonumber \\
a_{2} &\equiv
&\alpha_\kappa+2C_\kappa^{l}y_{\tau}^{2}\left(c_{x}^{2}s_{y}^{2}+2c_{\delta
}^{}c_{x}c_{y}s_{x}s_{y}s_{z}+c_{y}^{2}s_{x}^{2}s_{z}^{2}\right)
+2C_\kappa^\nu y_\nu^{2}c_{2}^{2}s_{1}^{2} ~;  \nonumber \\
a_{3} &\equiv &\alpha_\kappa+2C_\kappa^{l}y_{\tau
}^{2}c_{y}^{2}c_{z}^{2}+2C_\kappa^\nu y_{\nu
}^{2}c_{1}^{2}c_{2}^{2} ~. \label{anaMass}
\end{eqnarray}

For the running of leptonic mixing angles and CP-violating phases
($c_\delta^{} \equiv \cos \delta, ~s_\delta^{} \equiv \sin \delta,
~c_{ \left(\delta- \rho \right)}^{} \equiv\cos \left(\delta-\rho
\right), ~s_{\left(\delta- \rho \right)}^{}
\equiv\sin\left(\delta- \rho\right)$ and so on):
\begin{eqnarray}
\dot{\theta}_{x}
&=&C_\kappa^{l}y_{\tau}^{2}\left\{\frac{c_{\left(\rho -\sigma
\right)}^{}}{\zeta_{12}}\left[ c_{\left(\rho -\sigma
\right)}^{}c_{x}s_{x}\left(s_{y}^{2}-c_{y}^{2}s_{z}^{2}\right)
-\left(c_{\left(\delta +\rho -\sigma
\right)}^{}c_{x}^{2}-c_{\left(\delta -\rho +\sigma
\right)}^{}s_{x}^{2}\right) c_{y}s_{y}s_{z}\right] \right.  \nonumber \\
&&\ \ \ \ \ \ \ +\zeta_{12}\cdot s_{\left(\rho -\sigma \right)
}^{}\left[ s_{\left(\rho -\sigma
\right)}^{}c_{x}s_{x}\left(s_{y}^{2}-c_{y}^{2}s_{z}^{2}\right)
-\left(s_{\left(\delta +\rho -\sigma
\right)}^{}c_{x}^{2}+s_{\left(\delta -\rho +\sigma \right)
}^{}s_{x}^{2}\right) c_{y}s_{y}s_{z}\right]  \nonumber \\
&&\ \ \ \ \ \ \ -\left[
\frac{c_{\rho}^{}}{\zeta_{13}}\left(c_{\left(\delta -\rho
\right)}^{}s_{x}s_{y}-c_{\rho }^{}c_{x}c_{y}s_{z}\right)
-\zeta_{13}\cdot s_{\rho}^{}\left(s_{\left(\delta -\rho
\right)}^{}s_{x}s_{y}+s_{\rho }^{}c_{x}c_{y}s_{z}\right) \right]
c_{y}s_{x}s_{z}
\nonumber \\
&&\left. \ \ \ \ \ \ \ -\left[ \frac{c_{\sigma}^{}}{\zeta
_{23}}\left(c_{\left(\delta -\sigma \right)
}^{}c_{x}s_{y}+c_{\sigma}^{}c_{y}s_{x}s_{z}\right) -\zeta
_{23}\cdot s_{\sigma}^{}\left(s_{\left(\delta -\sigma \right)
}^{}c_{x}s_{y}-s_{\sigma
}^{}c_{y}s_{x}s_{z}\right) \right] c_{x}c_{y}s_{z}\right\}  \nonumber \\
&&+C_\kappa^\nu y_\nu^{2}\left\{-\left(\frac{c_{\left(\rho -\sigma
\right)}^{}c_{\left(\phi_{1}-\phi_{2}\right)
}^{}}{\zeta_{12}}-\zeta_{12}\cdot s_{\left(\rho -\sigma \right)
}^{}s_{\left(\phi
_{1}-\phi_{2}\right)}^{}\right) c_{2}s_{1}s_{2}\right.  \nonumber \\
&&\left. \ \ \ \ \ \ \ \ \ \ -\left(\frac{c_{\rho
}^{}c_{\phi_{1}}^{}}{\zeta_{13}}-\zeta_{13}\cdot s_{\rho
}^{}s_{\phi_{1}}^{}\right)
\frac{c_{1}c_{2}s_{2}s_{x}s_{z}}{c_{z}}+\left(\frac{c_{\sigma
}^{}c_{\phi_{2}}^{}}{\zeta_{23}}-\zeta_{23}\cdot s_{\sigma
}^{}s_{\phi_{2}}^{}\right)
\frac{c_{1}c_{2}^{2}c_{x}s_{1}s_{z}}{c_{z}}\right\}
\nonumber \\
&&+C_{l}^\nu y_\nu^{2} \left\{\left[ c_{\left(\rho
+\phi_{1}\right)}^{}s_{2}s_{x}-c_{\left(\sigma +\phi
_{2}\right)}^{}c_{2}c_{x}s_{1}\right]
\frac{c_{1}c_{2}s_{z}}{c_{z}}-c_{x}s_{x}\left(c_{2}^{2} s_{1}^{2}
-s_{2}^{2}\right) \right.  \nonumber \\
&&\left. \ \ \ \ \ \ \ \ \ \ \ \ \ -c_{\left(\rho -\sigma +\phi
_{1}-\phi_{2}\right)}^{}c_{2}s_{1}s_{2}\left(c_{x}^{2}-s_{x}^{2}\right)
\right\} ~; \label{dX}
\end{eqnarray}
\begin{eqnarray}
\dot{\theta}_{y} &=&C_\kappa^{l}y_{\tau}^{2}\left\{\left[
\frac{c_{\left(\delta -\rho
\right)}^{}}{\zeta_{13}}\left(c_{\left(\delta -\rho
\right)}^{}s_{x}s_{y}-c_{\rho }^{}c_{x}c_{y}s_{z}\right)
+\zeta_{13}\cdot s_{\left(\delta -\rho
\right)}^{}\left(s_{\left(\delta -\rho \right)
}^{}s_{x}s_{y}+s_{\rho}^{}c_{x}c_{y}s_{z}\right) \right]
c_{y}s_{x}\right.  \nonumber \\
&&\left. \ \ \ \ \ \ +\left[ \frac{c_{\left(\delta -\sigma
\right)}^{}}{\zeta_{23}}\left(c_{\left(\delta -\sigma \right)
}^{}c_{x}s_{y}+c_{\sigma}^{}c_{y}s_{x}s_{z}\right) +\zeta
_{23}\cdot s_{\left(\delta -\sigma
\right)}^{}\left(s_{\left(\delta -\sigma
\right)}^{}c_{x}s_{y}-s_{\sigma }^{}c_{y}s_{x}s_{z}\right) \right]
c_{x}c_{y}\right\}  \nonumber \\
&&+C_\kappa^\nu y_\nu^{2}\left\{\left(\frac{c_{\left(\delta -\rho
\right)}^{}c_{\phi_{1}}^{}}{\zeta_{13}}+\zeta_{13}\cdot
s_{\left(\delta -\rho \right) }^{}s_{\phi_{1}}^{}\right)
\frac{c_{1}c_{2}s_{2}s_{x}}{c_{z}}-\left(\frac{c_{\left(\delta
-\sigma \right)}^{}c_{\phi_{2}}^{}}{\zeta_{23}}+\zeta_{23}\cdot
s_{\left(\delta -\sigma \right)}^{}s_{\phi_{2}}^{}\right)
\frac{c_{1}c_{2}^{2}c_{x}s_{1}}{c_{z}}\right\}  \nonumber \\
&&+C_{l}^\nu y_\nu^{2}\left\{c_{y}s_{y}\left(-c_{1}^{2} c_{2}^{2}
c_{z}^{2} +s_{2}^{2}\left(s_{x}^{2} -c_{x}^{2} s_{z}^{2} \right)
+c_{2}^{2} s_{1}^{2} \left(c_{x}^{2}
-s_{x}^{2}s_{z}^{2}\right) \right) \right.  \nonumber \\
&&\ \ \ \ \ \ \ \ \ \
+2c_\delta^{}c_{x}c_{y}^{2}\left(c_{2}^{2}s_{1}^{2}-s_{2}^{2}\right)
s_{x}s_{z}+2\left(c_{\left(\rho
+\phi_{1}\right)}^{}c_{x}s_{2}+c_{\left(\sigma +\phi _{2}\right)
}^{}c_{2}s_{1}s_{x}\right) c_{1}c_{2}c_{y}c_{z}s_{y}s_{z}  \nonumber \\
&&\ \ \ \ \ \ \ \ \ \ +\left(c_{\left(\delta -\rho -\phi
_{1}\right)}^{}s_{2}s_{x}-c_{\left(\delta -\sigma -\phi
_{2}\right)}^{}c_{2}c_{x}s_{1}\right)
\frac{c_{1}c_{2}}{c_{z}}\left(c_{z}^{2}\left(c_{y}^{2}
-s_{y}^{2}\right) -s_{z}^{2}\right)  \nonumber \\
&&\ \ \ \ \ \ \ \ \ \ +2\left[ -c_{\left(\rho -\sigma +\phi
_{1}-\phi_{2}\right)}^{}c_{x}s_{x}s_{y}\left(1+s_{z}^{2}\right)
\right.  \nonumber \\
&&\left. \ \ \ \ \ \ \ \ \ \ \ \ \ \ \ \ \left.
+\left(c_{\left(\delta +\rho -\sigma +\phi_{1}-\phi_{2}\right)
}^{}c_{x}^{2}-c_{\left(\delta -\rho +\sigma -\phi_{1}+\phi
_{2}\right)}^{}s_{x}^{2}\right) c_{y}s_{z} \right]
c_{2}c_{y}s_{1}s_{2}\right\} ~;  \label{dY}
\end{eqnarray}
\begin{eqnarray}
\dot{\theta}_{z} &=&C_\kappa^{l}y_{\tau}^{2}\left\{-\left[
\frac{c_{\rho}^{}}{\zeta_{13}}\left(c_{\left(\delta -\rho
\right)}^{}s_{x}s_{y}-c_{\rho}^{}c_{x}c_{y}s_{z}\right) -\zeta
_{13}\cdot s_{\rho}^{}\left(s_{\left(\delta -\rho \right)
}^{}s_{x}s_{y}+s_{\rho
}^{}c_{x}c_{y}s_{z}\right) \right] c_{x}c_{y}c_{z}\right.  \nonumber \\
&&\ \ \ \ \ \ \ \left. +\left[ \frac{c_{\sigma}^{}}{\zeta
_{23}}\left(c_{\left(\delta -\sigma \right)
}^{}c_{x}s_{y}+c_{\sigma}^{}c_{y}s_{x}s_{z}\right) -\zeta
_{23}\cdot s_{\sigma}^{}\left(s_{\left(\delta -\sigma \right)
}^{}c_{x}s_{y}-s_{\sigma
}^{}c_{y}s_{x}s_{z}\right) \right] c_{y}c_{z}s_{x}\right\}  \nonumber \\
&&+C_\kappa^\nu y_\nu^{2}\left\{-\left(\frac{c_{\rho
}^{}c_{\phi_{1}}^{}}{\zeta_{13}}-\zeta_{13}\cdot s_{\rho
}^{}s_{\phi_{1}}^{}\right)
c_{1}c_{2}c_{x}s_{2}-\left(\frac{c_{\sigma}^{}c_{\phi_{2}}^{}}{\zeta_{23}}-\zeta
_{23}\cdot s_{\sigma}^{}s_{\phi
_{2}}^{}\right) c_{1}c_{2}^{2}s_{1}s_{x}\right\}  \nonumber \\
&&+C_{l}^\nu y_\nu^{2}\left\{\left(c_{x}^{2}s_{2}^{2}
-c_{2}^{2}\left(c_{1}^{2}-s_{1}^{2}s_{x}^{2}\right) \right)
c_{z}s_{z}-\left(c_{\left(\rho
+\phi_{1}\right)}^{}c_{x}s_{2}+c_{\left(\sigma
+\phi_{2}\right)}^{}c_{2}s_{1}s_{x}\right)
c_{1}c_{2}\left(c_{z}^{2}-s_{z}^{2}\right)
\right.  \nonumber \\
&&\ \ \ \ \ \ \ \ \ \ \ \ \left. +2c_{\left(\rho -\sigma +\phi
_{1}-\phi_{2}\right)
}^{}c_{2}c_{x}c_{z}s_{1}s_{2}s_{x}s_{z}\right\} ~;  \label{dZ}
\end{eqnarray}
\begin{eqnarray}
\dot\delta &=&C_\kappa^{l}y_{\tau}^{2}\left\{\frac{s_{\left(\rho
-\sigma \right)}^{}}{\zeta_{12}}\left[ c_{\left(\rho -\sigma
\right)}^{}\left(s_{y}^{2}-c_{y}^{2}s_{z}^{2}\right)
-\left(c_{\left(\delta +\rho -\sigma \right)
}^{}c_{x}^{2}-c_{\left(\delta -\rho +\sigma \right)
}^{}s_{x}^{2}\right) \frac{c_{y}s_{y}s_{z}}{c_{x}s_{x}}\right]
\right.
\nonumber \\
&&\ \ \ \ \ \ \ \ -\zeta_{12}\cdot c_{\left(\rho -\sigma \right)
}^{} \left[ s_{\left(\rho -\sigma
\right)}^{}\left(s_{y}^{2}-c_{y}^{2}s_{z}^{2}\right)
-\left(s_{\left(\delta +\rho -\sigma
\right)}^{}c_{x}^{2}+s_{\left(\delta -\rho +\sigma
\right)}^{}s_{x}^{2}\right)
\frac{c_{y}s_{y}s_{z}}{c_{x}s_{x}}\right]  \nonumber
\\
&&\ \ \ \ \ \ \ \ +\zeta_{13}^{-1}\cdot \left(c_{\left(\delta
-\rho \right)}s_{x}s_{y}-c_{\rho}^{}c_{x}c_{y}s_{z}\right) \left[
\frac{s_{\rho}^{}c_{y}}{c_{x}s_{z}}\left(c_{x}^{2}-s_{x}^{2}s_{z}^{2}\right)
-s_{\left(\delta -\rho \right)}^{}\frac{s_{x}}{s_{y}}
\left(c_{y}^{2}-s_{y}^{2}\right) \right]  \nonumber \\
&&\ \ \ \ \ \ \ \ +\zeta_{13}\cdot \left(s_{\left(\delta -\rho
\right)}^{}s_{x}s_{y}+s_{\rho}c_{x}c_{y}s_{z}\right) \left[
\frac{c_{\rho}^{}c_{y}}{c_{x}s_{z}}\left(c_{x}^{2}-s_{x}^{2}s_{z}^{2}\right)
+c_{\left(\delta -\rho \right)}^{}\frac{s_{x}}{s_{y}}
\left(c_{y}^{2}-s_{y}^{2}\right) \right]  \nonumber \\
&&\ \ \ \ \ \ \ \ -\zeta_{23}^{-1}\cdot \left(c_{\left(\delta
-\sigma \right)}c_{x}s_{y}+c_{\sigma}^{}c_{y}s_{x}s_{z}\right)
\left[\frac{s_{\sigma}^{}c_{y}}{s_{x}s_{z}}\left(s_{x}^{2}-c_{x}^{2}s_{z}^{2}\right)
+s_{\left(\delta -\sigma \right)}^{}\frac{c_{x}}{s_{y}}
\left(c_{y}^{2}-s_{y}^{2}\right) \right]  \nonumber \\
&&\ \ \ \ \ \ \ \left. -\zeta_{23}\cdot \left(s_{\left(\delta
-\sigma \right)}^{}c_{x}s_{y}-s_{\sigma}c_{y}s_{x}s_{z}\right)
\left[\frac{c_{\sigma}^{}c_{y}}{s_{x}s_{z}}\left(s_{x}^{2}-c_{x}^{2}s_{z}^{2}\right)
-c_{\left(\delta -\sigma \right)}^{}\frac{c_{x}}{s_{y}}
\left(c_{y}^{2}-s_{y}^{2}\right) \right] \right\}  \nonumber \\
&&+C_\kappa^\nu
y_\nu^{2}\left\{-\left(\frac{c_{\left(\phi_{1}-\phi_{2}\right)}^{}s_{\left(\rho
-\sigma \right) }^{}}{\zeta_{12}}+\zeta_{12}\cdot c_{\left(\rho
-\sigma \right) }^{}s_{\left(\phi_{1}-\phi_{2}\right)}^{}\right)
\right. \frac{c_{2}s_{1}s_{2}}{c_{x}s_{x} }\nonumber \\
&&\ \ \ \ \ \ \ \ \ \ \ +\left[ \frac{c_{\phi_{1}}^{}}{\zeta
_{13}}\left(s_{\rho}^{}c_{y}s_{y}\left(c_{x}^{2}-s_{x}^{2}s_{z}^{2}\right)
-s_{\left(\delta -\rho \right)}^{}\left(c_{y}^{2}-s_{y}^{2}\right)
c_{x}s_{x}s_{z}\right) \right.  \nonumber \\
&&\ \ \ \ \ \ \ \ \ \ \ \ \ \ \ \left. +\zeta_{13}\cdot s_{\phi
_{1}}^{}\left(c_{\rho}^{}c_{y}s_{y}\left(c_{x}^{2}-s_{x}^{2}s_{z}^{2}\right)
+c_{\left(\delta -\rho \right)}^{}\left(c_{y}^{2}-s_{y}^{2}\right)
c_{x}s_{x}s_{z}\right) \right] \frac{c_{1}c_{2}s_{2}}{c_{x}
c_{y}c_{z}s_{y}s_{z}}  \nonumber \\
&&\ \ \ \ \ \ \ \ \ \ \ +\left[ \frac{c_{\phi_{2}}^{}}{\zeta
_{23}}\left(s_{\sigma}^{}c_{y}s_{y}\left(s_{x}^{2}-c_{x}^{2}s_{z}^{2}\right)
+s_{\left(\delta -\sigma
\right)}^{}\left(c_{y}^{2}-s_{y}^{2}\right)
c_{x}s_{x}s_{z}\right) \right.  \nonumber \\
&&\ \ \ \ \ \ \ \ \ \ \ \ \ \ \ \left. \left. +\zeta_{23}\cdot
s_{\phi_{2}}^{}\left(c_{\sigma}^{}c_{y}s_{y}\left(s_{x}^{2}-c_{x}^{2}s_{z}^{2}\right)
-c_{\left(\delta -\sigma
\right)}^{}\left(c_{y}^{2}-s_{y}^{2}\right)
c_{x}s_{x}s_{z}\right) \right] \frac{c_{1}c_{2}^{2}s_{1}}
{c_{y}c_{z}s_{x}s_{y}s_{z}}\right\}  \nonumber \\
&&+C_{l}^\nu y_\nu^{2} \left\{-2s_\delta^{}
\left(c_{2}^{2}s_{1}^{2}-s_{2}^{2}\right)
\frac{c_{x}c_{y}s_{x}s_{z}}{s_{y}} +\left[ s_{\left(\rho +\phi
_{1}\right)}^{}s_{2}s_{x}\left(c_{x}^{2}\left(c_{z}^{2}-s_{z}^{2}\right)
+s_{x}^{2}s_{z}^{2}\right) \right. \right.
\nonumber \\
&&\ \ \ \ \ \ \ \ \ \ \ \left. \ +s_{\left(\sigma +\phi
_{2}\right)}^{}c_{2}c_{x}s_{1}\left(c_{x}^{2}s_{z}^{2}
+s_{x}^{2}\left(c_{z}^{2}-s_{z}^{2}\right)
\right) \right] \frac{c_{1}c_{2}}{c_{x}c_{z}s_{x}s_{z}}  \nonumber \\
&&\ \ \ \ \ \ \ \ \ \ \ \ -\left[ s_{\left(\delta -\rho -\phi
_{1}\right)}^{}s_{2}s_{x}\left(c_{z}^{2}-\left(c_{y}^{2}-s_{y}^{2}\right)
s_{z}^{2}\right) \right. \left. -s_{\left(\delta -\sigma
-\phi_{2}\right) }^{}c_{2}c_{x}s_{1}
\left(1-2c_{y}^{2}s_{z}^{2}\right) \right]
\frac{c_{1}c_{2}}{c_{y}c_{z}s_{y}}  \nonumber \\
&&\ \ \ \ \ \ \ \ \ \ \ \ -s_{\left(\rho -\sigma +\phi_{1}-\phi
_{2}\right)}^{}\left(c_{x}^{2}-s_{x}^{2}\right)
\frac{c_{2}s_{1}s_{2}}{c_{x}s_{x}}  \nonumber \\
&&\left. \ \ \ \ \ \ \ \ \ \ \ \ -2\left[ s_{\left(\delta +\rho
-\sigma +\phi_{1}-\phi_{2}\right)}^{}-2c_{\left(\rho -\sigma
+\phi_{1}-\phi_{2}\right)}^{}s_\delta^{}s_{x}^{2}\right]
\frac{c_{2}c_{y}s_{1}s_{2}s_{z}}{s_{y}}\right\} ~;  \label{dDelta}
\end{eqnarray}
\begin{eqnarray}
\dot{\rho} &=&C_\kappa^{l}y_{\tau}^{2}\left\{\frac{s_{\left(\rho
-\sigma \right)}^{}}{\zeta_{12}}\left[ c_{\left(\rho -\sigma
\right)}^{}c_{x}s_{x}\left(s_{y}^{2}-c_{y}^{2}s_{z}^{2}\right)
-\left(c_{\left(\delta +\rho -\sigma
\right)}^{}c_{x}^{2}-c_{\left(\delta -\rho +\sigma
\right)}^{}s_{x}^{2}\right) c_{y}s_{y}s_{z}\right]
\frac{s_{x}}{c_{x}}
\right.  \nonumber \\
&&\ \ \ \ \ \ \ -\zeta_{12}\cdot c_{\left(\rho -\sigma \right)
}^{}\left[ s_{\left(\rho -\sigma
\right)}^{}c_{x}s_{x}\left(s_{y}^{2}-c_{y}^{2}s_{z}^{2}\right)
-\left(s_{\left(\delta +\rho -\sigma
\right)}^{}c_{x}^{2}+s_{\left(\delta -\rho +\sigma
\right)}^{}s_{x}^{2}\right) c_{y}s_{y}s_{z}\right]
\frac{s_{x}}{c_{x}}  \nonumber
\\
&&\ \ \ \ \ \ \ +\left[
\frac{s_{\rho}^{}}{\zeta_{13}}\left(c_{\left(\delta -\rho
\right)}^{}s_{x}s_{y}-c_{\rho }^{}c_{x}c_{y}s_{z}\right)
+\zeta_{13}\cdot c_{\rho}^{}\left(s_{\left(\delta -\rho
\right)}^{}s_{x}s_{y}+s_{\rho }c_{x}c_{y}s_{z}\right) \right]
\frac{c_{y}\left(c_{x}^{2}c_{z}^{2}-s_{z}^{2}\right)}{c_{x}s_{z}}  \nonumber \\
&&\ \ \ \ \ \ \ \left. -\left[ \frac{s_{\sigma}^{}}{\zeta
_{23}}\left(c_{\left(\delta -\sigma \right)
}^{}c_{x}s_{y}+c_{\sigma}^{}c_{y}s_{x}s_{z}\right) +\zeta
_{23}\cdot c_{\sigma}^{}\left(s_{\left(\delta -\sigma \right)
}^{}c_{x}s_{y}-s_{\sigma}c_{y}s_{x}s_{z}\right) \right]
\frac{c_{y}c_{z}^{2}s_{x}}{s_{z}}\right\}
\nonumber \\
&&+C_\kappa^\nu y_\nu^{2}\left\{-\left[
\frac{c_{\left(\phi_{1}-\phi_{2}\right)}^{}s_{\left(\rho -\sigma
\right) }^{}}{\zeta_{12}}+\zeta_{12}\cdot c_{\left(\rho -\sigma
\right) }^{}s_{\left(\phi_{1}-\phi_{2}\right)}^{}\right]
\frac{c_{2}s_{1}s_{2}s_{x}}{c_{x}}\right.
\nonumber \\
&&\ \ \ \ \ \ \ \ \ \ \left.
+\left(\frac{c_{\phi_{1}}^{}s_{\rho}^{}}{\zeta_{13}}+\zeta_{13}\cdot
c_{\rho}^{}s_{\phi_{1}}^{}\right) \frac{c_{1}c_{2}s_{2}
\left(c_{x}^{2}c_{z}^{2} -s_{z}^{2}\right)}{c_{x}c_{z}s_{z}}
+\left(\frac{c_{\phi_{2}}^{}s_{\sigma}^{}}{\zeta_{23}}+\zeta
_{23}\cdot c_{\sigma}^{}s_{\phi_{2}}^{}\right)
\frac{c_{1}c_{2}^{2}c_{z}s_{1}s_{x}}{s_{z}}\right\}  \nonumber \\
&&+C_{l}^\nu y_\nu^{2}\left\{s_{\left(\rho +\phi_{1}\right) }^{}
\frac{c_{1}c_{2}s_{2}\left(1-c_{z}^{2}s_{x}^{2}\right)
}{c_{x}c_{z}s_{z}} +s_{\left(\sigma +\phi_{2}\right)
}^{}\frac{c_{1}c_{2}^{2}c_{z}s_{1}s_{x}}{s_{z}}+s_{\left(\rho
-\sigma +\phi_{1}-\phi_{2}\right)}^{}
\frac{c_{2}s_{1}s_{2}s_{x}}{c_{x}}\right\} ~;  \label{dRho}
\end{eqnarray}
\begin{eqnarray}
\dot{\sigma} &=&C_\kappa^{l}y_{\tau}^{2}\left\{\frac{s_{\left(\rho
-\sigma \right)}^{}}{\zeta_{12}}\left[ c_{\left(\rho -\sigma
\right)}^{}c_{x}s_{x}\left(s_{y}^{2}-c_{y}^{2}s_{z}^{2}\right)
-\left(c_{\left(\delta +\rho -\sigma
\right)}^{}c_{x}^{2}-c_{\left(\delta -\rho +\sigma
\right)}^{}s_{x}^{2}\right) c_{y}s_{y}s_{z}\right]
\frac{c_{x}}{s_{x}}
\right.  \nonumber \\
&&\ \ \ \ \ \ -\zeta_{12}\cdot c_{\left(\rho -\sigma \right)
}^{}\left[ s_{\left(\rho -\sigma
\right)}^{}c_{x}s_{x}\left(s_{y}^{2}-c_{y}^{2}s_{z}^{2}\right)
-\left(s_{\left(\delta +\rho -\sigma
\right)}^{}c_{x}^{2}+s_{\left(\delta -\rho +\sigma
\right)}^{}s_{x}^{2}\right) c_{y}s_{y}s_{z}\right]
\frac{c_{x}}{s_{x}}  \nonumber
\\
&&\ \ \ \ \ \ +\left[
\frac{s_{\rho}^{}}{\zeta_{13}}\left(c_{\left(\delta -\rho
\right)}^{}s_{x}s_{y}-c_{\rho }^{}c_{x}c_{y}s_{z}\right)
+\zeta_{13}\cdot c_{\rho}^{}\left(s_{\left(\delta -\rho
\right)}^{}s_{x}s_{y}+s_{\rho
}c_{x}c_{y}s_{z}\right) \right] \frac{c_{x}c_{y}c_{z}^{2}}{s_{z}}  \nonumber \\
&&\ \ \ \ \ \left. -\left[ \frac{s_{\sigma}^{}}{\zeta
_{23}}\left(c_{\left(\delta -\sigma \right)
}^{}c_{x}s_{y}+c_{\sigma}^{}c_{y}s_{x}s_{z}\right) +\zeta
_{23}c_{\sigma}^{}\left(s_{\left(\delta -\sigma \right)
}^{}c_{x}s_{y}-s_{\sigma}c_{y}s_{x}s_{z}\right) \right]
\frac{c_{y}\left(c_{z}^{2}s_{x}^{2}-s_{z}^{2}\right) }{s_{x}s_{z}}
\right\}  \nonumber \\
&&+C_\kappa^\nu y_\nu^{2}\left\{-\left(\frac{c_{
\left(\phi_{1}-\phi_{2}\right)}^{}s_{\left(\rho -\sigma \right)
}^{}}{\zeta_{12}}+\zeta_{12}\cdot c_{\left(\rho -\sigma \right)
}^{}s_{\left(\phi_{1}-\phi_{2}\right)}^{}\right)
\frac{c_{2}c_{x}s_{1}s_{2}}{s_{x}}\right.
\nonumber \\
&&\ \ \ \ \ \ \ \ \ \left. +\left(\frac{c_{\phi_{1}}^{}s_{\rho
}^{}}{\zeta_{13}}+\zeta_{13}\cdot c_{\rho
}^{}s_{\phi_{1}}^{}\right)
\frac{c_{1}c_{2}c_{x}c_{z}s_{2}}{s_{z}}+\left(\frac{c_{\phi
_{2}}^{}s_{\sigma}^{}}{\zeta_{23}}+\zeta_{23}\cdot c_{\sigma
}^{}s_{\phi_{2}}^{}\right) \frac{c_{1}c_{2}^{2}s_{1}\left(c_{z}^{2}s_{x}^{2}
-s_{z}^{2}\right)}{c_{z}s_{x}s_{z}}\right\}  \nonumber \\
&&+C_{l}^\nu y_\nu^{2}\left\{s_{\left(\rho +\phi_{1}\right) }^{}
\frac{c_{1}c_{2}c_{x}c_{z}s_{2}}{s_{z}}+s_{\left(\sigma
+\phi_{2}\right)}^{}\frac{c_{1}c_{2}^{2}s_{1}\left(c_{z}^{2}s_{x}^{2}
+s_{z}^{2}\right)}{c_{z}s_{x}s_{z}}-s_{\left(\rho -\sigma
+\phi_{1}-\phi_{2}\right)}^{}
\frac{c_{2}c_{x}s_{1}s_{2}}{s_{x}}\right\} ~.  \label{dSigma}
\end{eqnarray}
An outline of the derivation is given in Appendix B.  ~Concerning
these equations, two remarks are in order:

\begin{itemize}

\item As mentioned above, we can obtain exactly the same formulae
as in Ref.\cite{0312167} (and also in Ref.\cite{0305273} but with
a slightly different phase convention) if we set $y_\nu=0$. This
serves as a check of our derivation, at least for the part below
the seesaw threshold.

\item $\rho$ and $\sigma$ are only determined up to $n\pi$
($n=0,1,2,3,\cdots$) in Eqs.(\ref{mns1a}) and (\ref{mns1b}). Such
an ambiguity in $\rho$ and $\sigma $ also leads to an ambiguity in
$\phi_{1}$ and $\phi_{2}$ defined in Eqs.(\ref{Hnu}) and
(\ref{Unu}). However, both ambiguities cancel on the right hand
side of Eqs.(\ref{dX})-(\ref{dSigma}). So this ambiguity is
harmless, as it should be.

\end{itemize}
In addition to equations given above, a knowledge of the RGE
evolution behavior of $\theta_{1},\theta_{2},\phi_{1}$ and
$\phi_{2}$ will be helpful in our following discussions. So for
completeness, we have also derived one-loop RGEs for parameters in
$U_\nu$ under the condition that eigenvalues of $Y_\nu$ are
hierarchical. We find that RG corrections to $\theta_{1},
\theta_{2}, \theta_{3}, \delta_{\nu }, \phi_{1}$ and $\phi_{2}$
can be strongly enhanced by the factors $\zeta_{ij}^{-1}$ defined
in Eq.(\ref{ratio}), even if $M_{3}$ and $\Lambda_{\rm GUT}$ are
only one or two orders apart in magnitude. The full analytical
formulae are given in Appendix C.

\section{RG Correction to Neutrino Parameters With Three
Typical Mass Patterns: From $M_{\rm Z}$ to $\Lambda_{\rm GUT}$}

\noindent We have verified our analytical formulae in
Eqs.(\ref{anaMass})-(\ref{dSigma}) by comparing their numerical
solution with those obtained in the more conventional way, which
is to integrate RGEs for $Y_{l}$ and $ \kappa $ numerically, and
then to calculate left-handed neutrino masses, leptonic mixing
angles and CP-violating phases by diagonalizing these two matrices
at different energy scales.

In this section, we compare main features of the numerical result
with those predicted in Eqs.(\ref{anaMass})-(\ref{dSigma}). We try
to clarify what corrections are possible for neutrino parameters,
during the RG evolution from $M_{\rm Z}$ to $\Lambda_{\rm GUT}$.
Such a study is important in that, it helps us understand what
values are {\it possible} or even {\it preferred} for neutrino
parameters at the scale of Grand Unified theories. In the
numerical calculation, we follow a bottom-up procedure. We start
with the best fit values of neutrino parameters at the low energy,
and numerically integrate Eqs.(\ref{anaMass})-(\ref{dSigma}) to
obtain left-handed neutrino masses, leptonic mixing angles and
CP-violating phases at different high energy scales. Though such
an approach may not seem well motivated from the perspective of a
fundamental theory, it is quite advantageous for our present task.
In this way, we no longer have to tune parameters at the high
energy scale to meet low energy constraints.

At the present time, low energy experiments have measured leptonic
mixing angles and neutrino mass squared differences to a
reasonable degree of accuracy(best fit values and $3\sigma$
errors\cite{0405172}):
\begin{eqnarray} \label{data}
\theta_{x}\approx 33.2^\circ~_{-4.6^\circ}^{+4.8^\circ} ~;~~
\theta_{y}\approx 45.0^\circ~_{-9.3^\circ}^{+10.6^\circ} ~;~~
\theta_{z} < 13.0^\circ ~;  \nonumber \\
\Delta m_{21}^2 = m_2^2 - m_1^2 \approx
\left(7.9~_{-0.8}^{+1.0} \right) \times 10^{-5} {\rm eV}^2 ~;  \nonumber \\
\left|\Delta m_{31}^2\right| = \left|m_3^2 - m_1^2 \right| \approx
\left(2.2~_{-0.8}^{+1.1} \right) \times 10^{-3}{\rm eV}^{2} ~.
\end{eqnarray}
However, we still lack a lot of information. We do not know about
the smallest mixing angle $\theta_{z}$ and the absolute scale of
neutrino masses, except for a few upper bounds. And we know
nothing about leptonic CP-violating phases and the Yukawa coupling
matrix $Y_\nu$ at all. We can only speculate that $Y_\nu$ might
have some similarity to quark Yukawa coupling matrices. For such
reasons, it is not practical to scan the whole parameter space for
all possibilities.

On the other hand, earlier works\cite{9910231,9910420,0305273}
have emphasized that enhancing factors like $\zeta_{ij}^{-1}$ play
a significant role in the RG evolution of mixing angles and
CP-violating phases, if left-handed neutrinos are nearly
degenerate. So to be relevant and illustrative, we discuss in
detail the RG correction in theories with three typically
interesting neutrino mass patterns:
\begin{description}
\item[(i)] Normal Hierarchy: $m_{1}<<m_{2}<<m_{3}$.

With Eq.(\ref{data}), we find
\begin{equation}
m_{2}\simeq \sqrt{\Delta m_{21}^{2}}\simeq 0.009\mathrm{eV};\ \ \
\ \ m_{3}\simeq \sqrt{\left|\Delta m_{31}^{2}\right|}\simeq 0.047
\mathrm{eV.} \label{NormalMass}
\end{equation}
Then $m_{1}\lesssim 10^{-3}\mathrm{eV}$ is small enough to make
neutrino masses hierarchical:
\begin{eqnarray}
\zeta_{12}^{-1} &\approx &-1.25,\ \ \ \zeta_{13}^{-1}\approx
-1.0,\ \ \ \zeta_{23}^{-1}\approx -1.5\ \ \ :m_{1}=10^{-3}\mathrm{eV};  \nonumber \\
\zeta_{12}^{-1} &\approx &-1.0,\ \ \ \ \zeta_{13}^{-1}\approx
-1.0,\ \ \ \zeta_{23}^{-1}\approx -1.5\ \ \ :m_{1}\leq
10^{-4}\mathrm{eV}. \label{EnhancingNumberA}
\end{eqnarray}
Furthermore, from the determinant of Eq.(\ref{kappa})
\begin{equation}\label{SeesawScale}
M_3^{} =\frac{v_{}^2}{m_a^{}} \cdot \sqrt[3]{\varsigma y_\nu^2}
~;~~~ \varsigma \equiv \left(r_1^{} r_2^{}\right)^2 \left/
\left(\frac{m_x^{} m_y^{}}{ m_a^2} \frac{M_1^{} M_2^{}}
{M_3^2}\right)\right. ~,
\end{equation}
where $m_a^{} \equiv{\rm\bf max}\left[m_1,m_2,m_3\right]$, while
$m_x$ and $m_y$ denoting the other two lighter left-handed
neutrino masses. When $\varsigma \sim 1$, we have in the most
interesting case $y_\nu^{} \sim {\cal O}(1)$,
\begin{equation} \label{M3Value1}
M_{3}\sim \frac{174^{2}}{m_{3}}\mathrm{GeV}\simeq 6.4\times
10^{14}\mathrm{GeV}.
\end{equation}

\item[(ii)] Near Degeneracy: $m_{1}\lesssim m_{2}\lesssim m_{3}$.

Left-handed neutrinos are nearly degenerate if the absolute mass
scale is much larger than values given in Eq.(\ref{NormalMass}).
$\beta $ decay, $\beta \beta_{0\nu}$ decay, and cosmological and
astrophysical observations have all set upper bounds on certain
combinations of left-handed neutrino masses\cite {0412099}. A
rather stringent upper bound on nearly degenerate left-handed
neutrino masses is\cite{0412300}
\begin{equation}
m_{1,2,3}^{} <0.34\mathrm{eV.}
\end{equation}
If left-handed neutrino masses lie rightly beneath this bound, we
find that
\begin{eqnarray}
\zeta_{12}^{-1}\approx -5600,\ \ \ \zeta_{13}^{-1}\approx -201,\ \
\ \zeta_{23}^{-1}\approx -208 ~.  \label{EnhancingNumberB}
\end{eqnarray}
From Eq.(\ref{SeesawScale}), when $\varsigma \sim 1$ and $y_\nu^{}
\sim {\cal O}(1)$,
\begin{eqnarray} \label{M3Value2}
M_{3}\sim \frac{174^{2}}{m_{3}}\mathrm{GeV}\simeq 8.9\times
10^{13}\mathrm{GeV}\sim 10^{14}\mathrm{GeV}.
\end{eqnarray}
Note that the pattern $m_3\lesssim m_1 \lesssim m_2$ is also
possible, we shall consider this case in a different work.

\item[(iii)] Inverted Hierarchy: $m_{3}<<m_{1}\lesssim m_{2}$.

In this case
\begin{equation}
m_{1}\simeq \sqrt{| \Delta m_{31}^{2}|}\simeq 0.047 \mathrm{eV,}\
\ \ \ m_{2}\approx \sqrt{m_{1}^{2}+\Delta m_{21}^{2}}\simeq
0.048\mathrm{eV.}
\end{equation}
Then $m_{3}\lesssim 5\times 10^{-3}\mathrm{eV}$ is small enough to
make a hierarchy between itself and the other two masses:
\begin{eqnarray}
&& \zeta_{12}^{-1}\approx -112,\ \ \ \zeta_{13}^{-1}\approx 1.2,\
\ \ \zeta_{23}^{-1}\approx 1.2\ \ \ :m_{3}= 5\times
10^{-3}\mathrm{eV}; \nonumber \\
&& \zeta_{12}^{-1}\approx -111,\ \ \ \zeta_{13}^{-1}\approx 1.0,\
\ \ \zeta_{23}^{-1}\approx 1.0\ \ \ :m_{3}\leq 10^{-3}\mathrm{eV}.
\label{EnhancingNumberC}
\end{eqnarray}
Furthermore, when $\varsigma \sim 1$ and $y_\nu^{} \sim {\cal
O}(1)$,
\begin{equation}\label{M3Value3}
M_{3}\sim \frac{174^{2}}{m_{2}}\mathrm{GeV}\simeq 6.4\times
10^{14}\mathrm{GeV}.
\end{equation}
\end{description}
Note that in cases (i) and (iii), we shall always refer to the
first lines of Eqs.(\ref{EnhancingNumberA}) and
(\ref{EnhancingNumberC}) in our numerical calculation. And we
shall also need the value of $M_3$ (mostly in the factor $\ln
M_3$) in our discussions. Since varying $(\ln M_3)$ by one order
of magnitude requires changing the magnitude of $M_3$ by a factor
of ${\cal O} (10^4)$, the values of $M_3$ given in
Eqs.(\ref{M3Value1}), (\ref{M3Value2}) and (\ref{M3Value3}) are
"precise" enough for an order of magnitude estimation. However, it
should be stressed that although we always assume $y_\nu^2 \sim
{\cal O}(1)$, there can be significant errors in the subsequent
discussions if $\varsigma >>{ \cal O}(10^4)$ or $\varsigma <<{\cal
O}(10^{-4})$ in Eq.(\ref{SeesawScale}). But such errors are easily
corrected by taking into account the precise value of $\varsigma$.

In the following subsections, we firstly discuss the evolution
behavior of left-handed neutrino masses and of enhancing factors
$\zeta_{ij}^{-1}$ defined in Eq.(\ref{ratio}), then we study RG
corrections to leptonic mixing angles and CP-violating phases in
theories with each of the three neutrino mass patterns listed
above.

\subsection{RG Evolution of Neutrino Masses and Mass Ratios $\zeta_{ij}^{-1}$}

From Eq.(\ref{anaMass}), it is easy to obtain
\begin{equation}
\frac{m_{i}\left(\Lambda_{\rm GUT}\right)}{m_{i}\left(M_{\rm
Z}\right)} =\exp \left[ \frac{1}{16\pi ^{2}}\int_{t_{0}=\ln M_{\rm
Z}}^{t^{\prime }=\ln \Lambda_{\rm GUT}}a_i\left(t\right)
dt\right], \ \ \ i=1,2,3.
\end{equation}
Since $\alpha_\kappa$ dominates $a_i$ (for $i=1,2,3$) in general,
the RG evolution of left-handed neutrino masses is mainly governed
by a common scaling\cite{0110249,0305273}
\begin{equation}
\frac{m_{i}\left(\Lambda_{\rm GUT}\right)}{m_{i}\left(M_{\rm
Z}\right) }\simeq \exp \left[ \frac{1}{16\pi^{2}}\int_{t_{0}=\ln
M_{\rm Z}}^{t^{\prime}=\ln \Lambda_{\rm
GUT}}\alpha_\kappa\left(t\right) dt\right], \ \ \ i=1,2,3.
\end{equation}
There is a sudden change in the direction of the common scaling at
the point where $y_\nu\sim {\cal O}(1)$ is turned on, since
$y_\nu$ contributes to $\alpha_\kappa \left(t \right)$. This
feature is quite obvious in Figure \ref{FigMassCommonScaling},
where $m_{1}$ in cases (i) and (ii) and $m_{3}$ in case (iii) are
plotted as functions of the energy scale, normalized by their
values at $M_{\rm Z}$ .

Appreciable deviations from the common scaling may occur when
$y_{\tau}$ or $y_\nu$ is large. Below the seesaw threshold,
significant deviations are possible only in the MSSM when $\tan
\beta $ is large, in which case $y_{\tau} \sim {\cal O}(1)$. Above
the seesaw threshold, appreciable deviations are generally
possible since it is natural to have $y_\nu \sim {\cal O}(1)$. In
the following, we shall elaborate on this problem in a new but
more meaningful way, i.e. by studying the RG evolution of the
factors $\zeta_{ij}^{-1}$ (for $i<j, ~i,j=1,2,3$) defined in
Eq.(\ref{ratio}).

\subsubsection{Magnitudes of RG Corrections to $\zeta_{ij}^{-1}$}

As already mentioned, enhancing factors as $\zeta_{ij}^{-1}$ can
be very important in the RG evolution of mixing angles and
CP-violating phases. So it is necessary to study their own RG
evolution behavior. From Eqs.(\ref{ratio}) and (\ref{anaMass}), we
find
\begin{equation}
\frac{\dot{\zeta_{ij}^{-1}}}{\zeta_{ij}^{-1}}=-\frac{\dot{
\zeta}_{ij}} {\zeta_{ij}}= -\left(\zeta_{ij}^{-1}-\zeta_{
ij}\right) \frac{a_{i}-a_{j}}{2}\text{.} \label{dZeta}
\end{equation}
Obviously, $\zeta_{ij}$ would be constants if all left-handed
neutrino masses varied by an exactly common scaling, i.e.
$a_{i}=a_{j}$ (for $i,j=1,2,3$), which is of course not realistic.
To estimate magnitudes of possible corrections to $\zeta_{ij}$ (or
$\zeta_{ij}^{-1}$) from $M_{\rm Z}$ to $M_{3}$, we find
that\footnote{Throughout this work, we shall omit the difference
between $M_{\rm Z}$ and the SUSY breaking scale for simplicity.}
\begin{eqnarray}
\frac{\Delta \zeta_{ij}^{-1}}{\zeta_{ij}^{-1}} &\sim
&-\left(\zeta_{ij}^{-1}-\zeta_{ij}\right) \cdot
\frac{a_{i}-a_{j}}{2}\cdot \frac{\Delta t}{16\pi^{2}}\sim
-\left(\zeta_{ij}^{-1}-\zeta_{ij}\right)
\frac{C_\kappa^{l}y_{\tau}^{2}\ln \left(M_{3}/M_{\rm
Z}\right)}{16\pi^{2}}, \\ &\sim
&\left(\zeta_{ij}^{-1}-\zeta_{ij}\right) \cdot 10^{-5};\ \ \ \ \ \
\ \ \ \ \ \ \ \ \ \ \ \ \ \ \ \ \ \ \ \ \ :\mathrm{in\ the\ SM} \\
&\sim &-\left(\zeta_{ij}^{-1}-\zeta_{ij}\right) \cdot
\left(1+\tan^{2}\beta \right) \cdot 10^{-5},\ \ \ \ \ \
:\mathrm{in\ the \ MSSM}
\end{eqnarray}
where we have used Eq.(\ref{dZeta}) and have taken $y_{\tau}\sim
m_{\tau}\left(M_{\rm Z}\right)/\left(174
\mathrm{GeV}\right)\approx 0.01$ in the SM, and $y_{\tau}\approx
m_{\tau}\left(M_{\rm Z} \right)/ \left(\cos \beta \cdot 174
\mathrm{GeV}\right)\approx 0.01/\cos \beta$ in the MSSM.

With Eqs.(\ref{EnhancingNumberA}), (\ref{EnhancingNumberB}) and
(\ref{EnhancingNumberC}), we find that $\zeta_{ij}^{-1}$ vary
little from $M_{\rm Z}$ to $M_{3}$ in the SM, even in case (ii).
If Eq.(\ref{EnhancingNumberB}) is used, the variation of
$\zeta_{12}^{-1}$ is most significant: $| \Delta
\zeta_{12}^{-1}/\zeta _{12}^{-1}| \sim 5.6\times 10^{-2}$, while
variations of $\zeta_{13}^{-1}$ and $\zeta_{23}^{-1}$ are much
smaller: $ | \Delta \zeta_{13}^{-1}/\zeta_{13}^{-1}| \sim | \Delta
\zeta_{23}^{-1}/\zeta_{23}^{-1}| \sim 2\times 10^{-3}$.

In the MSSM, variations of $\zeta_{ij}^{-1}$ are amplified when $
\tan \beta $ is large. However, $\tan^2 \beta $ alone is not large
enough to generate appreciable variations of $\zeta_{ij}^{-1}$: $|
\Delta \zeta_{ij}^{-1}/ \zeta_{ij}^{-1} | \sim 2.5\times 10^{-2}$
in the case when $\tan \beta \sim 50$ but $| \zeta_{ij}^{-1}| \sim
{\cal O}(1)$. So corrections to $\zeta_{ij}^{-1}$ in case (i) are
always negligible from $M_{\rm Z}$ to $M_3$. Large variations of
$\zeta_{ij}^{-1}$ are possible only when $| \zeta_{ij}^{-1}| $ are
also large enough. In the limit $| \zeta_{ij}^{-1}|
\rightarrow\infty$, we find from Eq.(\ref{dZeta})
\begin{eqnarray}
\dot{\zeta}_{ij} \simeq \frac{a_{i}-a_{j}}{2}, \ \ \
&\Longrightarrow& \Delta \zeta_{ij} \sim \frac{a_{i}-a_{j}}
{2}\cdot \frac{\Delta t}{16\pi^{2}}\sim \frac{C_{\kappa
}^{l}y_{\tau}^{2}\ln \left(M_{3}/M_{\rm Z}\right)}{16\pi^{2}},
\label{dZeta2a} \\
&\Longrightarrow& \frac{\zeta_{ij}^{-1}\left(M_{\rm
Z}\right)}{\zeta _{ij}^{-1}\left(M_{3}\right)} = \frac{\zeta
_{ij}\left(M_{\rm Z}\right)+\Delta \zeta_{ij}}{\zeta
_{ij}\left(M_{\rm Z}\right)} \sim \frac{\left(1+\tan^{2}\beta
\right) \cdot 10^{-5}}{\zeta_{ij}\left(M_{\rm Z}\right)}.
\label{dZeta2}
\end{eqnarray}
When $\tan \beta \sim10$, only $\zeta_{12}^{-1}$ in case (ii) is
modified: $| \zeta_{12}^{-1}\left(M_{\rm Z} \right)/
\zeta_{12}^{-1}\left(M_{3} \right) | \sim 5.6$. In contrast,
significant variations of $\zeta_{ij}^{-1}$ are popular when $\tan
\beta \sim50$. Now there are roughly $|
\zeta_{12}^{-1}\left(M_{\rm Z}\right) /\zeta_{12}^{-1}\left(M_{3}
\right) | \sim 140$, $| \zeta_{13}^{-1} \left(M_{\rm Z} \right)
/\zeta_{13}^{-1}\left(M_{3}\right)| \sim |
\zeta_{23}^{-1}\left(M_{\rm Z} \right)/\zeta_{23}^{-1}
\left(M_{3}\right) | \sim 5$ in case (ii), and $|
\zeta_{12}^{-1}\left(M_{\rm Z}\right) /\zeta_{12}^{-1}
\left(M_{3}\right) | \sim 2.5$ in case (iii). It is spectacular
that $| \zeta_{12}^{-1}| $ is diminished to orders of magnitude
smaller at $M_{3}$ than at $M_{\rm Z}$ in case (ii). Numerically,
we find that the typical magnitude of $ \zeta_{12}^{-1}$ at
$M_{3}$ is of ${\cal O}(100)$, while $| \zeta_{13}^{-1}| $ and $|
\zeta_{23}^{-1}| $ are of a few tens\footnote{An interesting
exception to this simple estimation is that $\zeta_{23}^{-1}$ may
vary little from $M_{\rm Z}$ to $M_3$. Such a behavior occurs when
$\theta_x$ is swiftly diminished to stay with $\theta_z$ at very
near zero, while $\theta_y$ is kept at about $45^\circ$ during the
most part of the energy range from $M_{\rm Z}$ to $M_3$. One can
easily understand this point with Eqs.(\ref{dZeta2a}) and
(\ref{cZeta3}).}.

Since the common scaling of $k_{i}$ (for $i=1,2,3$) from $M_{\rm
Z}$ to $\Lambda_{\rm GUT}$ is only by a factor of ${\cal O}(1)$
and that $\Delta m_{21}^{2}\propto k_{2}^{2}-k_{1}^{2}=-\zeta_{12}
\left(k_{1}+k_{2}\right)^{2}$, the strong reduction in $|
\zeta_{12}^{-1}|$ means that $\Delta m_{21}^{2}$ is orders of
magnitude larger at $M_{3}$ than at $M_{\rm Z}$. This point
applies also to $| \Delta m_{32}^{2}| $ (or $|\Delta m_{31}^{2}|
$) when $| \zeta_{23}^{-1}| $ (or $ |\zeta_{13}^{-1}| $) is
strongly damped.

From $M_{3}$ to $\Lambda_{\rm GUT}$, $ y_\nu\sim {\cal O}(1)$
usually dominates corrections to $\zeta_{ij}^{-1}$ both in the SM
and in the MSSM with $\tan \beta \sim 10$, while $y_{\tau}$ is
comparable to $y_\nu$ in the MSSM when $\tan \beta \sim 50$. In
whichever case, the dominant contribution is of the order ${\cal
O}(y_\nu)$, so we can make an estimation as in Eq.(\ref{dZeta2})
\begin{equation}
\frac{\zeta_{ij}^{-1}\left(M_{3}\right)}{\zeta_{ij}^{-1}\left(\Lambda_{\rm
GUT}\right)}\sim 1+\frac{y_\nu^{2}\ln \left(\Lambda_{\rm
GUT}/M_{3}\right)}{16\pi^{2}\cdot \zeta_{ij}\left(M_{3}\right)
}\sim \frac{3\times 10^{-2}}{\zeta_{ij}\left(M_{3}\right)}:
~~~~~~~M_3\sim 10^{14} {\rm GeV}. \label{dZeta3}
\end{equation}
Obviously, corrections to $\zeta_{ij}^{-1}$ in case (i) are again
negligible from $M_3$ to $\Lambda_{\rm GUT}$. But it is remarkable
that substantial corrections to $\zeta_{ij}^{-1}$ are now possible
in the SM, both in cases (ii) and (iii). With
Eqs.(\ref{EnhancingNumberB}) and (\ref{EnhancingNumberC}) and also
considering the fact $| \zeta_{ij}^{-1}\left(M_{3}\right) |
\approx | \zeta_{ij}^{-1}\left(M_{\rm Z}\right) |$, we find $|
\zeta_{12}^{-1} \left(M_{3} \right)/ \zeta_{12}^{-1}
\left(\Lambda_{\rm GUT}\right)| \sim 170$, $|
\zeta_{13}^{-1}\left(M_{3}\right)/ \zeta_{ 13}^{-1}
\left(\Lambda_{\rm GUT} \right) | \sim | \zeta
_{23}^{-1}\left(M_{3}\right)/ \zeta_{23}^{-1} \left(\Lambda_{\rm
GUT}\right) | \sim 6$ in case (ii), and $|
\zeta_{12}^{-1}\left(M_{3}\right)/\zeta _{12}^{-1}
\left(\Lambda_{\rm GUT} \right)| \sim 3$ in case (iii). In the
MSSM when $\tan \beta \sim 10$, the situation is quite similar. In
this case only $\zeta_{12}^{-1}$ of case (ii) is moderately damped
from $M_{\rm Z}$ to $M_{3}$, but it still is a powerful enhancing
factor in Eq.(\ref{dZeta3}). For $\zeta_{13}^{-1}$ and
$\zeta_{23}^{-1}$ of case (ii) and for $\zeta_{12}^{-1}$ of case
(iii), the situation is the same as in the SM. In contrast, in the
MSSM when $\tan \beta \sim 50$, $| \zeta_{ij}^{-1}|$ in general
are strongly damped from $M_{\rm Z}$ to $M_3$. As a result, only
$\zeta_{12}^{-1}$ of case (ii) may vary significantly from $M_3$
to $\Lambda_{\rm GUT}$, by the ratio
$|\zeta_{12}^{-1}\left(M_{3}\right)/\zeta _{12}^{-1}
\left(\Lambda_{\rm GUT} \right)| \sim 3$. However, there can also
be appreciable modification in $\zeta_{13}^{-1}$ or
$\zeta_{23}^{-1}$ of case (ii) or in $\zeta_{12}^{-1}$ of case
(iii) if any of them is not so strongly damped from $M_{\rm Z}$ to
$M_3$.

\subsubsection{Signs of RG Corrections to $\zeta_{ij}^{-1}$}

Now we turn to discuss the signs of corrections to $\zeta
_{ij}^{-1}$. By Eq.(\ref{dZeta2a}), we find
\begin{eqnarray}
\zeta_{ij}\left(M_{3}\right) \approx \zeta_{ij}\left(M_{\rm
Z}\right) + \frac{a_{i}-a_{j}}{2}\cdot
\frac{\Delta t_{1}}{16\pi^{2}} ~; \\
\zeta_{ij}\left(\Lambda_{\rm GUT}\right) \approx
\zeta_{ij}\left(M_{3}\right) +\frac{a_{i}-a_{j}}{2}\cdot
\frac{\Delta t_{2}}{16\pi^{2}} ~,
\end{eqnarray}
where $\Delta t_{1}=\ln (M_{3}/M_{\rm Z})$, $\Delta t_{2}=\ln
(\Lambda_{\rm GUT}/M_{3})$ and
\begin{eqnarray}
\frac{a_{1}-a_{2}}{2} &=& -C_\kappa^{l}y_{\tau}^{2}\left[
\left(c_{x}^{2}-s_{x}^{2}\right)
\left(s_{y}^{2}-c_{y}^{2}s_{z}^{2}\right) +4c_{\delta
}^{}c_{x}c_{y}s_{x}s_{y}s_{z}\right] -C_\kappa^\nu y_{\nu
}^{2}\left(c_{2}^{2}s_{1}^{2}-s_{2}^{2}\right) ~;  \label{cZeta1} \\
\frac{a_{1}-a_{3}}{2} &=& -C_\kappa^{l}y_{\tau}^{2}\left[
c_{y}^{2}\left(c_{z}^{2}-c_{x}^{2}s_{z}^{2}\right)
-s_{x}^{2}s_{y}^{2}+2c_{\delta
}^{}c_{x}c_{y}s_{x}s_{y}s_{z}\right] -C_\kappa^\nu y_{\nu
}^{2}\left(c_{1}^{2}c_{2}^{2}-s_{2}^{2}\right) ~;  \label{cZeta2} \\
\frac{a_{2}-a_{3}}{2} &=& -C_\kappa^{l}y_{\tau}^{2}\left[
c_{y}^{2}\left(c_{z}^{2}-s_{x}^{2}s_{z}^{2}\right)
-c_{x}^{2}s_{y}^{2}-2c_{\delta
}^{}c_{x}c_{y}s_{x}s_{y}s_{z}\right] -C_\kappa^\nu y_{\nu
}^{2}\left(c_{1}^{2}-s_{1}^{2}\right)c_{2}^{2} ~.  \label{cZeta3}
\end{eqnarray}
In these equations, one should let $y_\nu=0$ at energies below the
seesaw threshold (or $M_{3}$ equivalently, in this work).
Left-handed neutrino masses will keep their order of sequence if
$(a_{i}-a_{j})/2$ are of the same signs as $\zeta_{ij}
\left(M_{\rm Z} \right)$ (for $i<j; ~i,j=1,2,3$, respectively).

From $M_{\rm Z}$ to $M_{3}$, signs of $(a_{i}-a_{j})/2$ are
determined by that of the $C_\kappa^{l}y_{\tau}^{2}$ term. As
estimated above, significant variations of $\zeta _{ij}^{-1}$ are
possible only in the MSSM, where $C_\kappa^{l}=1$. It has been
shown in Ref.\cite{0305273} that $\theta_{x}$ tends to evolve
toward zero whenever the correction is large; $\theta_{y}$ is
always smaller than but not far away from about $45^{\circ}$; and
$\theta_{z}$ is always smaller than about $15^{\circ}$. By this we
can conclude that the $C_\kappa^{l}y_{\tau}^{2}$ term in
Eqs.(\ref{cZeta1})-(\ref{cZeta3}) are usually \textit{negative},
and so that $| \zeta_{12}| $ of cases (ii) and (iii) is enlarged
with increasing energy scales in general. A positive correction to
$\zeta_{12}$ (which diminishes $|\zeta_{12}| $) is possible only
when $s_{z}$ is large and $c_\delta^{}$ is negative in
Eq.(\ref{cZeta1}), just as observed in Ref.\cite{0305273}.
However, it is numerically more difficult to arrange a positive
correction to $\zeta_{12}$ in the whole energy range from $M_{\rm
Z}$ to $M_{3}$ than a negative one. For the same reason,
$\zeta_{13}$ and $\zeta_{23}$ of case (ii) usually receive
corrections of their own signs during the RG evolution.

From $M_{3}$ to $\Lambda_{\rm GUT}$, the contribution from $y_\nu$
is important both in the SM and in the MSSM. In the SM, the
$y_\nu$ term dominates $ (a_{i}-a_{j})/2$. Since $\theta_{1}$ and
$\theta_{2}$ are totally arbitrary parameters, we can choose
$(a_{i}-a_{j})/2$ to be negative or positive at $M_{3}$ as we
like. In the MSSM, the $y_{\tau}$ term is also important when
$\tan \beta $ is large. But we still can change signs of
$(a_{i}-a_{j})/2$ by adjusting $\theta_{1}$ and $\theta_{2}$.
Furthermore, mixing angles and CP-violating phases often vary
dramatically at energies above $M_{3}$ when $| \zeta_{ij}^{-1}| $
are large. So $(a_{i}-a_{j})/2$ may also change their signs during
the evolution from $M_3$ to $\Lambda_{\rm GUT}$. In general, both
positive and negative $(a_{i}-a_{j})/2$ are possible at energy
scales above the seesaw threshold. However, as will be clear in
the following, signs of $\zeta_{ij}$ are not likely to be changed
despite this fact.

In Figure \ref{FigMassZeta}, we illustrate the typical evolution
behavior of  $\left(-\zeta_{ij}^{-1}\right) =| \zeta_{ij}^{-1}|$
in the SM and the MSSM. It is obvious that
$\left(-\zeta_{ij}^{-1}\right)$ vary in a way just as discussed
above.

\subsubsection{RG Evolution of $\zeta_{ij}^{-1}$ in Nearly Singular Situations}

As explained above, signs of $(a_{i}-a_{j})/2$ can be either
negative or positive from $M_3$ to $\Lambda_{\rm GUT}$, depending
mostly on $\theta_1$ and $\theta_2$. Immediately, there comes the
question of whether it is possible to change signs of $\zeta_{ij}$
at energies above $M_3$. We find that the answer is likely to be
negative.

When corrections to $\zeta_{ij}$ are positive, $|\zeta_{ij}|$ are
diminished and $|\zeta _{ij}^{-1}|$ may be enhanced to extremely
large values. Then the situation is nearly singular. However, we
find that $|\zeta _{ij}^{-1}|$ are always dramatically diminished
after they have reached some extremely large values. In such
cases, extremely high but very narrow peaks appear in the plot of
$|\zeta_{ij}^{-1}|$ against the energy scale. There seems to be a
protecting mechanism that keeps $|\zeta_{ij}^{-1}|$ from going to
infinity (or equivalently, keeps $\zeta_{ij}$ from going to zero).

One can understand this point most easily in the SM, in which
$y_{\tau}$ is small and so that Eqs.(\ref{cZeta1})-(\ref{cZeta3})
are dominated by the $C_\kappa^\nu y_{\nu }^{2}$ term. We also
need Eqs.(\ref{dTheta1}) and (\ref{dTheta2}) in the limit $\left
\vert \zeta_{ij}^{-1}\right \vert \rightarrow\infty$:
\begin{eqnarray}
\dot{\theta}_{1} &=&C_\kappa^\nu
y_\nu^{2}\left(-\frac{c_{\left(\phi_{1}-\phi_{2}\right)}^{2}s_{2}^{2}}{\zeta
_{12}}+\frac{c_{\phi_{1}}^{2}s_{2}^{2}}{\zeta_{13}}+\frac{c_{\phi
_{2}}^{2}c_{2}^{2}}{\zeta
_{23}}\right) c_{1}s_{1}+\cdot \cdot \cdot ;  \label{dTheta1b} \\
\dot{\theta}_{2} &=&C_\kappa^\nu
y_\nu^{2}\left(\frac{c_{\left(\phi_{1}-\phi_{2}\right)}^{2}s_{1}^{2}}{\zeta
_{12}}+\frac{c_{\phi_{1}}^{2}c_{1}^{2}}{\zeta_{13}}\right)
c_{2}s_{2}+\cdot \cdot \cdot . \label{dTheta2b}
\end{eqnarray}
It is crucial that, with $\zeta_{ij}^{-1}$ given in
Eqs.(\ref{EnhancingNumberB}) and (\ref{EnhancingNumberC}), the
dominant correction to $\theta_{2}$ is always \textit{negative},
while $\theta_{1}$ can either be enlarged by the $\zeta_{12}^{-1}$
term or be reduced by terms of $ \zeta_{13}^{-1}$ and
$\zeta_{23}^{-1}$. As a result, $\theta_{2}$ is always driven
toward $0^{\circ}$ when $| \zeta_{12}^{-1}| $ or $|
\zeta_{13}^{-1}| $ is large. In case (ii), since all three factors
$| \zeta_{ij}^{-1}| $ are large, $\theta_{1}$ can be driven toward
either $0^{\circ}$ or $90^{\circ}$, depending on the competition
among different terms. In case (iii), since only
$|\zeta_{12}^{-1}|$ is significant, $\theta_{1}$ is always driven
toward $90^{\circ}$.

Corrections to $\zeta_{ij}$ in case (i) are always negligible, so
we only need to consider cases (ii) and (iii). We shall firstly
discuss case (ii) in detail.

In Eqs.(\ref{cZeta1}), (\ref{cZeta2}) and (\ref{cZeta3}), a nearly
singular situation is most likely to occur when any of $\theta_{1}
\sim 0^{\circ}$ or $90^{\circ}$ or $\theta_{2}\sim 90^{\circ}$ is
satisfied: if there is only $\theta_{1}\sim 0^{\circ}$, a peak of
$| \zeta_{12}^{-1}| $ may appear; if there is only $\theta_{1}\sim
90^{\circ}$, peaks of $| \zeta_{13}^{-1}|$ and $|
\zeta_{23}^{-1}|$ may appear; and if there is only $\theta_2 \sim
90^{\circ}$, peaks of $| \zeta_{12}^{-1} |$, $| \zeta_{13}^{-1} |$
and $| \zeta_{13}^{-1} |$ are all possible. But since in case (ii)
\begin{equation}
\zeta_{13}^{-1}\approx\frac{1}{\zeta_{12}+\zeta_{23}} ~,
\end{equation}
a peak of $|\zeta_{13}^{-1}|$ should always coexist with peaks of
$|\zeta_{12}^{-1}|$ and $|\zeta_{23}^{-1}|$. Such a situation is
in fact quite rare.

Now we can discuss how signs of $\zeta_{ij}$ are protected.

Firstly, if $\zeta_{12}$ is driven to very near zero and
$|\zeta_{12}^{-1}|$ develops a peak, terms led by
$\zeta_{12}^{-1}$ dominate Eqs.(\ref{dTheta1b}) and
(\ref{dTheta2b}). Then $\theta_{2}$ is dramatically driven to near
$0^{\circ}$, while $\theta_{1}$ is swiftly enhanced toward
$90^{\circ}$. The smaller $| \zeta_{12}| $ is, the more efficient
this mechanism can take effect. As a result, $(a_{1}-a_{2})/2$ in
Eq.(\ref{cZeta1}) is quickly driven to be negative, leaving no
chance for $\zeta_{12}$ to reach zero or even to get its sign
changed.

Secondly, if $\zeta_{23}$ is driven to very near zero and
$|\zeta_{23}^{-1}|$ develops a peak for some reasons, the
$\zeta_{23}^{-1}$ term dominates Eq.(\ref{dTheta1b}). Then
$\theta_{1}$ is dramatically driven toward $0^{\circ}$. The
smaller $|\zeta _{23}|$ is, the more efficient this mechanism can
take effect. Since $\theta_{2}\sim 90^{\circ}$ is not preferred in
Eq.(\ref{dTheta2b}), $c_{2}^{2}\neq 0$ in general. As a result, $
(a_{2}-a_{3})/2$ in Eq.(\ref{cZeta2}) is quickly driven to be
negative, leaving no chance for $\zeta_{23}$ to reach zero or even
become positive.

Finally, if $|\zeta_{13}^{-1}|$ develops a peak (so there are also
peaks of $|\zeta_{12}^{-1}|$ and $|\zeta_{23}^{-1}|$), then
$\zeta_{12}^{-1}$ and $\zeta_{13}^{-1}$ in Eq.(\ref{dTheta2b}) can
dramatically drive $\theta_{2}$ to $0^{\circ}$. Since terms led by
$\zeta_{12}^{-1}$ and $\zeta_{13}^{-1}$ in Eq.(\ref{dTheta1b}) are
suppressed by $s_{2}^{2}$, the $\zeta_{23}^{-1}$ term then
dominates Eq.(\ref{dTheta1b}) and drives $\theta_{1}$ to
$0^{\circ}$. Then $(a_i-a_j)/2$ in Eqs.(\ref{cZeta1}),
(\ref{cZeta2}) and (\ref{cZeta3}) all become negative quickly. As
a result, none of $\zeta_{12}$, $\zeta_{13}$ and $\zeta_{23}$ will
vanish or even get its sign changed.

So it is not likely that signs of $\zeta_{ij}$ can be changed.
However, we should stress that such a discussion is only a way to
understand how signs of $\zeta_{ij}$ can always be preserved,
while the latter point is in fact not proved.

In case (iii), only $|\zeta_{12}^{-1}|$ is important. A nearly
singular situation is possible if $\theta_{1}\sim 0^{\circ}$ or
$\theta_{2}\sim 90^{\circ}$. At the peak of $|\zeta_{12}^{-1}|$,
$\theta_{2}$ is dramatically driven to $0^{\circ}$ and
$\theta_{1}$ is driven to $90^{\circ}$. Then there will be no more
peaks.

In the MSSM when $\tan \beta \sim 50$, $y_{\tau}$ is comparable to
$y_\nu$ and the contribution from $y_{\tau}$ is not negligible in
general. But numerically, we find that the mechanism discussed
above still works as long as $y_\nu$ is large.

Numerical examples illustrating these features are given in Figure
\ref{FigMassPeaks}.

\subsection{The Normal Hierarchy Case}

When left-handed neutrino masses are hierarchical, $|
\zeta_{ij}^{-1}| $ are of ${\cal O}(1)$. So there are no enhancing
factors in the evolution of mixing angles in
Eqs.(\ref{dX})-(\ref{dZ}), while the evolution of CP-violating
phases in Eqs.(\ref{dDelta})-(\ref{dSigma}) can still be enhanced
by the factor $s_{z}^{-1}$. In the limit $s_{z}\longrightarrow 0$:
\begin{eqnarray}
\dot\delta\approx \dot{\rho}\approx \dot{\sigma} &=&C_{\kappa
}^{l}y_{\tau}^{2}\left[ \left(\frac{c_{\left(\delta -\rho
\right)}^{}s_{\rho}^{}
}{\zeta_{13}}+\zeta_{13}c_{\rho}^{}s_{\left(\delta -\rho
\right)}^{}\right) -\left(\frac{c_{\left(\delta -\sigma
\right)}^{}s_{\sigma}^{}}{\zeta_{23}}+\zeta_{23}c_{\sigma
}^{}s_{\left(\delta -\sigma \right)}^{}\right) \right]
\frac{c_{x}c_{y}s_{x}s_{y}}{s_{z}}  \nonumber
\\
&&+C_\kappa^\nu y_\nu^{2}\left[ \left(\frac{c_{\phi
_{1}}^{}s_{\rho}^{}}{\zeta_{13}}+\zeta_{13}c_{\rho
}^{}s_{\phi_{1}}^{}\right)
c_{x}s_{2}+\left(\frac{c_{\phi_{2}}^{}s_{\sigma}^{}
}{\zeta_{23}}+\zeta_{23}c_{\sigma}^{}s_{\phi_{2}}^{}\right)
c_{2}s_{1}s_{x}\right] \frac{c_{1}c_{2}}{s_{z}}  \nonumber \\
&&+C_{l}^\nu y_\nu^{2}\left(s_{\left(\rho +\phi_{1}\right)
}^{}c_{x}s_{2}+s_{\left(\sigma +\phi_{2}\right)
}^{}c_{2}s_{1}s_{x}\right) \frac{c_{1}c_{2}}{s_{z}}+\cdots ,
\label{NormalDRS}
\end{eqnarray}
where we have neglected all terms that are not enhanced by
$s_z^{-1}$. It is remarkable that dominant corrections to $\delta
,\rho$ and $\sigma $ are exactly the same. It is also interesting
that the term proportional to $C_\kappa^{l}y_{\tau}^{2}$ vanishes
in the limit $ \zeta_{13}\approx \zeta_{23}\approx -1$, while
those led by $y_\nu^{2}$ become proportional to $\left(C_{l}^\nu
-C_\kappa^\nu \right)$. Since $ C_\kappa^\nu=C_{l}^\nu=1$ in the
MSSM, the contribution from $y_\nu$ also vanishes in the limit $
\zeta_{13}\approx \zeta_{23}\approx -1$ in this case.

\subsubsection{RG Corrections in the SM}

In the SM, $y_{\tau}\approx m_{\tau}\left(M_{\rm Z}\right)
/(174\mathrm{GeV})\approx 0.01$, and the typical contribution is
\begin{equation}
\frac{y_{\tau}^{2}}{16\pi^{2}}\ln \frac{\Lambda_{\rm GUT}}{M_{\rm
Z}}\left(\mathrm{rad}\right) \sim 10^{-5} \left( \mathrm{rad}
\right) \sim \left(10^{-3}\right)^{\circ}. \label{SM1}
\end{equation}
Sine we are mostly interested in cases with a large $y_\nu$, we
shall consider only $y_\nu\sim {\cal O}(1)$ in this work. When
$y_\nu\sim {\cal O}(1)$, the contribution is
\begin{equation}
\frac{y_\nu^{2}}{16\pi^{2}}\ln \frac{\Lambda_{\rm
GUT}}{M_{3}}\left(\mathrm{rad}\right) \sim \left\{\begin{array}{c}
0.014\left(\mathrm{rad}\right) \sim 1.1^{\circ}:\ \ \ \ M_{3}\sim
10^{15}
\mathrm{GeV} \\
0.029\left(\mathrm{rad}\right) \sim 1.9^{\circ}:\ \ \ \ M_{3}\sim
10^{14} \mathrm{GeV}
\end{array}
\right. .  \label{SM2}
\end{equation}
Apart from these, all other miscellaneous terms (besides
$\zeta_{ij}^{-1}$) on the right-hand side of
Eqs.(\ref{dX})-(\ref{dSigma}) usually damp the values given above
strongly. Through out this work, we shall assume that the net
effect of all these terms is equivalent to a factor of ${\cal
O}(0.1)$. Though such an assumption is mainly based on our
numerical experience and is far from precise, it serves as a crude
estimation and can help us understand the most important part of
RG corrections. With this assumption, we estimate that only
corrections of $\left(10^{-4}\right)^{\circ}$ are possible for
mixing angles running from $M_{\rm Z}$ to $M_3$, while corrections
of ${\cal O} \left(0.1^{\circ}\right) $ are possible from $M_3$ to
$\Lambda_{\rm GUT}$, if $y_\nu\sim {\cal O}(1)$.

For CP-violating phases, significant corrections are possible in
the energy range from $M_3$ to $\Lambda_{\rm GUT}$, since there is
the enhancing factor $s_{z}^{-1}\gtrsim5$.

Firstly, the RG correction to $\theta_{z}$ is negligible in the
whole range from $M_{\rm Z}$ to $\Lambda_{\rm GUT}$, when
$\theta_{z}$ is of ${\cal O} \left(1^\circ \sim 10^{\circ}\right)
$ at $M_{\rm Z}$. In this case, $s_z^{-1}\sim {\cal O} \left(5
\sim 50\right)$. From Eq.(\ref{SM2}) and the ${\cal O}
\left(0.1\right) $ factor explained above, corrections to
CP-violating phases are of ${\cal O} \left(0.5^{\circ}\sim
5^{\circ} \right)$.

Secondly, when $\theta_{z}$ is of ${\cal O} \left(0.1^{\circ }\sim
1^{\circ}\right)$ at $M_{\rm Z}$, the RG correction to it is not
negligible from $M_{3}$ to $\Lambda_{\rm GUT}$. However, the order
of magnitude of $s_{z}^{-1}$ is not changed by the correction. So
in this case, corrections to CP-violating phases are of ${\cal O}
\left(5^{\circ}\sim 50^{\circ} \right)$, given that $y_\nu \sim
{\cal O}(1)$.

Finally, though a value much smaller than possible radiative
corrections may not seem natural for $\theta_{z}$, we shall
consider the case $\theta_{z}<<0.1^{\circ}$ for completeness.
Since the radiative correction can enlarge $\theta_{z}$ to ${\cal
O} \left(0.1^{\circ}\sim 1^{\circ}\right) $ at energies above
$M_{3}$, corrections to CP-violating phases in this energy range
are roughly the same as in the second case. But if $\theta_{z}$ is
not magnified above $M_3$, the corrections can be extraordinarily
large. At energies below $M_{3}$, corrections from $y_{\tau}$ to
CP-violating phases are still negligible when $\theta _{z}$ is of
$\left(0.01^\circ \sim 0.1^{\circ}\right)$. But when
$\theta_{z}<0.01^{\circ}$, i.e. $s_{z}^{-1}>0.5\times 10^{4}$, the
contribution from $y_{\tau}$ can be so strongly enhanced as to
become appreciable or even important.

In Figure \ref{FigCase1SM}, we illustrate each of the
possibilities discussed above. It is obvious that the phases
$\delta, \rho$ and $\sigma $ in the figure always vary by
approximately the \textit{same} size, just as predicted in
Eq.(\ref{NormalDRS}).

\subsubsection{RG Corrections in the MSSM when $\tan \beta \sim 10$}

In the MSSM when $\tan \beta \sim 10$, $y_{\tau}\approx
m_{\tau}\left(M_{\rm Z}\right)/(\cos \beta \cdot 174
\mathrm{GeV})\approx 0.1$, and the typical contribution from which
is
\begin{equation}
\frac{y_{\tau}^{2}}{16\pi^{2}}\ln \frac{\Lambda_{\rm GUT}}{M_{\rm
Z}}\left(\mathrm{rad}\right) \sim 0.1^{\circ}. \label{MSSM10}
\end{equation}
The contribution form $y_\nu$ is the same as given in
Eq.(\ref{SM2}). Then corrections to mixing angles are of ${\cal
O}(0.01^{\circ})$ in the range from $M_{\rm Z}$ to $M_3$, but are
still of ${\cal O}(0.1^{\circ})$ from $M_3$ to $\Lambda_{\rm
GUT}$. For CP-violating phases, RG corrections from $M_{\rm Z}$ to
$ M_{3}$ can already be appreciable when $\theta_{z}$ is of $
{\cal O} \left(0.1^{\circ}\sim 1^{\circ}\right) $. This is in vast
contrast to the SM case, where to make\ the contribution from
$y_{\tau}$ appreciable, $\theta_{z}$ should be about one hundred
times smaller than what is given here.

Now we have a novel possibility that there can be large
corrections to CP-violating phases both at energies {\it above}
and {\it below} $M_{3}$. When $\theta_{z}\sim {\cal O}
\left(0.01^{\circ}\right)$ at $M_{\rm Z}$, the RG correction does
not change the order of magnitude of $\theta_{z}$ in the energy
range from $M_{\rm Z}$ to $M_3$, so $s_{z}^{-1}\sim {\cal O}
\left(5\times 10^{3}\right) $ can enhance corrections to
CP-violating phases to be of ${\cal O} \left(50^{\circ}\right)$.
At energies above $M_{3}$ , $\theta_{z}$ may acquire a correction
of ${\cal O}(0.1^{\circ})$ in general, but it is still of ${\cal
O}(0.01)$ at $M_3$. As a result, CP-violating phases can vary
dramatically above $M_3$, and corrections of ${\cal O}(50^\circ)$
are readily possible for them. In contrast, in the MSSM when $\tan
\beta \sim 50$, $\theta_{z}$ usually is magnified to be well above
$0.1^{\circ}$ in the energy range from $M_{\rm Z}$ to $M_3$. So
corrections (from terms led by $y_\nu^2$) to CP-violating phases
are less enhanced and the phases vary little at energies above
$M_{3}$. To conclude, it is only in the MSSM when $\tan \beta \sim
10$ that large corrections to CP-violating phases are most
probable, both at energies {\it below} and {\it above} $M_3$. This
observation is supported in Figure \ref{FigCase1MSSM}.

\subsubsection{RG Corrections in the MSSM when $\tan \beta \sim 50$}

In the MSSM when $\tan \beta \sim 50$, $y_{\tau} \approx m_{\tau}
\left(M_{\rm Z}\right) /(\cos \beta \cdot 174 \mathrm{GeV})\approx
0.5$, and the correction from which is
\begin{equation}
\frac{y_{\tau}^{2}}{16\pi^{2}}\ln \frac{\Lambda_{\rm GUT}}{M_{\rm
Z}}\left(\mathrm{rad} \right) \sim 2.9^{\circ}. \label{MSSM50}
\end{equation}
The correction from $y_\nu$ is still the same as given in
Eq.(\ref{SM2}). As a result, corrections of ${\cal O}
(0.3^{\circ})$ are possible for mixing angles running from $M_{\rm
Z}$ to $M_3$. From $M_3$ to $\Lambda_{\rm GUT}$, corrections to
mixing angles are still of ${\cal O}(0.1^{\circ})$. But for
CP-violating phases, RE corrections can be extraordinarily large
now.

Firstly, the RG correction does not change the order of magnitude
of $s_z^{-1}$ when $\theta_{z}$ is of ${\cal O}
\left(0.1^{\circ}\sim 10^{\circ}\right) $ at $M_{\rm Z}$. Then
corrections to CP-violating phases are of ${\cal O}(1.5^{\circ}),
~{\cal O}(15^{\circ})$ and ${\cal O}(150^{ \circ})$ when
$\theta_{z} \sim {\cal O}(10^{\circ}), ~{\cal O}(1^{\circ})$ and
${\cal O}(0.1^{\circ})$ respectively. When possible corrections
are as large as ${\cal O}(150^{\circ})$, CP-violating phases are
often driven to near their (pseudo-) fixed points: the phases keep
varying dramatically until the right hand side of
Eq.(\ref{NormalDRS}) vanishes, and then they become stable against
the energy scale. RG corrections to CP-violating phases can be
largely damped because of this behavior.

Secondly, when $\theta_{z}<<0.1^{\circ}$ at $M_{\rm Z}$, the RG
correction can enlarge $\theta_{z}$ to be of ${\cal O}
\left(0.1^{\circ} \sim 1^{\circ}\right)$ along the way from
$M_{\rm Z}$ to $M_{3}$. However, extraordinarily large corrections
to CP-violating phases can arise in a very narrow energy range at
near $M_{\rm Z}$, if $\theta_{z}$ is extremely small at the
beginning. In this case, (pseudo-) fixed points of CP-violating
phase are often swiftly reached, and then the phases evolve little
with the energy scale. Above $M_{3}$, the phases start to vary
again when the contribution from $y_\nu$ is turned on. However,
the dominant contribution still comes from $y_{\tau}$. This is
because $\theta_{z}$ is generally enlarged to ${\cal O}
\left(0.1^{\circ}\sim 1^{\circ}\right)$ below $M_{3}$, and so that
the contribution from $y_\nu$ is less enhanced.

Numerical examples illustrating these points are given in Figure
\ref{FigCase1MSSM}.

Apart from the magnitudes, signs of RG corrections to CP-violating
phases are also important. In Eq.(\ref{NormalDRS}), the signs are
determined by those of $C_\kappa^{l}$, $ C_\kappa^\nu$,
$C_{l}^\nu$, $\zeta_{13}$ and $\zeta_{23}$, and by values of
CP-violating phases. It may not worth while discussing all
possibilities, but one case is simple and interesting. In the SM,
since $C_\kappa^\nu=1/2$,$\ C_{l}^\nu=-3/2$ and since $\zeta_{13}$
and $ \zeta_{23}$ are negative, terms led by
 $y_\nu^2$ in Eq.(\ref{NormalDRS}) are always negative when both
$\rho +\phi_{1}$ and $\sigma +\phi_{2}$ are in the range
$\left(0\sim \pi \right) $, but the contribution from these terms
is positive when the range is $\left(\pi \sim 2\pi \right) $. One
can check this point with Figure \ref{FigCase1SM}.

\subsection{The Near Degeneracy Case}

When neutrino masses are nearly degenerate, corrections to mixing
angles and CP-violating phases can be resonantly enhanced since
enhancing factors $|\zeta_{ij}^{-1}|$ in
Eq.(\ref{EnhancingNumberB}) are large. Before the discussion of
any specific models, we remark that

\begin{itemize}
\item In Eqs.(\ref{dX})-(\ref{dSigma}), corrections to $\theta_x,
~\delta, ~\rho$ and $\sigma$ are enhanced by $\zeta _{12}^{-1}$,
$\zeta _{13}^{-1}$ and $\zeta _{23}^{-1}$, while corrections to
$\theta_y$ and $\theta_z$ are only enhanced by $\zeta _{13}^{-1}$
and $\zeta _{23}^{-1}$.

\item In Eqs.(\ref{dX})-(\ref{dSigma}), the factors $\zeta
_{ij}^{-1}$ appear in terms led by $y_\nu$ via three common
combinations: $\zeta_{12}^{-1}\cdot c_{2} s_{1} s_{2}
c_{\left(\phi_{1}-\phi_{2}\right)}$, $\zeta_{13}^{-1}\cdot c_{1}
c_{2} s_{2} c_{\phi_{1}}$ and $\zeta_{23}^{-1}\cdot c_{1}
c_{2}^{2} s_{1} c_{\phi_{2}}$. So there will be no resonantly
enhanced contribution from $y_\nu$ if $\theta_{1}=\theta_{2}=0$
(or $\theta_{1}=\theta_{2}=90^\circ$ etc).

\item Just like $c_{\phi_{1}}$, $c_{\phi_{2}}$ and
$c_{\left(\phi_{1}-\phi_{2}\right)}$ in the contribution from
$y_\nu$, certain functions of $\delta, \rho $ and $\sigma $ can
also be factorized out together with the enhancing factors
$\zeta_{ij}^{-1}$, both in contributions from $y_{\tau}$ and from
$y_\nu$. The result is collected in Table \ref{TableCase2a}. With
this, it is easy to find out what CP-violating phases can best
damp the resonantly enhanced correction to a specific mixing angle
or CP-violating phase.

\item In the contribution from $y_\nu$ in
Eqs.(\ref{dX})-(\ref{dSigma}), the association of
$\zeta_{ij}^{-1}$ with CP-violating phases is rather simple. So it
is easy to tell signs of corrections enhanced by different factors
$\zeta_{ij}^{-1}$. We collect the result in Table
\ref{TableCase2b}, where we have assumed that minus signs in
Eqs.(\ref{dX})-(\ref{dSigma}) belong to phase factors.
\end{itemize}

In the following we discuss RG corrections to mixing angles and
CP-violating phases both in the SM and in the MSSM.

\subsubsection{RG Corrections in the SM}

From $M_{\rm Z}$\ to $M_{3}$, RG corrections are determined by the
contribution from $y_{\tau}$. With the help of
Eqs.(\ref{EnhancingNumberB}) and (\ref{SM1}) and also considering
the overall ${\cal O} \left(0.1\right)$ factor explained below
Eq.(\ref{SM2}), we estimate that the correction (enhanced by
$\zeta_{12}^{-1}$) to $\theta_{x}$ is of ${\cal O} \left(0.5^{
\circ} \right)$, and corrections (enhanced by $\zeta_{13}^{-1}$
and $\zeta_{23}^{-1}$) to $\theta_{y}$ and $\theta_{z}$ are of
${\cal O} \left(0.02^{\circ}\right)$.

From $M_{3}$\ to $\Lambda_{\rm GUT}$, the contribution from
$y_\nu\sim {\cal O}(1)$ is dominant and the magnitude is given in
Eq.(\ref{SM2}). Since $\zeta_{ij}^{-1}(M_{3}) \approx
\zeta_{ij}^{-1}(M_{\rm Z})$, the correction (enhanced by
$\zeta_{12}^{-1}$) to $\theta_{x}$ is of ${\cal O}
\left(1100^{\circ}\right) $, and corrections (enhanced by
$\zeta_{13}^{-1}$ and $\zeta_{23}^{-1}$) to $\theta_{y}$ and
$\theta_{z}$ are of ${\cal O} \left(40^{\circ}\right)$. The ${\cal
O} \left(40^{\circ}\right) $ correction to $\theta_{z}$ is a quite
generous gift: a correction of ${\cal O} \left(1^{\circ}\right)$
could be interesting when $\theta_z$ is really small at $M_{\rm
Z}$\cite{0301234,0305273,0404081}. We shall discuss this problem
in more detail in the MSSM when $\tan \beta \sim50$ (within this
subsection, i.e. also for the near degeneracy case). For
$\theta_{x}$, however, the ${\cal O} \left(1100^{ \circ}\right)$
correction is overestimated. As we have discussed in the first
subsection and also is vivid in Figure \ref{FigMassZeta}, in a
large part of the energy range from $M_{3}$ to $\Lambda_{\rm
GUT}$, $| \zeta_{12}^{-1}| $ is more than an order of magnitude
smaller\ than it is at $M_{3}$. So the correction enhanced by
$\zeta_{12}^{-1}$ should be about an order smaller in magnitude
than estimated above.

In cases when there are nearly singular situations, i.e. when $|
\zeta _{ij}^{-1}| $ develop peaks, the ``protective mechanism"
discussed in the first subsection becomes important again in
damping extraordinarily enhanced corrections to mixing angles and
CP-violating phases. As mentioned in the beginning of this
subsection, corrections from $y_\nu$ that can be enhanced by
$|\zeta_{ij}^{-1}|$ are always controlled by $s_{1}$ and $s_{2}$:
i.e. $\zeta_{12}^{-1}$ by $ s_{1}s_{2}$, $\zeta_{13}^{-1}$ by
$s_{2}$ and $\zeta_{23}^{-1}$ by $s_{1}$. $\theta_{1}$ is
dramatically driven to near zero whenever $| \zeta _{23}^{-1}| $
develops a peak, and $\theta_{2}$ is dramatically driven to near
zero whenever $|\zeta_{12}^{-1}| $ or $|\zeta _{13}^{-1}| $
develops a peak. As a result, enhancing effect of all peaks of
$|\zeta_{ij}^{-1}|$ are strongly damped. However, the contribution
from $y_\nu$ can still be very large in such a case (numerically,
we find that a value of ${\cal O}(100^{\circ}$) is popular).

Signs of corrections to mixing angles are determined by
CP-violating phases $\delta, \rho ,\sigma ,\phi_{1}$ and
$\phi_{2}$, and the competition among different contributions led
by $\zeta _{12}^{-1}$, $\zeta_{13}^{-1}$ and $\zeta_{23}^{-1}$.
For an interesting example, we discuss in detail the case when all
of the five phases are in the range $\left(0,\pi/2\right) $. As
discussed above, only the contribution from $y_\nu$ at energies
above $M_3$ is important.

\begin{itemize}
\item For $\theta_{x}$: In Eq.(\ref{dX}), the contributions from
$y_\nu$ led by $\zeta_{12}^{-1}$ and $\zeta_{13}^{-1}$ are
\textit{positive}, while that led by $\zeta_{23}^{-1}$ is
\textit{negative}. Since $| \zeta_{12}^{-1}| $ is generally larger
than $|\zeta_{13}^{-1}|$ and $|\zeta _{23}^{-1}|$, the net
correction to $\theta_{x}$ is positive in general, at near
$M_{3}$. However, since $\theta_{2}$ usually is more swiftly
driven to near zero than $\theta_{1}$, negative contributing terms
led by $\zeta_{23}^{-1}$ in Eq.(\ref{dX}) often have a chance to
catch up with the positive contributions led by $\zeta_{12}^{-1}$
and $\zeta_{13}^{-1}$. As a result, sometimes $\theta_{x}$ may
make a change of its evolution direction and evolve toward zero.

\item For $\theta_{y}$: In Eq.(\ref{dY}), the part of the
correction from $y_\nu$ led by $\zeta_{13}^{-1}$ is
\textit{negative} but that led by $\zeta_{23}^{-1}$ is
\textit{positive}. Since $| \zeta_{13}^{-1}| $ and $ |
\zeta_{23}^{-1}| $ are comparable to each other in general, the
two parts of the correction usually cancel each other to a large
extent. So the net correction to $\theta_{y}$ is relatively small.
Furthermore, since $\theta_{2}$ is often more swiftly diminished
than $\theta_{1}$, the part led by $\zeta_{23}^{-1}$ often is
dominant and the correction to $\theta_y$ is positive.

\item For $\theta_{z}$: In Eq.(\ref{dZ}), the correction from
$y_\nu$ to $\theta_z$ is \textit{positive}. So $\theta_{z}$ only
{\it increases} with the energy scale in this case. This feature
leads to a quite interesting possibility that $\theta_{z}$ is
comparable to the other two angles at $\Lambda_{\rm GUT}$, but it
is diminished when the energy scale is decreased.

\end{itemize}

Numerical examples illustrating these features are given in Figure
\ref{FigCase2SM}. In the figure, $\theta_y$ is decreased above
$M_3$. This is because $\delta>\pi/2$ in that energy range.

For CP-violating phases, from $M_{\rm Z}$\ to $M_{3}$, the factors
$| \zeta_{ij}^{-1}| $ alone can enhance corrections from $
y_{\tau}$ to be of ${\cal O} \left(0.5^{\circ}\right)$. Large
corrections to CP-violating phases are possible only when
$\theta_{z}$ is small enough. To make this point clear, we find
(similar to Eq.(\ref{NormalDRS})) in the limit $s_z \rightarrow
0$:
\begin{eqnarray}
\dot\delta\approx \dot{\rho}\approx \dot{\sigma} &=&C_{\kappa
}^{l}y_{\tau}^{2}\left[\frac{c_{\left(\delta -\rho
\right)}^{}s_{\rho}^{}}{\zeta_{13}} -\frac{c_{\left(\delta -\sigma
\right)}^{}s_{\sigma}^{}}{\zeta_{23}}\right]
\frac{c_{x}c_{y}s_{x}s_{y}}{s_{z}}   \nonumber \\
&&+C_\kappa^\nu y_\nu^{2}\left[\frac{c_{\phi _{1}}^{}s_{\rho}^{}
c_{x}s_{2}}{\zeta_{13}}+ \frac{c_{\phi_{2}}^{}s_{\sigma}^{}
c_{2}s_{1}s_{x}}{\zeta_{23}} \right] \frac{c_{1}c_{2}}{s_{z}}   \nonumber \\
&&+C_{l}^\nu y_\nu^{2}\left(s_{\left(\rho +\phi_{1}\right)
}^{}c_{x}s_{2}+s_{\left(\sigma +\phi_{2}\right)
}^{}c_{2}s_{1}s_{x}\right) \frac{c_{1}c_{2}}{s_{z}}+ \cdots \ ,
\label{DgDRS}
\end{eqnarray}
where we have omitted terms that are not enhanced by $s_z^{-1}$
(including those led by $\zeta_{12}^{-1}$). The contribution from
$y_\tau$ can be enhanced to ${\cal O} \left(0.02^{\circ}\right)$
by $\zeta_{13}^{-1}$ and $\zeta_{23}^{-1}$, so $\theta_{z}\sim
{\cal O} \left(0.1^{\circ}\right) $ is small enough to help
generating corrections of ${\cal O} \left(10^{\circ}\right) $.
However, since $|\zeta_{13}^{-1}|$ and $|\zeta _{23}^{-1}|$ are
comparable to each other in general, corrections from $y_\tau$ are
strongly damped if $\rho \approx \sigma $.

From $M_{3}$ to $\Lambda_{\rm GUT}$, $s_{z}^{-1}$ in general is
not an enhancing factor since a correction of ${\cal O}
\left(40^{\circ }\right) $ is possible to $\theta_{z}$. With
$\zeta_{ij}^{-1}$ in Eq.(\ref{EnhancingNumberB}) being the only
enhancing factors, the situation for CP-violating phases is much
like that for mixing angles discussed above. So in this energy
range, corrections of ${\cal O} \left(100^{\circ}\right)$ are
possible for CP-violating phases. However, there is still a
notable difference between the evolution of CP-violating phases
and that of mixing angles. At energies right above $M_{3}$,
$\theta_{z}$ is still very small. So the enhancing effect of
$s_{z}^{-1}$ combined with that of $\zeta_{13}^{-1}$ and
$\zeta_{23}^{-1}$ in Eq.(\ref{DgDRS}) can drive CP-violating
phases to vary almost abruptly. Corrections of this origin are
distinguishable since they are enhanced by $s_z^{-1}$ and thus are
exactly the same for $\delta, \rho$ and $\sigma$. Only after
$\theta_{z}$ has grown large enough, will the evolution of
CP-violating phases slow down and acquire a strength comparable to
that of mixing angles.

The evolution of CP-violating phases is also illustrated in Figure
\ref{FigCase2SM}. In the figure, the difference between the RG
evolution of mixing angles and that of CP-violating phases (as
discussed above) is obvious.

\subsubsection{RG Corrections in the MSSM when $\tan \beta \sim 10$}

In this case, the contribution of $y_{\tau}$ is given in
Eq.(\ref{MSSM10}). From $M_{\rm Z}$ to $M_{3}$, $\zeta_{12}^{-1}$
in Eq.(\ref{EnhancingNumberB}) can enhance the correction to
$\theta_{x}$ to be of $ {\cal O} \left(50^{\circ}\right)$, while
corrections to $\theta_{y}$ and $\theta_{z}$ (enhanced by
$\zeta_{13}^{-1}$ and $\zeta_{23}^{-1}$) are of ${\cal O}
\left(2^{ \circ}\right)$. Here, the ${\cal O}
\left(50^{\circ}\right) $ correction to $\theta_{x}$ is usually
\textit{negative} and will be damped when $\theta_{x}$ is near
zero. A positive correction to $\theta_{x}$ is also possible. But
such a correction is always suppressed by factors of ${\cal O}
\left(s_{z}\right) $. Considering the upper bound for
$s_{z}\left(\lesssim 1/5\right) $, a positive correction to
$\theta_{x}$ should be smaller than ${\cal O}
\left(10^{\circ}\right) $. The smaller $\theta_{z}$ is, the
smaller such a correction would be. If we take $ \theta_{z}\sim
2^{\circ}$, the correction should be smaller than ${\cal O}
\left(2^{\circ}\right)$. For $\theta_{y}$, the situation is
similar to that of $\theta_x$. The dominant terms are usually
negative, while a positive correction is possible but is always
suppressed by $s_{z}$. For $\theta_{z}$, the correction can be
either positive or negative, depending on values of CP-violating
phases and on the competition among the terms led by
$\zeta_{13}^{-1}$ and $\zeta_{23}^{-1}$.

From $M_{3}$\ to $\Lambda_{\rm GUT}$, $y_\nu$ dominates the
corrections and the situation for mixing angles is quite similar
to that in the SM. A notable difference is that, now one can start
the evolution with a small $\theta_{x}$ at $M_3$, since it has a
good chance of being damped from $M_{\rm Z}$ to $M_{3}$.

For CP-violating phases, $\zeta_{12}^{-1}$ alone can enhance the
corrections to ${\cal O} \left(50^{\circ}\right) $, in the energy
range from $M_{\rm Z}$ to $M_{3}$. But this correction is strongly
damped when $\rho \approx\sigma$. Alternative large corrections
can come from terms led by $\zeta_{13}^{-1}$ and
$\zeta_{23}^{-1}$, in cases when $\theta_{z}$ is very small. The
corresponding RGEs are the same as those in Eq.(\ref{DgDRS}).
Since dominant corrections to $\delta, \rho $ and $\sigma $ are
exactly the same, the relation $\rho \approx \sigma$ can be easily
retained in this case. CP-violating phases are often driven to
near their (pseudo-) fixed points when $\theta_{z}$ is extremely
small at $M_{\rm Z}$.

From $M_{3}$ to $\Lambda_{\rm GUT}$, corrections to CP-violating
phases are dominated by $y_\nu$ and the situation is quite similar
to that in the SM.

For a numerical illustration and also as a check of the third
remark at the beginning of this subsection, we stress that
$\theta_{x}$ and the phases $\delta $, $\rho $ and $\sigma $
cannot acquire their largest possible corrections simultaneously.
For example, one needs $\rho \approx \sigma $ to make the
$\zeta_{12}^{-1}$ term in Eq.(\ref{dX}) least suppressed, so that
the correction to $\theta_{x}$ can be largest. However, in Table
\ref{TableCase2a} and Eqs.(\ref{dDelta})-(\ref{dSigma}), the
condition $\rho \approx \sigma $ damps corrections (enhanced by
$\zeta_{12}^{-1}$) to CP-violating phases most strongly. Also,
this $\rho \approx \sigma $ condition damps the RG correction to
$\theta_{z}$ in the range from $M_{\rm Z}$ to $M_{3}$. One can
understand this point with the help of Eq.(\ref{dZ}) and the fact
that $\zeta_{13}^{-1}$ and $\zeta_{23}^{-1}$ are usually
comparable to each other. These results are all illustrated in
Figure \ref{FigCase2MSSM10}. Note that, since we take $\rho
\approx \sigma $ in the calculation, the correction to
$\theta_{z}$ is largely damped and $\theta_{z}$ is kept unchanged
until $M_{3}$. As a result, CP-violating phases vary almost
abruptly at the point where $y_\nu$ is turned on, just as in the
case of the SM.

\subsubsection{RG Corrections in the MSSM when $\tan \beta \sim 50$}

In this case, the contribution of $y_{\tau}$ is given in
Eq.(\ref{MSSM50}). Along with the usual ${\cal O}(0.1)$ factor
explained below Eq.(\ref{SM2}), $\zeta_{12}^{-1}$ in
Eq.(\ref{EnhancingNumberB}) can enhance the correction to
$\theta_{x}$ to ${\cal O} \left(1700^{\circ}\right)$, in the
energy range from $M_{\rm Z}$ to $M_{3}$. The other two factors
$\zeta_{13}^{-1}$ and $\zeta_{23}^{-1}$ in
Eq.(\ref{EnhancingNumberB}) can enhance corrections to
$\theta_{y}$ and $\theta_{z}$ to ${\cal O}
\left(60^{\circ}\right)$. However, these values are overestimated.
We have shown in the first subsection that, $\vert
\zeta_{ij}^{-1}\vert$ usually are strongly reduced from $M_{\rm
Z}$ to $M_{3}$. In Figure \ref{FigMassZeta}, $\zeta_{12}^{-1}$ is
more than an order smaller in magnitude than it is at $M_{\rm Z}$,
during a large part of the energy range from $M_{\rm Z}$ to
$M_{3}$. So in general, the correction to $\theta_{x}$ should also
be an order of magnitude smaller than estimated above. The
variation of $\vert \zeta_{13}^{-1}\vert$ and
$\vert\zeta_{23}^{-1}\vert$ is more moderate and the strength of
the reduction is more uniform in the whole energy range than those
of $\vert\zeta_{12}^{-1}\vert$. For such reasons, we re-estimate
that the correction to $\theta_{x}$ is roughly of ${\cal O}
\left(170^{\circ}\right) $, while those to $\theta_{ y}$ and
$\theta_{z}$ are roughly of ${\cal O} \left(30^{ \circ} \right)$.

Furthermore, (pseudo-) fixed points are always possible for mixing
angles and CP-violating phases whenever corrections are very
large. When certain angles or phases are near their fixed points,
RG corrections to them are strongly damped. As a result, real
corrections to mixing angles and CP-violating phases depend not
only on values estimated above, but also on how (pseudo-) fixed
points can be reached. This is also true for RG corrections above
$M_{3}$.

From $M_{3}$\ to $\Lambda_{\rm GUT}$, contributions from $y_\nu$
and  $y_{\tau}$ are both important. As estimated in the first
subsection, since $| \zeta_{12}^{-1}| $ is generally orders of
magnitude smaller at $M_{3}$ than at $M_{\rm Z}$, and since $|
\zeta_{13}^{-1}| $ and $| \zeta_{23}^{-1}| $ are about $5$ times
smaller at $M_{3}$ than at $M_{\rm Z}$, the correction to $\theta
_{x}$ is roughly of ${\cal O} \left(100^{\circ}\right) $, while
those to $\theta_{y}$ and $\theta_{z}$ are of ${\cal O}
\left(10^{\circ}\right) $.

\begin{itemize}
\item For $\theta_{x}$: From $M_{\rm Z}$\ to $M_{3}$, the dominant
contribution in Eq.(\ref{dX}) (that enhanced by $\zeta_{12}^{-1}$,
but not suppressed by $s_{z}$) is always {\it negative}. But if
$\theta_{z}$ is large, there can be a positive correction. The
correction to $\theta_{x}$ becomes largest when $|c_{\left(\rho
-\sigma \right)}| \sim 1$. Since $\rho =\sigma =0$ is stable
against RG corrections, this condition is easy to retain.
Furthermore, $|c_{\left(\rho -\sigma \right)}|\sim 1$ can also
lead to a large correction to $\theta_x$ in the energy range from
$M_{3}$\ to $\Lambda_{\rm GUT}$, if there is in addition $|
c_{\left(\phi_{1}-\phi_{2}\right)}| \sim 1$. In contrast, if we
want the correction to $\theta_{x}$ to be small, we need both
$c_{\left(\rho -\sigma \right)}\sim 0$ and
$c_{\rho}c_{\phi_{1}}\sim c_{\sigma}c_{\phi_{2}}\sim 0$. These
requirements can only be partially satisfied by $\delta =\rho
=\phi_{1}=0$ (up to $\pi $) and $\sigma =\phi_{2}=\pi/2$ (up to
$\pi $), which are also stable against RG corrections.

\item For $\theta_{y}$: From $M_{\rm Z}$\ to $M_{3}$, dominant
contributions in Eq.(\ref{dY}) (those enhanced by
$\zeta_{13}^{-1}$, $ \zeta_{23}^{-1}$, but not suppressed by
$s_{z}$) are also \textit{negative}. Though a positive correction
seems possible in Eq.(\ref{dY}) when $\theta_{z}$ is large, a
numerical example is hard to find. The correction to $\theta_{y}$
becomes largest when $| c_{\left(\delta -\rho \right)}| \sim |
c_{\left(\delta -\sigma \right)}| \sim 1$ below $M_{3}$, and $|
c_{\left(\delta -\rho \right)}c_{\phi_{1}}| \sim | c_{\left(\delta
-\sigma \right)}c_{\phi_{2}}| \sim 1$ (these two terms should be
in opposite signs) above $M_{3}$. These conditions can be
satisfied, e.g. by $\delta =\rho =\sigma =\phi_{1}=0$ and
$\phi_{2}=\pi$, which are stable against RG corrections. In
contrast, the correction is smallest both at energies above and
below $M_{3}$ when $c_{\left(\delta -\rho \right)}\sim
c_{\left(\delta -\sigma \right)}\sim 0$. This condition can be
satisfied, e.g. by $\delta =0$ and $\rho =\sigma =\pi/2$, which
are stable from $M_{\rm Z}$ to $\Lambda_{\rm GUT}$ if there is
$\phi_{1}=\phi _{2}=\pi/2$ in addition.

\item For $\theta_{z}$: The sign of the dominant correction to
$\theta_{z}$ depends on CP-violating phases and on the competition
among terms led by $\zeta_{13}^{-1}$ and $\zeta_{23}^{-1}$ in
Eq.(\ref{dZ}). From $M_{\rm Z}$ to $M_{3}$, the ${\cal O}
\left(30^{\circ}\right) $ correction to $\theta_{z}$\ is quite
spectacular: it means that a too small value is no longer natural
for $\theta_{z}$. However, there is still a notable exception: if
$c_{\rho}\sim c_{\sigma}\sim 0$ in Eq.(\ref{dZ}), the correction
to $\theta_{z}$ is strongly damped both at energies above and
below $M_{3}$, and so $\theta_{z}$ can be kept at a small value.
Only in this case, can a tiny $\theta_{z}$ be probable. However,
for this to happen, $\rho$ and $\sigma$ must also be stable
against RG corrections. In Eqs.(\ref{dDelta})-(\ref{dSigma}) and
Eqs.(\ref{dDeltaNu})-(\ref{dPhi2}), we find that if $s_{\left(\rho
-\sigma \right)}\sim c_{\left(\delta -\rho \right)}\sim
c_{\left(\delta -\sigma \right)}\sim c_{\phi_{1}}\sim
c_{\phi_{2}}\sim 0$, corrections to $\delta , \rho, \sigma,
\phi_{1}$ and $\phi_{2}$ can all be strongly damped, and so that
$\delta, \rho, \sigma, \phi_{1}$ and $\phi_{2}$ are all stable
against RG corrections. Up to $\pi $, these conditions mean $
\delta =0$ and $\rho =\sigma =\phi_{1}=\phi_{2}=\pi/2$. In
contrast, the correction to $\theta_{z}$ can be large both at
energies above and below $M_{3}$, if $c_{\rho}c_{\left(\delta
-\rho \right)}\approx -c_{\sigma}c_{\left(\delta -\sigma \right)}$
and $c_{\rho}c_{\phi_{1}}\approx c_{\sigma}c_{\phi_{2}}$ , and all
of them being of ${\cal O}(1)$ in magnitude. A simple but
interesting phase configuration that can satisfy these conditions
is $\delta =\rho + \pi/4=\sigma -\pi/4=\pi/2$ and
$\phi_{1}=\phi_{2}-\pi =0$ (which, however, is not stable against
RG corrections).
\end{itemize}

For CP-violating phases, possible corrections are roughly of
${\cal O} \left(100^{\circ}\right)$ both at energies above and
below $M_{3}$, if only the factors $\zeta _{ij}^{-1}$ are taken
into account. But if the smallness of $\theta_{z}$ is retained
during a small energy range, there can be extraordinarily large
corrections to CP-violating phases. As in previous cases, such
corrections often drive CP-violating phases to near their
(pseudo-) fixed points dramatically. However, CP-violating phases
usually need to be kept at special values if one wants to damp all
large corrections to $\theta_{z}$, just as we have mentioned
above. So in such special cases, there should not be any large
corrections to CP-violating phases, though $\theta_z$ may be tiny
in a wide energy range such as from $M_{\rm Z}$ to
$\Lambda_{GUT}$.

\begin{itemize}
\item For $\delta $: Eq.(\ref{dDelta}) for the running of $\delta
$ is rather complicated. But much simplified approximation can be
obtained in the limit $ \theta_{z}\rightarrow 0$, which is given
in Eq.(\ref{DgDRS}). A notable feature is that, only terms led by
$\zeta_{13}^{-1}$ and $\zeta_{23}^{-1}$ are possibly enhanced by
$s_{z}^{-1}$. The contribution from these terms can be dominant
when $\theta_{z}$ is small enough ($ \theta_{z}<$ $|
\zeta_{13}^{-1}/\zeta_{12}^{-1} | \sim |
\zeta_{23}^{-1}/\zeta_{12}^{-1} | \sim {\cal O}
\left(1^{\circ}\right) $).

\item For $\rho $ and $\sigma $: Eqs.(\ref{dRho}) and
(\ref{dSigma}) for the running of $\rho $ and $\sigma $ are also
quite complicated. But the dominant contribution from $ y_\nu$ is
now simple. We can predict signs of corrections enhanced by
different factors $\zeta_{ij}^{-1}$ with the help of Table
\ref{TableCase2b}, where the association of enhancing factors
$\zeta_{ij}^{-1}$ with CP-violating phases is clearly shown.
Furthermore, much simplified approximations of Eqs.(\ref{dRho})
and (\ref{dSigma}) can be obtained in the limit $\theta
_{z}\rightarrow 0$ and the results are also the same as given in
Eq.(\ref{DgDRS}).
\end{itemize}

In Figure \ref{FigCase2MSSM50}, we illustrate cases in which the
correction to a specific mixing angle is mostly enhanced or
damped, just as discussed above. But the corrections shown are not
largest in general. There can be larger corrections when specially
chosen CP-violating phases are used. However, this is not our main
concern here. What we want to demonstrate is that a good
prediction of RG corrections can often be made with the help of
Eqs.(\ref{dX})-(\ref{dSigma}). For CP-violating phases, the
situation is quite complicated and few general conclusions
regarding their RG evolution can be reached. In Figure
\ref{FigCase2MSSM50}, we only give an example to illustrate how
the RG evolution of CP-violating phases may be affected by
$\theta_{z}$.

\subsection{The Inverted Hierarchy Case}

What is special with the inverted hierarchy case is that only $|
\zeta_{12}^{-1}|$ is moderately large among the factors $\vert
\zeta_{ij}^{-1}\vert$ (for $i<j; ~i,j=1,2,3$) defined in
Eq.(\ref{ratio}). So more interesting corrections can be expected
than in the normal hierarchy case. However, since only $|
\zeta_{12}^{-1}|$ is large, the situation will not be so
complicated as the near degeneracy case. Similar to
Eq.(\ref{NormalDRS}), we find in the limit $| \zeta_{12}^{-1}|
>>1$ and $s_{z}\rightarrow 0$:
\begin{equation}
\dot{\theta}_{x}=\frac{1}{\zeta_{12}}\left[ C_\kappa^{l}y_{\tau
}^{2}c_{\left(\rho -\sigma \right)}^{2}s_{y}^{2}-C_\kappa^{\nu
}y_\nu^{2}c_{\left(\rho -\sigma
\right)}^{}c_{\left(\phi_{1}-\phi_{2}\right)
}^{}\frac{c_{2}s_{1}s_{2}}{c_{x}s_{x}}\right] c_{x}s_{x}+\cdots ,
\label{InvertedX}
\end{equation}
and
\begin{eqnarray}
\dot\delta &=&\frac{1}{\zeta_{12}}\left[ C_\kappa^{l}y_{\tau
}^{2}c_{\left(\rho -\sigma \right)}^{}s_{\left(\rho -\sigma
\right)}^{}s_{y}^{2}-C_\kappa^\nu
y_\nu^{2}c_{\left(\phi_{1}-\phi_{2}\right)}^{}s_{\left(\rho
-\sigma \right)}^{}\frac{c_{2}s_{1}s_{2}
}{c_{x}s_{x}}\right] +{\cal O} \left(s_{z}^{-1}\right) +\cdots ;  \nonumber \\
\dot{\rho} &=&\frac{1}{\zeta_{12}}\left[ C_\kappa^{l}y_{\tau
}^{2}c_{\left(\rho -\sigma \right)}^{}s_{\left(\rho -\sigma
\right)}^{}s_{y}^{2}-C_\kappa^\nu
y_\nu^{2}c_{\left(\phi_{1}-\phi_{2}\right)}^{}s_{\left(\rho
-\sigma \right)}^{}\frac{c_{2}s_{1}s_{2}}{c_{x}s_{x}}\right]
s_{x}^{2}+{\cal O} \left(s_{z}^{-1}\right) +\cdots ; \nonumber
\\
\dot{\sigma} &=&\frac{1}{\zeta_{12}}\left[ C_\kappa^{l}y_{\tau
}^{2}c_{\left(\rho -\sigma \right)}^{}s_{\left(\rho -\sigma
\right)}^{}s_{y}^{2}-C_\kappa^\nu
y_\nu^{2}c_{\left(\phi_{1}-\phi_{2}\right)}^{}s_{\left(\rho
-\sigma \right)}^{}\frac{c_{2}s_{1}s_{2}}{c_{x}s_{x}}\right]
c_{x}^{2}+{\cal O} \left(s_{z}^{-1}\right) +\cdots ,
\label{InvertedDRS}
\end{eqnarray}
where the ${\cal O} \left(s_{z}^{-1}\right)$ terms are exactly the
same for $\dot\delta$, $ \dot{\rho}$ and $\dot{\sigma}$, and have
been given in Eq.(\ref{NormalDRS}). It is notable that terms led
by $\zeta_{12}^{-1}$ are {\it not} enhanced by $ s_{z}^{-1}$. As a
result, there are no more extraordinary corrections in this case
than in the normal hierarchy case with extremely small
$\theta_{z}$. However, the inclusion of a moderately large
$\zeta_{12}^{-1}$ leads to two non-trivial consequences: (a) the
correction to $\theta_x$ can now be much larger than in the normal
hierarchy case, and (b) there can be much larger corrections to
CP-violating phases than in the normal hierarchy case, when
$s_z^{-1}$ is not an efficient enhancing factor.

\subsubsection{RG Correction in the SM}

In the normal hierarchy case, enhancing factors $|
\zeta_{ij}^{-1}| \sim {\cal O}(1)$ and RG corrections to mixing
angles are negligible (except for $\theta_{z}$ when it is
extremely tiny): $ {\cal O}((10^{-4})^\circ) $ in the energy range
from $M_{\rm Z}$ to $M_{3}$ and ${\cal O} \left(0.1^{\circ}\right)
$ from $M_{3}$ to $\Lambda_{\rm GUT}$. In the present case when $|
\zeta_{12}^{-1}|$ is large, only the correction to $\theta_{x}$ is
enhanced: ${\cal O} \left(0.01^{\circ}\right) $ below $M_{3}$ and
${\cal O} \left(10^{\circ}\right) $ above $M_{3}$. Note that the
correction to $\theta_{x}$ becomes largest when $| c_{\left(\rho
-\sigma \right) }c_{\left(\phi_{1}-\phi_{2}\right)}| \sim 1$ in
Eq.(\ref{InvertedX}).

Corrections to CP-violating phases are dominated by
$\zeta_{12}^{-1}$ when $|\zeta_{12}^{-1}|>>s_{z}^{-1}$ (or
equivalently, $\theta_{z}>>| \zeta_{12}| \sim 0.01\approx
0.5^{\circ}$). In this case, the situation for CP-violating phases
is the same as that for $\theta_{x}$ . So the contribution from
$y_{\tau}$ is negligible and large corrections of $ {\cal O}
\left(10^{\circ}\right)$ are possible only in the energy range
from $M_{3}$ to $\Lambda_{\rm GUT}$. In Eq.(\ref{InvertedDRS}),
contributions to CP-violating phases become largest when
$|c_{\left(\phi_{1} -\phi_{2} \right)}s_{\left(\rho -\sigma
\right)}| \sim 1$.

When $\theta_{z} << 0.5^{\circ}$, corrections to CP-violating
phases are dominated by $ s_{z}^{-1}$ and the situation is similar
to the normal hierarchy case.

In Figure \ref{FigCase3SM}, we illustrate the typical evolution
behavior of $\theta_{x}, \theta_{z}, \delta, \rho$ and $\sigma$ in
the SM. The competition between contributions from
$\zeta_{12}^{-1}$ and $s_{z}^{-1}$ is obvious.

\subsubsection{RG Correction in the MSSM}

In the MSSM when $\tan \beta \sim 10$, the contribution from
$y_{\tau}$ is about $100(\approx\tan^{2}\beta )$ times larger than
that in the SM but the contribution from $y_\nu$ is the same. So
corrections to $\theta_{y}$ and $\theta_{z}$ are of ${\cal O}
\left(0.01^{\circ}\right)$ in the energy range from $M_{\rm Z}$ to
$ M_{3}$ and of ${\cal O} \left(0.1^{\circ}\right) $ from $M_{3}$
to $\Lambda_{\rm GUT}$, while the correction to $\theta_{x}$
(enhanced by $\zeta_{12}^{-1}$) is of ${\cal O}
\left(1^{\circ}\right) $ below $M_{3}$ and of $ {\cal O}
\left(10^{\circ}\right)$ above $M_{3}$. Corrections to
CP-violating phases are the same as that to $\theta_x$ when
$|\zeta_{12}^{-1}|>>s_z^{-1}$ (or equivalently,
$\theta_{z}>>0.5^{\circ}$).

In the MSSM when $\tan \beta \sim 50$, the correction from
$y_{\tau}$ is about $2500$ times larger than that in the SM but
the contribution from $y_\nu$ is the same. So corrections to
$\theta _{y}$ and $\theta_{z}$ are of ${\cal O}
\left(0.1^{\circ}\sim 1^{\circ}\right) $ both at energies above
and below $ M_{3}$, while the correction to $\theta_{x}$ is of
${\cal O} \left(30^{\circ} \right) $ in the range from $M_{\rm Z}$
to $M_{3}$ but is still of ${\cal O} \left(10^{\circ}\right) $
from $M_{3}$ to $\Lambda_{\rm GUT}$. Note that in
Eq.(\ref{InvertedX}), the $y_\tau$ correction to $\theta_x$ is
always {\it negative} in the MSSM, while the sign of the
contribution from $y_\nu$ depends on simple phase factors.
Corrections to CP-violating phases are the same as that to
$\theta_x$ when $\zeta_{12}^{-1}$ is dominant.

When $\theta_{z} << 0.5^{\circ}$, corrections to CP-violating
phases are dominated by $s_{z}^{-1}$ and the situation is similar
to the normal hierarchy case. This is true no matter $\tan \beta$
is small or large.

In Figure \ref{FigCase3MSSM}, we illustrate the typical evolution
behavior of $\theta_{x}, \theta_{z}, \delta, \rho$ and $\sigma$ in
the MSSM. In the calculation, we take $\rho \approx \sigma $ to
make the correction to $\theta_{x}$ significant, and $\vert \rho -
\sigma \vert \sim 90^{\circ} $ to make corrections to $\theta_{z},
\delta, \rho $ and $\sigma $ significant.

\section{Summary}

In this work, we have derived one-loop renormalization group
equations for left-handed neutrino masses, leptonic mixing angles
and CP-violating phases, both in the SM and the MSSM extended with
three right-handed neutrinos. At energies above the seesaw
threshold, we show explicitly the contribution from the Yukawa
coupling matrix that connects left- and right-handed neutrinos.
For simplicity, we have assumed hierarchical eigenvalues of this
matrix in our derivation, so our analytical results may not be
applicable when the eigenvalues are not hierarchical. And since we
have also simplified the task by decoupling all right-handed
neutrinos at a common scale, the discussion may have to be
modified when the RG evolution between right-handed neutrino
thresholds is important\cite{0006229,0203233}.

Based on these equations, we study possible RG corrections related
to three typically interesting neutrino mass patterns: normal
hierarchy, near degeneracy and inverted hierarchy.

We firstly study the RG evolution of the factors $\zeta_{ij}$ (for
$i<j; ~i,j=1,2,3$) defined in Eq.(\ref{ratio}). We find that
$\vert \zeta_{ij}^{-1} \vert$ can be significantly damped from
$M_{\rm Z}$ to $\Lambda_{\rm GUT}$, both in the near degeneracy
case and in the inverted hierarchy case. It is also possible that
$\vert \zeta_{ij}^{-1} \vert$ may develop extremely high and
narrow peaks, so that the situation is nearly singular. However,
signs of $\zeta_{ij}$ are not likely to be changed, neither is the
order of sequence of left-handed neutrino masses.

In the normal hierarchy case, RG corrections from $M_{\rm Z}$ to
$\Lambda_{\rm GUT}$ are always negligible for mixing angles,
except for $\theta_z$ when it is extremely small. Appreciable or
even significant RG corrections to CP-violating phases are
possible only when $\theta_z<{\cal O}(1^\circ)$. In the SM,
dominant RG corrections to CP-violating phases generally arise in
the energy range from $M_3$ to $\Lambda_{\rm GUT}$. In the MSSM
when $\tan\beta$ is large, dominant RG corrections generally arise
from $M_{\rm Z}$ to $M_3$. Only in the MSSM when $\tan\beta$ is
about $10$, can large corrections to CP-violating phases arise
both at energies above and below $M_3$.

In the near degeneracy case, possible large corrections to mixing
angles and CP-violating phases are plethora. Mixing angles and
CP-violating phases are often driven to near their (pseudo-) fixed
points, since corrections are usually very large. Interesting
mixing angles at high energy scales are often possible. For
example, it is natural to find a large $\theta_z$ (comparable to
$\theta_x$, $\theta_y$) at $\Lambda_{\rm GUT}$.

In the inverted hierarchy case, only $\zeta_{12}^{-1}$ and
$s_z^{-1}$ are significant enhancing factors. So the situation is
much like that in the normal hierarchy case. However, because of
the large $\zeta_{12}^{-1}$, the correction to $\theta_x$ can be
large, and significant RG corrections to CP-violating phases are
possible even when $\theta_z^{-1}$ is not an efficient enhancing
factor.

To conclude, since RG corrections play a significant role in
relating the low- and high-energy  physics, an analytical
understanding of the RG evolution behavior of neutrino parameters
is necessary and important. Following earlier works, we have
extended this understanding beyond the seesaw threshold by
deriving RGEs for left-handed neutrino masses, leptonic mixing
angles and CP-violating phases running at energies above the
heaviest right-handed neutrino mass, under a few reasonable
simplifications. The significance of these equations are
demonstrated by studying the RG correction related to three
especially interesting neutrino mass patterns. We expect that our
work will be very useful for building realistic neutrino mass
models at high energy scales.

\section*{Acknowledgments}

The author is indebted to Professor Zhi-zhong Xing for reading the
manuscript with great care and patience, and also for his valuable
comments and numerous corrections. This work was supported in part
by the National Nature Science Foundation of China.

\vspace{0.5cm}

\noindent {\bf Note added}

\noindent When this work is being completed, we notice that
another paper about the similar topic is released by S. Antusch et
al\cite{0501272}. However, our strategy and approach are
apparently different from theirs. The two works are complementary
to each other.

\bigskip
%\bigskip
%\newpage
%%%%%%%%%%%%%%%%%%%%%%%%%%%%%%%%%%%%%%%%%%%%%%%%%%%%%%%%%%%%%%%%%%%%%%%%%%%5
\section*{Appendix}
%\bigskip
\bigskip
\appendix

\renewcommand{\thesection}{\Alph{section}}
\renewcommand{\thesubsection}{\Alph{section}.\arabic{subsection}}
\def\theequation{\Alph{section}.\arabic{equation}}
\renewcommand{\thetable}{\arabic{table}}
\renewcommand{\thefigure}{\arabic{figure}}
\setcounter{section}{0} \setcounter{equation}{0}
\setcounter{footnote}{0}

\section{One-loop RGEs for leptonic Yukawa coupling matrices}

Let $Y_u, ~Y_d, ~Y_l$ and $Y_\nu$ denote Yukawa coupling matrices
of up-type quarks, down-type quarks, charged leptons and the one
that connects left- and right-handed neutrinos, respectively. At
energies below the seesaw threshold, one-loop RGEs for $Y_l$ and
$\kappa$ are ($H_{f}\equiv Y_{f}Y_{f}^{\dag}$ for $f=u,d,l,\nu $):
\begin{eqnarray}
\dot{Y}_{l} &=&\left(\hat\alpha_{l}^{}+\hat{N}_{l}\right) Y_{l};\
\ \ \ \ \ \ \ \ \ \ \ \ \ \hat{N}_{l}=C_{l}^{l}H_{l}; \label{dYl} \\
\dot\kappa &=&\hat{\alpha}_\kappa\kappa +\hat{N}_\kappa\kappa
+\kappa \hat{N}_\kappa^{T};\ \ \ \
\hat{N}_\kappa=C_\kappa^{l}H_{l}. \label{dKappa}
\end{eqnarray}
For coefficients, we have in the MSSM
\begin{eqnarray}
&& \hat\alpha_{l} =-\left(\frac{9}{5}g_{1}^{2}+3g_{2}^{2}\right)
+{\rm Tr} \left(3H_{d}+H_{l}\right) ;\ \ C_{l}^{l}=3;  \nonumber \\
&& \hat{\alpha}_\kappa =-\left(\frac{6}{5} g_{1}^{2}
+6g_{2}^{2}\right) +{\rm Tr}\left(6H_{u}\right) ;\ \ \ \ \ \ \ \ \
C_\kappa^{l}=1; \label{AlphaK1}
\end{eqnarray}
and in the SM
\begin{eqnarray}
&& \hat\alpha_{l} =-\left(\frac{9}{4}g_{1}^{2} +\frac{9}{4}
g_{2}^{2}\right) + {\rm Tr} \left(3H_{u}+3H_{d}+H_{l}\right) ;\ \
\ \ \ C_{l}^{l}=\frac{3}{2} ;  \nonumber \\
&& \hat{\alpha}_\kappa =-\left(3g_{2}^{2}-\lambda \right) +{\rm
Tr} \left(6H_{u} +6H_{d}+2H_{l}\right) ;\ \ \ \ \ \ \
C_\kappa^{l}=-\frac{3}{2}. \label{AlphaK2}
\end{eqnarray}
At energies above the heaviest right-handed neutrino mass,
one-loop RGEs for $Y_{l},Y_\nu$ and $M_{R}$ are
\begin{eqnarray}
&&\dot{Y}_{l} =\left(\alpha_{l}^{}+N_{l}\right) Y_{l};\ \ \ \ \ \
\ \ \ \ \ \ \ \ N_{l}=C_{l}^{l}H_{l}+C_{l}^\nu H_\nu;  \label{runYl} \\
&&\dot{Y}_\nu =\left(\alpha_\nu^{}+N_\nu\right) Y_{\nu };\ \ \ \ \
\ \ \ \ \ \ \ N_\nu=C_\nu^{l}H_{l}+C_\nu^{\nu }H_\nu; \label{runYnu} \\
&&\dot{M}_{R} =N_{R}M_{R}+M_{R}N_{R}^{T};\ \ \ \ \ N_{R}=C_{R}
(Y_\nu^{\dag}Y_\nu)^T.  \label{cR}
\end{eqnarray}
In the MSSM,
\begin{eqnarray}
\alpha_{l} &=&-\left(\frac{9}{5}g_{1}^{2}+3g_{2}^{2}\right) +{\rm
Tr} \left(3H_{d}+H_{l}\right) ; \\
\alpha_\nu &=&-\left(\frac{3}{5}g_{1}^{2}+3g_{2}^{2}\right) +
{\rm Tr} \left(3H_{u}+H_\nu\right) ; \\
C_{l}^{l} &=&3;\ C_{l}^\nu=1;\ C_\nu^{l}=1;\ C_\nu^\nu=3;\
C_{R}=2.
\end{eqnarray}
In the SM,
\begin{eqnarray}
\alpha_{l}
&=&-\left(\frac{9}{4}g_{1}^{2}+\frac{9}{4}g_{2}^{2}\right) + {\rm
Tr}\left(3H_{u}+3H_{d}+H_{l}+H_\nu\right) ; \\
\alpha_\nu &=&-\left(\frac{9}{20}g_{1}^{2}+
\frac{9}{4}g_{2}^{2}\right) + {\rm Tr}\left( 3H_{u}+3H_{d} +H_{l}+H_\nu\right) ; \\
C_{l}^{l} &=&\frac{3}{2};\ C_{l}^\nu=-\frac{3}{2};\
C_\nu^{l}=-\frac{3}{2};\ C_\nu^\nu=\frac{3}{2};\ C_{R}=1.
\end{eqnarray}
If we make use of $\kappa $ also at energies above the seesaw
threshold, we can derive from Eqs.(\ref{kappa}), (\ref{runYnu})
and (\ref{cR})
\begin{eqnarray}
&&\dot\kappa =\alpha_\kappa^{}\kappa +N_\kappa\kappa
+\kappa N_\kappa^{T} \ ; \label{runKappa} \\
&&\alpha_\kappa^{} =2\alpha_\nu^{} ~; ~~~N_\kappa
=N_\nu-\tilde{N}_{R}= C_\kappa^{l}H_{l}+C_{ \kappa}^\nu H_\nu~,
\label{nKappa}
\end{eqnarray}
where $\tilde{N}_{R}=C_{R}H_\nu$ , with $C_{R}$ being the same as
that in Eq.(\ref{cR}). Also, $C_\kappa^{l}$ is the same as that in
Eq.(\ref{dKappa}). In the MSSM, $C_\kappa^\nu=1$; and in the SM,
$C_\kappa^\nu=1/2$.

\section{Derivation of RGEs for Individual Parameters}

In the same way as in Refs.\cite{9910420,0305273}, we need to
calculate $T_{\rm MNS}=U_{\rm MNS}^{\dag}\dot{U}_{\rm MNS}$ to
find out RGEs for leptonic mixing angles and CP-violating phases.
From Eq.(\ref{MNS}),
\begin{eqnarray} \label{Tmns}
T_{\rm MNS} &=&U_{\rm MNS}^{\dag}\left(U_{l}^{\dag}\dot{U}_{\kappa
}+\dot{U_{l}^{\dag}}U_\kappa\right) =U_\kappa^{\dag
}\dot{U}_{\kappa
}+U_{\rm MNS}^{\dag}\dot{U_{l}^{\dag}}U_{l}U_{\rm MNS}  \nonumber \\
&=&T_\kappa-U_{\rm MNS}^{\dag}T_{l}U_{\rm MNS}\ ,
\end{eqnarray}%
where%
\begin{equation}
T_\kappa\equiv U_\kappa^{\dag}\dot{U}_\kappa=-T_\kappa^\dagger ~;\
\ \ \ T_{l}\equiv U_{l}^{\dag}\dot{U}_{l}=-T_l^\dagger ~.
\end{equation}%
For $T_\kappa$, we find from Eqs.(\ref{runKappa}), (\ref{mns1a}),
(\ref{HL}) and (\ref{Hnu})
\begin{eqnarray}
&&\dot{\kappa^{\prime}}+T_\kappa\kappa^{\prime}+\kappa
^{\prime}T_\kappa^{T}=2\alpha_\kappa\kappa^{\prime
}+\tilde{N}_{\kappa
}\kappa^{\prime}+\kappa^{\prime}\tilde{N}_\kappa^{T}\ ; \label{MatrixEquationKappa} \\
&&\tilde{N}_\kappa=U_\kappa^{\dag}N_\kappa U_{\kappa
}=C_\kappa^{l}U_{\rm MNS}^{\dag}H_{l}^{\prime}U_{\rm MNS}
+C_{\kappa }^\nu U_{ \nu}H_\nu^{\prime}U_\nu^{\dag}\ .
\label{tildeNkappa}
\end{eqnarray}%
In Eq.(\ref{MatrixEquationKappa}), for diagonal elements
($i=1,2,3$)
\begin{equation} \label{runKappaMass}
\dot{k}_{i}=2\left(\alpha_\kappa+{\rm Re}\tilde{N}_{\kappa
.ii}\right) k_{i}\ ;\ \ \ {\rm Im}T_{\kappa .ii}={\rm
Im}\tilde{N}_{\kappa .ii}=0\ ;
\end{equation}%
and for off-diagonal elements ($i,j=1,2,3$)
\begin{eqnarray} \label{Tkappa}
&&T_{\kappa .ij}k_{j}-k_{i}T_{\kappa .ij}^{\ast }=\tilde{
N}_{\kappa .ij} k_{j}+k_{i}\left(\tilde{N}_{\kappa
}^{T}\right)_{ij}=\tilde{ N}_{\kappa
.ij}k_{ j}+k_{i} \tilde{N}_{\kappa .ji} \ ,  \nonumber \\
&\Longrightarrow &\left\{\begin{array}{c} {\rm Re}T_{\kappa
.ij}=-\displaystyle \frac{k_{i}+k_{j}}{k_{i}-k_{j}} \  {\rm
Re}\tilde{N} _{\kappa .ij}\equiv -\zeta_{ ij}^{-1} \ {\rm
Re}\tilde{N}_{\kappa .ij}\ ; \\
{\rm Im}T_{\kappa .ij}=-\displaystyle \frac{k_{i}-k_{j}}{k_{i}
+k_{j}} \  {\rm Im}\tilde{N}_{\kappa .ij}\equiv -\zeta_{ij} \ {\rm
Im}\tilde{N}_{\kappa .ij}\ .
\end{array}%
\right.
\end{eqnarray}%
For $T_{l}$, we find from Eq.(\ref{runYl})
\begin{equation}
\dot{H}_{l}=2\alpha_{l}H_{l}+N_{l}H_{l}+H_{l}N_{l}~.
\end{equation}%
Then from Eqs.(\ref{HL}) and (\ref{Hnu})%
\begin{eqnarray}
&&\dot{H_{l}^{\prime}}+T_{l}H_{l}^{\prime}+H_{l}^{\prime
}T_{l}^{\dag}=2\alpha_{l}H_{l}^{\prime }+\tilde{N}_{l}H_{l}^{
\prime}+H_{l}^{\prime} \tilde{N}_{l}\ ; \label{MatrixEquationHL} \\
&&\tilde{N}_{l}=U_{l}^{ \dag}N_{l}U_{l}=C_{l}^{l}H_{l}^{\prime
}+C_{l}^\nu\left(U_{\rm MNS}U_\nu\right) H_\nu^{
\prime}\left(U_{\rm MNS}U_\nu\right)^{\dag}\ . \label{tildeNl}
\end{eqnarray}%
In Eq.(\ref{MatrixEquationHL}), for diagonal elements ($i=1,2,3$)
\begin{equation}
\dot{H}_{l.ii}^{\prime}=2\left(\alpha _{l}+\tilde{N}_{l.ii}\right)
H_{l.ii}^{\prime}\ ,
\end{equation}%
and for off-diagonal elements  ($i,j=1,2,3$)
\begin{eqnarray} \label{Tl}
T_{l.ij}H_{l.jj}^{\prime}+H_{l.ii}^{\prime}\left(T_{l}^{\dag
}\right)_{ij} &=&\tilde{N}_{l.ij}H_{l.jj}^{\prime
}+H_{l.ii}^{\prime}\tilde{N}
_{l.ij}  \nonumber \\
\Longrightarrow T_{l.ij} &=&-\frac{H_{l.ii}^{\prime
}+H_{l.jj}^{\prime}}{H_{l.ii}^{\prime}-H_{l.jj}^{\prime }} \
\tilde{N}_{l.ij}\ .
\end{eqnarray}%
Note that $T_{l.ii}$ (for $i=1,2,3$) are arbitrary, since $U_{l}$
is only determined up to a diagonal phase matrix on its right.

Furthermore, in order to derive RGEs for parameters in $Y_\nu$, we
find from Eq.(\ref{runYnu})
\begin{equation}
\dot{H}_\nu=2\alpha_\nu H_\nu+N_\nu H_\nu+H_\nu N_\nu ~.
\end{equation}%
Then from Eq.(\ref{Hnu})
\begin{eqnarray}
&&\dot{H_\nu^{\prime}}+\left(T_\nu+U_\nu^{\dag}T_\kappa U_{\nu
}\right) H_\nu^{\prime}+H_\nu^{\prime}\left(T_\nu+U_{\nu
}^{\dag}T_\kappa U_\nu\right)^\dagger =2\alpha_\nu H_{\nu
}^{\prime}+\tilde{N}_\nu H_\nu^{\prime} +H_\nu^{\prime}
\tilde{N}_\nu\ ; \label{MatrixEquationHnu} \\
&&\tilde{N}_\nu =C_{ \nu}^{l}\left(U_{\rm MNS}U_\nu\right)^{
\dag}H_{l}^{\prime}\left(U_{\rm MNS}U_\nu\right)+ C_\nu^{\nu
}H_\nu^{\prime}\ ; \ \ \ \ \ \ \ \ \ \ \
 T_\nu \equiv U_\nu^{\dag}\dot{U}_\nu\ . \label{tildeNu}
\end{eqnarray}%
In Eq.(\ref{MatrixEquationHnu}), for diagonal elements ($i=1,2,3$)
\begin{equation}
H_{\nu .ii}^{\prime}=2\left(\alpha_\nu+\tilde{N}_{\nu .ii}\right)
H_{\nu .ii}^{\prime}\ ;
\end{equation}%
and for off-diagonal elements ($i,j=1,2,3$)
\begin{eqnarray} \label{Tnu}
\left(T_\nu+U_\nu^{\dag}T_\kappa U_\nu\right)_{ij}H_{\nu
.jj}^{\prime}-H_{\nu .ii}^{\prime}\left(T_\nu+U_\nu^{\dag
}T_\kappa U_\nu\right)_{ij}=\tilde{N}_{\nu .ij}H_{\nu
.jj}^{\prime}+H_{\nu.ii}^{\prime}\tilde{N}_{\nu .ij} \ , &&  \nonumber \\
\Longrightarrow T_{\nu .ij}=-\frac{H_{\nu .ii}^{\prime}+H_{\nu
.jj}^{\prime}}{H_{\nu .ii}^{\prime}-H_{\nu .jj}^{\prime }} \
\tilde{N}_{\nu.ij}-\left(U_\nu^{\dag}T_\kappa U_\nu\right)_{ij}\ .
&&\
\end{eqnarray}
Just like the diagonal elements of $T_l$,  $T_{\nu.ii}$ (for
$i=1,2,3$) are also arbitrary since $U_\nu$ is only determined up
to a diagonal phase matrix on its right.

To calculate $\dot{\theta}_x, ~\dot{\theta}_y, ~\dot{\theta}_z,
~\dot{\delta}, ~\dot{\rho}$ and $\dot{\sigma}$ from $T_{\rm
MNS}^{}$ , an auxiliary diagonal phase matrix is required on the
left hand side of $U_{\rm MNS}^{}$ as defined in Eq.(\ref{mns1b}),
i.e. we have to use a more general parametrization of the MNS
matrix in Eq.(\ref{Tmns}):
\begin{equation}%\label{}
U_{\rm MNS}^{} = P\cdot V ~;~~~ P\equiv \left(\matrix{
e_{}^{i\alpha} &  &  \cr & e_{}^{i\beta} & \cr &  &
e_{}^{i\gamma}} \right) ~,
\end{equation}
where $V$ denoting the original $U_{\rm MNS}^{}$ defined in
Eq.(\ref{mns1b}). Then from Eq.(\ref{tildeNkappa})
\begin{equation}%\label{}
\tilde{N}_\kappa=C_\kappa^{l}V_{}^\dag H_{l}^{\prime}V +C_{\kappa
}^\nu U_\nu^{} H_\nu^{\prime}U_\nu^{\dag}\ .
\end{equation}
Together with Eqs.(\ref{runKappaMass}) and (\ref{Tkappa}), this
equation shows that $T_\kappa^{}$ is independent of the phase
matrix $P$. Furthermore, from Eqs.(\ref{tildeNl}) and (\ref{Tl}),
the product $\left(P_{}^\ast T_l^{} P\right)$ is also independent
of $P$. So in an equivalence of Eq.(\ref{Tmns})
\begin{equation} \label{B19}
\dot{V}V^\dag = V T_\kappa^{} V^\dag - P_{}^\ast \left(T_l^{}
+P_{}^\ast \dot{P}\right)P ~,
\end{equation}
the equations of off-diagonal elements are obviously independent
of the matrix $P$. They are all together six linearly independent
equations of $\dot{\theta}_x, ~\dot{\theta}_y, ~\dot{\theta}_z,
~\dot{\delta}, ~\dot{\rho}$ and $\dot{\sigma}$, and can thus
determine these six quantities unambiguously. For the diagonal
elements, $(P_{}^\ast \dot{P} )_{ii}^{} ~(i=1,2,3)$ are only
determined up to arbitrary $T_{ l.ii}^{} ~(i=1,2,3)$ , but this is
of no problem since $\alpha$, $\beta$ and $\gamma$ are not
physical by definition. We may choose whatever value for $T_{
l.ii}^{} ~(i=1,2,3)$ as we like in the calculation, or may simply
ignore the equations for the diagonal elements in Eq.(\ref{B19}).

In contrast, we can see from Eqs.(\ref{tildeNu}) and (\ref{Tnu})
that to extract $\dot{\theta}_{1}, ~\dot{\theta}_{2},
~\dot{\theta}_{3}, ~\dot{\delta}_\nu, ~\dot{\phi}_{1}$ and $\dot{
\phi}_{2}$ from $T_\nu^{}$, an auxiliary phase matrix on the right
hand side of $U_\nu$ (just as the phase matrix $P$ on the left
hand side of $U_{\rm MNS}$) is not necessary. We can use the
$U_\nu$ defined in Eq.(\ref{Unu}) directly during the calculation.
There are totally six linearly independent equations of
$\dot{\theta}_{1}, ~\dot{ \theta} _{2}, ~\dot{\theta}_{3},
~\dot\delta_\nu, ~\dot{\phi}_{1}$ and $\dot{ \phi}_{2}$ in
Eq.(\ref{Tnu}), so these quantities can also be determined
unambiguously.

%\newpage
\section{Full Expressions of $\dot{\theta}_{1},\dot{\theta}
_{2},\dot{\theta}_{3},\dot\delta_\nu,\dot{\phi}_{1}$ and
$\dot{\phi}_{2}$}

\setcounter{equation}{0}

\begin{center}
(Note that for what ever $F: ~\dot{F}\equiv 16\pi^2
\displaystyle\frac{d F}{d t}; ~t={\rm ln} \mu$, with $\mu$ being
the energy scale.)
\end{center}

\begin{eqnarray}
\dot{\theta}_{1}
&=&C_\kappa^{l}y_{\tau}^{2}\left\{\frac{c_{\left(\phi_{1}-\phi_{2}\right)}^{}}{\zeta_{12}}\left[
c_{\left(\rho -\sigma
\right)}^{}c_{x}s_{x}\left(s_{y}^{2}-c_{y}^{2}s_{z}^{2}\right)
-\left(c_{\left(\delta +\rho -\sigma
\right)}^{}c_{x}^{2}-c_{\left(\delta -\rho +\sigma
\right)}^{}s_{x}^{2}\right) c_{y}s_{y}s_{z}\right] \frac{c_{1}s_{2}}{c_{2}}\right.  \nonumber \\
&&\ \ \ \ \ \ -\zeta_{12}\cdot s_{\left(\phi_{1}-\phi_{2}\right)
}^{} \left[ s_{\left(\rho -\sigma \right)
}^{}c_{x}s_{x}\left(s_{y}^{2}-c_{y}^{2}s_{z}^{2}\right)
-\left(s_{\left(\delta +\rho -\sigma
\right)}^{}c_{x}^{2}+s_{\left(\delta -\rho +\sigma
\right)}^{}s_{x}^{2}\right) c_{y}s_{y}s_{z}\right]
\frac{c_{1}s_{2}}{c_{2}}
\nonumber \\
&&\ \ \ \ \ \ +\left[
\frac{c_{\phi_{1}}}{\zeta_{13}}\left(c_{\left(\delta -\rho
\right)}^{}s_{x}s_{y}-c_{\rho }^{}c_{x}c_{y}s_{z}\right)
+\zeta_{13}\cdot s_{\phi_{1}}\left(s_{\left(\delta -\rho \right)
}^{}s_{x}s_{y}+s_{\rho}^{}c_{x}c_{y}s_{z}\right) \right]
\frac{c_{y}c_{z}s_{1}s_{2}}{c_{2}}  \nonumber \\
&&\ \ \ \ \ \ \left. -\left[ \frac{c_{\phi_{2}}}{\zeta
_{23}}\left(c_{\left(\delta -\sigma\right)}^{}c_{x}s_{
y}+c_{\sigma }^{}c_{y}s_{x}s_{z}\right) +\zeta_{23}\cdot
s_{\phi_{2}}\left(s_{\left(\delta -\sigma\right)}^{}c_{
x}s_{y}-s_{\sigma }^{}c_{y}s_{x}s_{z}\right) \right]
c_{y}c_{z}\right\}  \nonumber \\
&&+C_\kappa^\nu
y_\nu^{2}\left\{-\left(\frac{c_{\left(\phi_{1}-\phi_{2}\right)}^{2}}{\zeta_{12}}+\zeta_{12}\cdot
s_{\left(\phi
_{1}-\phi_{2}\right)}^{2}\right) c_{1}s_{1}s_{2}^{2}\right.  \nonumber \\
&&\left. \ \ \ \ \ \ \ \ +\left(\frac{c_{\phi_{1}}^{2}}{\zeta
_{13}}+\zeta_{13}\cdot s_{\phi_{1}}^{2}\right)
c_{1}s_{1}s_{2}^{2}+\left(\frac{c_{\phi_{2}}^{2}}{\zeta
_{23}}+\zeta_{23}\cdot s_{\phi_{2}}^{2}\right)
c_{1}c_{2}^{2}s_{1}\right\}  \nonumber \\
&&+C_\nu^{l}y_{\tau}^{2}\left\{c_{1}s_{1}\left[
c_{x}^{2}s_{y}^{2}-c_{y}^{2}\left(c_{z}^{2}-s_{x}^{2}s_{z}^{2}\right)
\right] +2c_{\delta }^{}c_{1}c_{x}c_{y}s_{1}s_{x}s_{y}s_{z}\right.
\nonumber \\
&&\ \ \ \ \ \ \ \ +\left[ c_{\left(\rho +\phi_{1}\right)
}^{}c_{x}s_{1}s_{2}-c_{\left(\sigma +\phi_{2}\right)
}^{}\left(c_{1}^{2}-s_{1}^{2}\right) c_{2}s_{x}\right]
\frac{c_{y}^{2}c_{z}s_{z}}{c_{2}
}  \nonumber \\
&&\ \ \ \ \ \ \ \ -\left[ c_{\left(\delta -\rho -\phi_{1}\right)
}^{}s_{1}s_{2}s_{x}+c_{\left(\delta -\sigma -\phi_{2}\right)
}^{}c_{2}c_{x}\left(c_{1}^{2}-s_{1}^{2}\right) \right] \frac{c_{y}c_{z}s_{y}}{c_{2}}  \nonumber \\
&&\ \ \ \ \ \ \ \ -c_{\left(\rho -\sigma
+\phi_{1}-\phi_{2}\right)}^{}
\frac{c_{1}c_{x}s_{2}s_{x}}{c_{2}}\left(s_{y}^{2}-c_{y}^{2}s_{z}^{2}\right)
\nonumber \\
&&\left. \ \ \ \ \ \ \ \ +\left[ c_{\left(\delta +\rho -\sigma
+\phi_{1}-\phi_{2}\right)}^{}c_{x}^{2}-c_{\left(\delta -\rho
+\sigma -\phi_{1}+\phi_{2}\right)}^{}s_{x}^{2}\right]
\frac{c_{1}c_{y}s_{2}s_{y}s_{z}}{c_{2}}\right\} ~.
\label{dTheta1}
\end{eqnarray}
\begin{eqnarray}
\dot{\theta}_{2}
&=&C_\kappa^{l}y_{\tau}^{2}\left\{-\frac{c_{\left(\phi_{1}-\phi_{2}\right)}^{}}{\zeta_{12}}\left[
c_{\left(\rho -\sigma
\right)}^{}c_{x}s_{x}\left(s_{y}^{2}-c_{y}^{2}s_{z}^{2}\right)
-\left(c_{\left(\delta +\rho -\sigma
\right)}^{}c_{x}^{2}-c_{\left(\delta -\rho +\sigma
\right)}^{}s_{x}^{2}\right) c_{y}s_{y}s_{z}\right]
s_{1}\right.  \nonumber \\
&&\ \ \ \ \ \ \ +\zeta_{12}\cdot s_{\left(\phi_{1}-\phi
_{2}\right)}^{} \left[ s_{\left(\rho -\sigma \right)
}^{}c_{x}s_{x}\left(s_{y}^{2}-c_{y}^{2}s_{z}^{2}\right)
-\left(s_{\left(\delta +\rho -\sigma
\right)}^{}c_{x}^{2}+s_{\left(\delta -\rho +\sigma \right)
}^{}s_{x}^{2}\right) c_{y}s_{y}s_{z}\right] s_{1}  \nonumber \\
&&\left. \ \ \ \ \ \ \ +\left[ \frac{c_{\phi_{1}}^{}}{\zeta
_{13}}\left(c_{\left(\delta -\rho \right)
}^{}s_{x}s_{y}-c_{\rho}^{}c_{x}c_{y}s_{z}\right) +\zeta _{13}\cdot
s_{\phi_{1}}^{}\left(s_{\left(\delta -\rho \right)
}^{}s_{x}s_{y}+s_{\rho
}^{}c_{x}c_{y}s_{z}\right) \right] c_{1}c_{y}c_{z}\right\}  \nonumber \\
&&+C_\kappa^\nu
y_\nu^{2}\left\{\left(\frac{c_{\left(\phi_{1}-\phi_{2}\right)}^{2}}{\zeta_{12}}+\zeta_{12}\cdot
s_{\left(\phi_{1}-\phi_{2}\right)}^{2}\right)
c_{2}s_{1}^{2}s_{2}+\left(\frac{c_{\phi_{1}}^{2}}{\zeta
_{13}}+\zeta_{13}\cdot s_{\phi_{1}}^{2}\right)
c_{1}^{2}c_{2}s_{2}\right\}  \nonumber \\
&&+C_\nu^{l}y_{\tau}^{2}\left\{-c_{2}s_{2}\left[
\left(c_{x}^{2}s_{1}^{2}-s_{x}^{2}\right)
s_{y}^{2}+c_{y}^{2}\left(c_{1}^{2}c_{z}^{2}-\left(c_{x}^{2}-s_{1}^{2}s_{x}^{2}\right)
s_{z}^{2}\right) \right] \right.  \nonumber \\
&&\ \ \ \ \ \ \ \ \
-2c_\delta^{}c_{2}c_{x}c_{y}\left(1+s_{1}^{2}\right)
s_{2}s_{x}s_{y}s_{z}-\left[ c_{\left(\rho
+\phi_{1}\right)}^{}c_{x}\left(c_{2}^{2}-s_{2}^{2}\right)
-2c_{\left(\sigma +\phi_{2}\right) }^{}c_{2}s_{1}s_{2}s_{x}\right]
c_{1}c_{y}^{2}c_{z}s_{z}
\nonumber \\
&&\ \ \ \ \ \ \ \ \ +\left[ c_{\left(\delta -\rho -\phi
_{1}\right)}^{}\left(c_{2}^{2}-s_{2}^{2}\right)
s_{x}+2c_{\left(\delta -\sigma -\phi_{2}\right)
}^{}c_{2}c_{x}s_{1}s_{2}\right] c_{1}c_{y}c_{z}s_{y}
\nonumber \\
&&\ \ \ \ \ \ \ \ \ -c_{\left(\rho -\sigma +\phi_{1}-\phi
_{2}\right)}^{}c_{x}s_{1}\left(c_{2}^{2}-s_{2}^{2}\right)
s_{x}\left(s_{y}^{2}-c_{y}^{2}s_{z}^{2}\right)  \nonumber \\
&&\left. \ \ \ \ \ \ \ \ \ +\left[ c_{\left(\delta +\rho -\sigma
+\phi_{1}-\phi_{2}\right)}^{}c_{x}^{2}-c_{\left(\delta -\rho
+\sigma -\phi_{1}+\phi_{2}\right)}^{}s_{x}^{2}\right]
c_{y}s_{1}\left(c_{2}^{2}-s_{2}^{2}\right) s_{y}s_{z}\right\} ~.
\label{dTheta2}
\end{eqnarray}
\begin{eqnarray}
\dot{\theta}_{3}
&=&C_\kappa^{l}y_{\tau}^{2}\left\{-\frac{c_{\left(\delta_\nu+\phi_{1}-\phi_{2}\right)
}^{}}{\zeta_{12}}\left[ c_{\left(\rho -\sigma \right)
}^{}c_{x}s_{x}\left(s_{y}^{2}-c_{y}^{2}s_{z}^{2}\right)
-\left(c_{\left(\delta +\rho -\sigma
\right)}^{}c_{x}^{2}-c_{\left(\delta -\rho +\sigma
\right)}^{}s_{x}^{2}\right) c_{y}s_{y}s_{z}\right]
\frac{c_{1}}{c_{2}}\right.
\nonumber \\
&&\ \ \ \ \ \ \ +\zeta_{12}\cdot s_{\left(\delta_\nu+\phi
_{1}-\phi_{2}\right)}^{}\left[ s_{\left(\rho -\sigma \right)
}^{}c_{x}s_{x}\left(s_{y}^{2}-c_{y}^{2}s_{z}^{2}\right)
-\left(s_{\left(\delta +\rho -\sigma
\right)}^{}c_{x}^{2}+s_{\left(\delta -\rho +\sigma
\right)}^{}s_{x}^{2}\right) c_{y}s_{y}s_{z}\right]
\frac{c_{1}}{c_{2}}
\nonumber \\
&&\left. \ \ \ \ \ \ \ -\left[ \frac{c_{\left(\delta_\nu+\phi
_{1}\right)}^{}}{\zeta_{13}}\left(c_{\left(\delta -\rho
\right)}^{}s_{x}s_{y}-c_{\rho}^{}c_{x}c_{y}s_{z}\right)
+\zeta_{13}\cdot s_{\left(\delta_\nu+\phi_{1}\right)
}^{}\left(s_{\left(\delta -\rho \right)
}^{}s_{x}s_{y}+s_{\rho}^{}c_{x}c_{y}s_{z}\right) \right]
\frac{c_{y}c_{z}s_{1}}{c_{2}}\right\}  \nonumber \\
&&+C_\kappa^\nu y_\nu^{2}\left\{\left[ \left(\zeta
_{12}^{-1}+\zeta_{12}\right) c_{\delta_\nu}^{}+\left(\zeta
_{12}^{-1}-\zeta_{12}\right) c_{\left(\delta_\nu+2\phi _{1}-2\phi
_{2}\right)}^{}\right]\frac{c_{1}s_{1}s_{2}}{2} \right.  \nonumber \\
&&\ \ \ \ \ \ \ \ \ \left. -\left[ \left(\zeta_{13}^{-1}+\zeta
_{13}\right) c_{\delta_\nu}^{}+\left(\zeta_{13}^{-1}-\zeta
_{13}\right) c_{\left(\delta_\nu+2\phi_{1}\right)}^{}
\right]\frac{c_{1}s_{1}s_{2}}{2} \right\}  \nonumber \\
&&+C_\nu^{l}y_{\tau}^{2}\left\{c_{3}s_{3}\left[
c_{2}^{2}\left(s_{x}^{2}s_{y}^{2}+c_{x}^{2}c_{y}^{2}s_{z}^{2}\right)
-\left(c_{1}^{2}-s_{1}^{2}s_{2}^{2}\right)
\left(c_{x}^{2}s_{y}^{2}+c_{y}^{2}s_{x}^{2}s_{z}^{2}\right)
-c_{y}^{2}c_{z}^{2}\left(s_{1}^{2}-c_{1}^{2}s_{2}^{2}\right)
\right] \right.
\nonumber \\
&&\ \ \ \ \ \ \ \ \
-2c_\delta^{}c_{3}c_{x}c_{y}\left(c_{1}^{2}+c_{2}^{2}-s_{
1}^{2}s_{2}^{2}\right) s_{3}s_{x}s_{y}s_{z}-2c_{\delta_{\nu
}}^{}c_{1}c_{3}^{ 2} s_{1}s_{2} \left(c_{x}^{2} s_{
y}^{2}-c_{y}^{2} \left(c_{z}^{2}-s_{x}^{2}s_{z}^{2}\right)
\right)  \nonumber \\
&&\ \ \ \ \ \ \ \ \ +2\left[ c_{\left(\rho +\phi_{1}\right)
}^{}c_{2}c_{x}s_{2}-c_{\left(\sigma +\phi_{2}\right)
}^{}s_{1}\left(1+s_{2}^{2}\right) s_{x}\right]
c_{1}c_{3}c_{y}^{2}c_{z}s_{3}s_{z}  \nonumber
\\
&&\ \ \ \ \ \ \ \ \ +2c_{\left(\rho -\sigma +\phi_{1}-\phi
_{2}\right)}^{}c_{2}c_{3}c_{x}s_{1}s_{2}s_{3}s_{x}\left(s_{y}^{2}-c_{y}^{2}s_{z}^{2}\right)  \nonumber \\
&&\ \ \ \ \ \ \ \ \ -2\left[ c_{\left(\delta -\rho -\phi
_{1}\right)}^{}c_{2}s_{2}s_{x}+c_{\left(\delta -\sigma -\phi
_{2}\right)}^{}c_{x}s_{1}\left(1+s_{2}^{2}\right) \right]
c_{1}c_{3}c_{y}c_{z}s_{3}s_{y}  \nonumber \\
&&\ \ \ \ \ \ \ \ \ -2\left[ c_{\left(\delta +\rho -\sigma +\phi
_{1}-\phi_{2}\right)}^{}c_{x}^{2}-c_{\left(\delta -\rho +\sigma
-\phi_{1}+\phi_{2}\right)}^{}s_{x}^{2}\right]
c_{2}c_{3}c_{y}s_{1}s_{2}s_{3}s_{y}s_{z}
\nonumber \\
&&\ \ \ \ \ \ \ \ \ +c_{\left(\delta_\nu+\rho +\phi_{1}\right)}^{}
\frac{c_{x}c_{y}^{2}c_{z}s_{1}s_{z}}{c_{2}}\left(c_{2}^{2}\left(c_{3}^{2}-s_{3}^{2}\right) -s_{2}^{2}\right)  \nonumber \\
&&\ \ \ \ \ \ \ \ \ +2\left[ c_{\left(\delta_\nu-\sigma -\phi
_{2}\right)}^{}c_{1}^{2}-c_{\left(\delta_\nu+\sigma +\phi
_{2}\right)}^{}s_{1}^{2}\right]
c_{3}^{2}c_{y}^{2}c_{z}s_{2}s_{x}s_{z}
\nonumber \\
&&\ \ \ \ \ \ \ \ \ -c_{\left(\delta_\nu+\rho -\sigma +\phi
_{1}-\phi_{2}\right)}^{}\frac{c_{1}c_{x}s_{x}}{c_{2}}\left(c_{2}^{2}\left(c_{3}^{2}-s_{3}^{2}\right)
-s_{2}^{2}\right) \left(s_{y}^{2}-c_{y}^{2}s_{z}^{2}\right)  \nonumber \\
&&\ \ \ \ \ \ \ \ \ -c_{\left(\delta -\delta_\nu-\rho -\phi
_{1}\right)}^{}\frac{c_{y}c_{z}s_{1}s_{x}s_{y}}{c_{2}}\left(c_{2}^{2}\left(c_{3}^{2}-s_{3}^{2}\right) -s_{2}^{2}\right)  \nonumber \\
&&\ \ \ \ \ \ \ \ \ +2\left[ c_{\left(\delta +\delta_{\nu }-\sigma
-\phi_{2}\right)}^{}c_{1}^{2}-c_{\left(\delta -\delta_\nu-\sigma
-\phi_{2}\right)}^{}s_{1}^{2}\right]
c_{3}^{2}c_{x}c_{y}c_{z}s_{2}s_{y}
\nonumber \\
&&\ \ \ \ \ \ \ \ \ +\left[ c_{\left(\delta +\delta_\nu+\rho
-\sigma +\phi_{1}-\phi_{2}\right)}^{}c_{x}^{2}-c_{\left(\delta
-\delta_\nu-\rho +\sigma
-\phi_{1}+\phi_{2}\right)}^{}s_{x}^{2}\right]
\frac{c_{1}c_{y}s_{y}s_{z}}{c_{2}}\left(c_{2}^{2}\left(c_{3}^{2}-s_{3}^{2}\right) -s_{2}^{2}\right)  \nonumber \\
&&\ \ \ \ \ \ \ \ \ \left. -4c_\delta^{}c_{\delta_{\nu
}}^{}c_{1}c_{3}^{2}c_{x}c_{y}s_{1}s_{2}s_{x}s_{y}s_{z}\right\} ~.
\label{dTheta3}
\end{eqnarray}
\begin{eqnarray}
\dot\delta_\nu
&=&C_\kappa^{l}y_{\tau}^{2}\left\{-\frac{1}{\zeta_{12}}\left[
s_{\left(\phi_{1}-\phi_{2}\right)
}^{}c_{3}\left(s_{1}^{2}-c_{1}^{2}s_{2}^{2}\right)
s_{3}-s_{\left(\delta_\nu+\phi_{1}-\phi_{2}\right)}^{}c_{1}s_{1}s_{2}\left(c_{3}^{2}-s_{3}^{2}\right)
\right] \right.  \nonumber \\
&&\ \ \ \ \ \ \ \ \ \ \ \ \ \ \cdot \left[ \left\{c_{\left(\rho
-\sigma
\right)}^{}c_{x}s_{x}\left(s_{y}^{2}-c_{y}^{2}s_{z}^{2}\right)
-\left(c_{\left(\delta +\rho -\sigma
\right)}^{}c_{x}^{2}-c_{\left(\delta -\rho +\sigma
\right)}^{}s_{x}^{2}\right) c_{y}s_{y}s_{z}\right\} \frac{1
}{c_{2}c_{3}s_{1}s_{2}s_{3}}\right]  \nonumber \\
&&\ \ \ \ \ \ \ -\zeta_{12}\left[ c_{\left(\phi_{1}-\phi
_{2}\right)}^{}c_{3}\left(s_{1}^{2}-c_{1}^{2}s_{2}^{2}\right)
s_{3}-c_{\left(\delta_\nu+\phi_{1}-\phi_{2}\right)
}^{}c_{1}s_{1}s_{2}\left(c_{3}^{2}-s_{3}^{2}\right) \right]  \nonumber \\
&&\ \ \ \ \ \ \ \ \ \ \ \ \ \ \cdot \left[ \left\{s_{\left(\rho
-\sigma
\right)}^{}c_{x}s_{x}\left(s_{y}^{2}-c_{y}^{2}s_{z}^{2}\right)
-\left(s_{\left(\delta +\rho -\sigma
\right)}^{}c_{x}^{2}+s_{\left(\delta -\rho +\sigma
\right)}^{}s_{x}^{2}\right) c_{y}s_{y}s_{z}\right\} \frac{1
}{c_{2}c_{3}s_{1}s_{2}s_{3}}\right]  \nonumber \\
&&\ \ \ \ \ \ \ +\frac{1}{\zeta_{13}}\left[ s_{\phi
_{1}}^{}c_{3}\left(c_{1}^{2}-s_{1}^{2}s_{2}^{2}\right)
s_{3}+s_{\left(\delta_\nu+\phi_{1}\right)
}^{}c_{1}s_{1}s_{2}\left(c_{3}^{2}-s_{3}^{2}\right) \right]
\left(c_{\left(\delta -\rho \right)}^{}s_{x}s_{y}-c_{\rho
}^{}c_{x}c_{y}s_{z}\right)
\frac{c_{y}c_{z}}{c_{1}c_{2}c_{3}s_{2}s_{3}}
\nonumber \\
&&\ \ \ \ \ \ \ -\zeta_{13}\left[
c_{\phi_{1}}^{}c_{3}\left(c_{1}^{2}-s_{1}^{2}s_{2}^{2}\right)
s_{3}+c_{\left(\delta_{\nu
}+\phi_{1}\right)}^{}c_{1}s_{1}s_{2}\left(c_{3}^{2}-s_{3}^{2}\right)
\right] \left(s_{\left(\delta -\rho
\right)}^{}s_{x}s_{y}+s_{\rho}^{}c_{x}c_{y}s_{z}\right)
\frac{c_{y}c_{z}}{c_{1}c_{2}c_{3}s_{2}s_{3}}
\nonumber \\
&&\ \ \ \ \ \ \ \left. +\left[ \frac{s_{\phi_{2}}^{}}{\zeta
_{23}}\left(c_{\left(\delta -\sigma \right)
}^{}c_{x}s_{y}+c_{\sigma}^{}c_{y}s_{x}s_{z}\right) -\zeta
_{23}\cdot c_{\phi_{2}}^{}\left(s_{\left(\delta -\sigma
\right)}^{}c_{x}s_{y}-s_{\sigma}^{}c_{y}s_{x}s_{z}\right) \right]
\frac{c_{y}c_{z}}{c_{1}s_{1}}\right\}
\nonumber \\
&&+C_\kappa^\nu y_\nu^{2}\left\{\frac{c_{\left(\phi
_{1}-\phi_{2}\right)}^{}}{\zeta_{12}}\left[ s_{\left(\phi
_{1}-\phi_{2}\right)}^{}\left(s_{1}^{2}-c_{1}^{2}s_{2}^{2}\right)
-s_{\left(\delta_\nu+\phi _{1}-\phi_{2}\right)
}^{}\frac{c_{1}s_{1}s_{2}}{c_{3}s_{3}}\left(c_{3}^{2}-s_{3}^{2}\right) \right] \right.  \nonumber \\
&&\ \ \ \ \ \ \ \ \ \ -\zeta_{12}\cdot s_{\left(\phi_{1}-\phi
_{2}\right)}^{}\left[ c_{\left(\phi_{1}-\phi_{2}\right)
}^{}\left(s_{1}^{2}-c_{1}^{2}s_{2}^{2}\right) -c_{\left(\delta
_\nu+\phi_{1}-\phi_{2}\right)
}^{}\frac{c_{1}s_{1}s_{2}}{c_{3}s_{3}}\left(c_{3}^{2}-s_{3}^{2}\right) \right]  \nonumber \\
&&\ \ \ \ \ \ \ \ \ \ +\frac{c_{\phi_{1}}^{}}{\zeta_{13}}\left[
s_{\phi_{1}}^{}\left(c_{1}^{2}-s_{1}^{2}s_{2}^{2}\right)
+s_{\left(\delta_\nu+\phi_{1}\right)
}^{}\frac{c_{1}s_{1}s_{2}}{c_{3}s_{3}}\left(c_{3}^{2}-s_{3}^{2}\right) \right]  \nonumber \\
&&\ \ \ \ \ \ \ \ \ \ \left. -\zeta_{13}\cdot s_{\phi
_{1}}^{}\left[
c_{\phi_{1}}^{}\left(c_{1}^{2}-s_{1}^{2}s_{2}^{2}\right)
+c_{\left(\delta_\nu+\phi_{1}\right)}^{}\frac{c_{1}s_{1}s_{2}}{c_{3}s_{3}}
\left(c_{3}^{2}-s_{3}^{2}\right) \right] -\left(\zeta
_{23}^{-1}-\zeta_{23}\right) c_{\phi_{2}}^{}s_{\phi
_{2}}^{}c_{2}^{2}\right\}  \nonumber
\\
&&+C_\nu^{l}y_{\tau}^{2}\left\{2s_{\delta_\nu}^{}\left[
c_{x}^{2}s_{y}^{2}-c_{y}^{2}\left(c_{z}^{2}-s_{x}^{2}s_{z}^{2}\right)
+2c_\delta^{}c_{x}c_{y}s_{x}s_{y}s_{z}\right]
\frac{c_{1}c_{3}s_{1}s_{2}
}{s_{3}}\right.  \nonumber \\
&&\ \ \ \ \ \ \ -s_{\left(\rho
+\phi_{1}\right)}^{}\frac{c_{x}c_{y}^{2}c_{z}s_{z}}{c_{1}c_{2}s_{2}}\left[
s_{1}^{2}s_{2}^{2}+c_{1}^{2}\left(c_{2}^{2}-s_{2}^{2}\right)
\right] +s_{\left(\sigma
+\phi_{2}\right)}^{}\frac{c_{y}^{2}c_{z}s_{x}s_{z}}{c_{1}s_{1}}\left(c_{1}^{2}-s_{1}^{2}\right)  \nonumber \\
&&\ \ \ \ \ \ \ -s_{\left(\rho -\sigma
+\phi_{1}-\phi_{2}\right)}^{}
\frac{c_{x}s_{x}}{c_{2}s_{1}s_{2}}\left[
c_{2}^{2}s_{1}^{2}+\left(c_{1}^{2}-s_{1}^{2}\right)
s_{2}^{2}\right] \left(s_{y}^{2}-c_{y}^{2}s_{z}^{2}\right)  \nonumber \\
&&\ \ \ \ \ \ \ -s_{\left(\delta -\rho
-\phi_{1}\right)}^{}\frac{c_{y}c_{z}s_{x}s_{y}}{c_{1}c_{2}s_{2}}\left[
s_{1}^{2}s_{2}^{2}+c_{1}^{2} \left(c_{2}^{2}-s_{2}^{2}\right)
\right] -s_{\left(\delta -\sigma -\phi_{2}\right)
}^{}\frac{c_{x}c_{y}c_{z}s_{y}}{c_{1}s_{1}}\left(c_{1}^{2}-s_{1}^{2}\right)  \nonumber \\
&&\ \ \ \ \ \ \ -s_{\left(\delta_\nu+\rho
+\phi_{1}\right)}^{}\frac{c_{x}c_{y}^{2}c_{z}s_{1}s_{z}}{c_{2}c_{3}s_{3}}\left[
c_{2}^{2}-s_{2}^{2} \left(c_{3}^{2}-s_{3}^{2}\right) \right]
-2\left[ s_{\left(\delta_\nu-\sigma
-\phi_{2}\right)}^{}c_{1}^{2}-s_{\left(\delta_\nu+\sigma
+\phi_{2}\right)}^{}s_{1}^{2}\right] \frac{c_{3}c_{y}^{2}c_{z}s_{2}s_{x}s_{z}}{s_{3}}  \nonumber \\
&&\ \ \ \ \ \ \ +s_{\left(\delta_\nu+\rho -\sigma +\phi
_{1}-\phi_{2}\right)
}^{}\frac{c_{1}c_{x}s_{x}}{c_{2}c_{3}s_{3}}\left[
c_{2}^{2}-s_{2}^{2}\left(c_{3}^{2}-s_{3}^{2}\right) \right] \left(s_{y}^{2}-c_{y}^{2}s_{z}^{2}\right)  \nonumber \\
&&\ \ \ \ \ \ \ -s_{\left(\delta -\delta_\nu-\rho -\phi
_{1}\right)
}^{}\frac{c_{y}c_{z}s_{1}s_{x}s_{y}}{c_{2}c_{3}s_{3}}\left[
c_{2}^{2}-s_{2}^{2}\left(c_{3}^{2}-s_{3}^{2}\right) \right]  \nonumber \\
&&\ \ \ \ \ \ \ -2\left[ s_{\left(\delta +\delta_\nu-\sigma
-\phi_{2}\right)}^{}c_{1}^{2}+s_{\left(\delta -\delta_{\nu
}-\sigma -\phi_{2}\right)}^{}s_{1}^{2}\right]
\frac{c_{3}c_{x}c_{y}c_{z}s_{2}s_{y}}{s_{3}}  \nonumber \\
&&\ \ \ \ \ \ \ +\left[ c_{\left(\rho -\sigma +\phi_{1}-\phi
_{2}\right)}^{}s_\delta^{}+c_\delta^{}s_{\left(\rho -\sigma
+\phi_{1}-\phi_{2}\right)}^{}\left(c_{x}^{2}-s_{x}^{2}\right)
\right] \left[ c_{2}^{2}s_{1}^{2}+\left(c_{1}^{2}-s_{1}^{2}\right)
s_{2}^{2}\right] \frac{c_{y}s_{y}s_{z}}{c_{2}s_{1}s_{2}}  \nonumber \\
&&\ \ \ \ \ \ \ \left. -\left[ s_{\left(\delta +\delta_{\nu }+\rho
-\sigma +\phi_{1}-\phi_{2}\right)}^{}c_{x}^{2}+s_{\left(\delta
-\delta_\nu-\rho +\sigma -\phi_{1}+\phi_{2}\right)
}^{}s_{x}^{2}\right] \left[
c_{2}^{2}-s_{2}^{2}\left(c_{3}^{2}-s_{3}^{2}\right) \right]
\frac{c_{1}c_{y}s_{y}s_{z}}{c_{2}c_{3}s_{3}}\right\} ~.
\label{dDeltaNu}
\end{eqnarray}
\begin{eqnarray}
\dot{\phi}_{1}
&=&C_\kappa^{l}y_{\tau}^{2}\left\{\frac{s_{\left(\phi_{1}-\phi_{2}\right)}^{}}{\zeta_{12}}\left[
c_{\left(\rho -\sigma
\right)}^{}c_{x}s_{x}\left(s_{y}^{2}-c_{y}^{2}s_{z}^{2}\right)
-\left(c_{\left(\delta +\rho -\sigma
\right)}^{}c_{x}^{2}-c_{\left(\delta -\rho +\sigma
\right)}^{}s_{x}^{2}\right) c_{y}s_{y}s_{z}\right] \frac{c_{2}s_{1}}{s_{2}}\right.  \nonumber \\
&&\ \ \ \ \ \ \ \ \ +\zeta_{12}\cdot c_{\left(\phi_{1}-\phi
_{2}\right)}^{}\left[ s_{\left(\rho -\sigma \right)
}^{}c_{x}s_{x}\left(s_{y}^{2}-c_{y}^{2}s_{z}^{2}\right)
-\left(s_{\left(\delta +\rho -\sigma
\right)}^{}c_{x}^{2}+s_{\left(\delta -\rho +\sigma
\right)}^{}s_{x}^{2}\right) c_{y}s_{y}s_{z}\right]
\frac{c_{2}s_{1}}{s_{2}}
\nonumber \\
&&\ \ \ \ \ \ \ \ \ -\left[ \frac{s_{\phi_{1}}^{}}{\zeta
_{13}}\left(c_{\left(\delta -\rho \right)
}^{}s_{x}s_{y}-c_{\rho}^{}c_{x}c_{y}s_{z}\right) -\zeta _{13}\cdot
c_{\phi_{1}}^{}\left(s_{\left(\delta -\rho \right)
}^{}s_{x}s_{y}+s_{\rho}^{}c_{x}c_{y}s_{z}\right) \right]
\frac{c_{y}c_{z}\left(c_{1}^{2}c_{2}^{2}-s_{2}^{2}\right)}{c_{1}c_{2}s_{2}}  \nonumber \\
&&\ \ \ \ \ \ \ \ \ \left. -\left[
\frac{s_{\phi_{2}}^{}}{\zeta_{23}} \left(c_{\left(\delta -\sigma
\right)}^{}c_{x}s_{y}+c_{\sigma }^{}c_{y}s_{x}s_{z}\right)
-\zeta_{23}\cdot c_{\phi _{2}}^{}\left(s_{\left(\delta -\sigma
\right) }^{}c_{x}s_{y}-s_{\sigma}^{}c_{y}s_{x}s_{z}\right) \right]
\frac{c_{y}c_{z}s_{1}}{c_{1}}\right\}
\nonumber \\
&&+C_\kappa^\nu y_\nu^{2}\left\{-\left(\zeta
_{12}^{-1}-\zeta_{12}\right) c_{\left(\phi_{1}-\phi_{2}\right)
}^{}s_{\left(\phi_{1}-\phi_{2}\right)
}^{}c_{2}^{2}s_{1}^{2}-\left(\zeta_{13}^{-1}-\zeta_{13}\right)
c_{\phi_{1}}^{}s_{\phi_{1}}^{}\left(c_{1}^{2}c_{2}^{2}-s_{2}^{2}\right) \right.  \nonumber \\
&&\ \ \ \ \ \ \ \ \ \left. +\left(\zeta_{23}^{-1}-\zeta
_{23}\right)
c_{\phi_{2}}^{}s_{\phi_{2}}^{}c_{2}^{2}s_{1}^{2}\right\}  \nonumber \\
&&+C_\nu^{l}y_{\tau}^{2}\left\{\left(s_{\left(\delta -\rho
-\phi_{1}\right)}^{}s_{x}s_{y}+s_{\left(\rho +\phi_{1}\right)
}^{}c_{x}c_{y}s_{z}\right) \frac{c_{y}c_{z}\left(c_{1}^{2}+s_{1}^{2}s_{2}^{2}\right)}{c_{1}c_{2}s_{2}}\right.  \nonumber \\
&&\ \ \ \ \ \ \ \ \ \ -\left(s_{\left(\delta -\sigma -\phi
_{2}\right)}^{}c_{x}s_{y}-s_{\left(\sigma +\phi_{2}\right)
}^{}c_{y}s_{x}s_{z}\right) \frac{c_{y}c_{z}s_{1}}{c_{1}}+\left[
s_{\left(\rho -\sigma +\phi_{1}-\phi_{2}\right)
}^{}c_{x}s_{x}\left(s_{y}^{2}-c_{y}^{2}s_{z}^{2}\right) \right.  \nonumber \\
&&\ \ \ \ \ \ \ \ \left. \left. -\left(s_{\left(\delta +\rho
-\sigma +\phi_{1}-\phi_{2}\right)}^{}c_{x}^{2}+s_{\left(\delta
-\rho +\sigma -\phi_{1}+\phi_{2}\right)}^{}s_{x}^{2}\right)
c_{y}s_{y}s_{z}\right] \frac{c_{2}s_{1}}{s_{2}}\right\} ~.
\label{dPhi1}
\end{eqnarray}
\begin{eqnarray}
\dot{\phi}_{2}
&=&C_\kappa^{l}y_{\tau}^{2}\left\{\frac{s_{\left(\phi_{1}-\phi_{2}\right)}^{}}{\zeta_{12}}\left[
c_{\left(\rho -\sigma
\right)}^{}c_{x}s_{x}\left(s_{y}^{2}-c_{y}^{2}s_{z}^{2}\right)
-\left(c_{\left(\delta +\rho -\sigma
\right)}^{}c_{x}^{2}-c_{\left(\delta -\rho +\sigma
\right)}^{}s_{x}^{2}\right) c_{y}s_{y}s_{z}\right] \frac{s_{2}}{c_{2}s_{1}}\right\}  \nonumber \\
&&\ \ \ \ \ \ \ +\zeta_{12}\cdot c_{\left(\phi_{1}-\phi
_{2}\right)}^{} \left[ s_{\left(\rho -\sigma \right)
}^{}c_{x}s_{x}\left(s_{y}^{2}-c_{y}^{2}s_{z}^{2}\right)
-\left(s_{\left(\delta +\rho -\sigma
\right)}^{}c_{x}^{2}+s_{\left(\delta -\rho +\sigma
\right)}^{}s_{x}^{2}\right) c_{y}s_{y}s_{z}\right]
\frac{s_{2}}{c_{2}s_{1}}
\nonumber \\
&&\ \ \ \ \ \ \ +\left[ \frac{s_{\phi_{1}}^{}}{\zeta
_{13}}\left(c_{\left(\delta -\rho \right)
}^{}s_{x}s_{y}-c_{\rho}^{}c_{x}c_{y}s_{z}\right) -\zeta _{13}\cdot
c_{\phi_{1}}^{}\left(s_{\left(\delta -\rho \right)
}^{}s_{x}s_{y}+s_{\rho}^{}c_{x}c_{y}s_{z}\right) \right]
\frac{c_{y}c_{z}s_{2}}{c_{1}c_{2}}
\nonumber \\
&&\ \ \ \ \ \ \ \left.+\left[ \frac{s_{\phi_{2}}^{}}{\zeta
_{23}}\left(c_{\left(\delta -\sigma \right)
}^{}c_{x}s_{y}+c_{\sigma}^{}c_{y}s_{x}s_{z}\right) -\zeta
_{23}\cdot c_{\phi_{2}}^{}\left(s_{\left(\delta -\sigma
\right)}^{}c_{x}s_{y}-s_{\sigma}^{}c_{y}s_{x}s_{z}\right) \right]
\frac{c_{y}c_{z}\left(c_{1}^{2}-s_{1}^{2}\right)}{c_{1}s_{1}}
\right\} \nonumber \\
&&+C_\kappa^\nu y_\nu^{2}\left\{-\left(\zeta
_{12}^{-1}-\zeta_{12}\right) c_{\left(\phi_{1}-\phi_{2}\right)
}^{}s_{\left(\phi_{1}-\phi_{2}\right)}^{}s_{2}^{2}+
\left(\zeta_{13}^{-1}-\zeta_{13}\right) c_{\phi_{1}}^{}s_{\phi
_{1}}^{}s_{2}^{2}\right.  \nonumber \\
&&\ \ \ \ \ \ \ \ \left.-\left(\zeta_{23}^{-1}-\zeta _{23}\right)
c_{\phi_{2}}^{}s_{\phi_{2}}^{}c_{2}^{2}\left(c_{1}^{2}-s_{1}^{2}\right) \right\}  \nonumber \\
&&+C_\nu^{l}y_{\tau}^{2}\left\{\left(s_{\left(\rho -\phi
_{1}\right)}^{}c_{x}s_{1}s_{2}+s_{\left(\sigma +\phi
_{2}\right)}^{}c_{2}s_{x}\right)
\frac{c_{y}^{2}c_{z}s_{z}}{c_{1}c_{2}s_{1}} -s_{\left(\rho -\sigma
+\phi_{1}-\phi_{2}\right)}^{}\frac{c_{x}s_{2}s_{x}\left(s_{y}^{2}-c_{y}^{2}s_{z}^{2}\right)}{c_{2}s_{1}}
\right.  \nonumber \\
&&\ \ \ \ \ \ \ \ \ +\left(s_{\left(\delta -\rho -\phi
_{1}\right)}^{}s_{1}s_{2}s_{x}-s_{\left(\delta -\sigma -\phi
_{2}\right)
}^{}c_{2}c_{x}\right) \frac{c_{y}c_{z}s_{y}}{c_{1}c_{2}s_{1}}  \nonumber \\
&&\ \ \ \ \ \ \ \ \ \left. +\left(s_{\left(\delta +\rho -\sigma
+\phi_{1}-\phi_{2}\right)}^{}c_{x}^{2}+s_{\left(\delta -\rho
+\sigma -\phi_{1}+\phi_{2}\right)}^{}s_{x}^{2}\right)
\frac{c_{y}s_{2}s_{y}s_{z}}{c_{2}s_{1}}\right\} ~.  \label{dPhi2}
\end{eqnarray}

%\newpage
%%%%%%%%%%%%%%%%%%%%%%%%%%%%%%%%%%%%%%%%%%%%%%%%%%%%%%%%%%%%%%%%%%%%%%%%%%%%%%%%%%%%%

%\newpage
%%%%%%%%%%%%%%%%%%%%%%%%%%%%%%%%%%%%%%%%%%%%%%%%%%%%%%%%%%%%%%%%%%%%%%%%%%%%%%%%%%%
%%%%%%%%%%%%%%%%%%%%%%%%%%%%%%%%%%%%%%%%%%%%%%%%%%%%%%%%%%%%%%%%%%%%%%%%%%%%%%%%%%%
\begin{table}[tbp]
\caption{Functions of $\delta $, $\rho $ and $\sigma $ that can be
factorized out together with $\zeta_{ij}^{-1}$ (for $i<j;
~i,j=1,2,3$) defined in Eq.(\ref{ratio}), both in contributions
from $y_{\tau}$ and $y_\nu$ in Eqs.(\ref{dX})-(\ref{dSigma}). Note
that terms led by $\zeta_{13}^{-1}$ and $\zeta_{23}^{-1}$ in
Eq.(\ref{dDelta}) are complicated, so that the last two rows in
the table do not apply to $\dot\delta$.} \label{TableCase2a}

\begin{center}
\begin{tabular}{|c|cccc|}
%\hline
 & $\dot{\theta}_{x}$ & $\dot{\theta}_{y}$ &
$\dot{\theta}_{z}$ & $\dot\delta$, $\dot{\rho}$, $\dot{\sigma}$
\\ \hline $\zeta_{12}^{-1}$ & $c_{\left(\rho -\sigma
\right)}^{}$ & 0 & 0 & $ s_{\left(\rho -\sigma \right)}^{}$
\\ \hline $\zeta_{13}^{-1}$ & $c_{\rho}^{}$ & $c_{\left(\delta
-\rho \right)}^{}$ & $c_{\rho}^{}$ & $s_{\rho}^{}$ \\ \hline
$\zeta_{23}^{-1}$ & $c_{\sigma}^{}$ & $c_{\left(\delta -\sigma
\right)}^{}$ & $c_{\sigma}^{}$ & $s_{\sigma}^{}$ \\
%\hline
\end{tabular}
\end{center}
\end{table}

%%%%%%%%%%%%%%%%%%%%%%%%%%%%%%%%%%%%%%%%%%%%%%%%%%%%%%%%%%%%%%%%%%%%%%%%%%%%%%%%%%%
\begin{table}[tbp]
\caption{Association of CP-violating phases with enhancing factors
$\zeta_{ij}^{-1}$ (for $i<j; ~i,j=1,2,3$) defined in
Eq.(\ref{ratio}), in the contribution from $y_\nu$ in
Eqs.(\ref{dX})-(\ref{dSigma}). Note that terms led by
$\zeta_{13}^{-1}$ and $\zeta_{23}^{-1}$ in Eq.(\ref{dDelta}) are
complicated, so that the last two rows in the table do not apply
to $\dot\delta$.} \label{TableCase2b}

\begin{center}
\begin{tabular}{|c|cccc|}
%\hline
 & $\dot{\theta}_{x}$ & $\dot{\theta}_{y}$ &
$\dot{\theta}_{z}$ & $\dot\delta$, $\dot{\rho}$, $\dot{\sigma}$
\\ \hline $\zeta_{12}^{-1}$ & $-c_{\left(\rho -\sigma
\right)}^{}c_{\left(\phi_{1}-\phi_{2}\right)}^{}$ & 0 & 0 &
$-c_{\left(\phi_{1}-\phi_{2}\right)}^{}s_{\left(\rho -\sigma
\right)}^{}$ \\ \hline $\zeta_{13}^{-1}$ &
$-c_{\rho}^{}c_{\phi_{1}}^{}$ & $c_{\left(\delta -\rho
\right)}^{}c_{\phi_{1}}^{}$ & $-c_{\rho
}^{}c_{\phi_{1}}^{}$ & $c_{\phi_{1}}^{}s_{\rho}^{}$ \\
\hline $\zeta_{23}^{-1}$ & $c_{\sigma}^{}c_{\phi_{2}}^{}$ &
$-c_{\left(\delta -\sigma \right)}^{}c_{\phi_{2}}^{}$ &
$-c_{\sigma}^{}c_{\phi_{2}}^{}$ & $c_{\phi_{2}}^{}s_{\sigma
}^{}$ \\ %\hline
\end{tabular}
\end{center}
\end{table}

%\newpage
%%%%%%%%%%%%%%%%%%%%%%%%%%%%%%%%%%%%%%%%%%%%%%%%%%%%%%%%%%%%%%%%%%%%%%%%%%%%%%%%%%%
%%%%%%%%%%%%%%%%%%%%%%%%%%%%%%%%%%%%%%%%%%%%%%%%%%%%%%%%%%%%%%%%%%%%%%%%%%%%%%%%%%%
\begin{figure}[tbp]
\begin{center}
\includegraphics[width=8cm,height=8cm,angle=0]{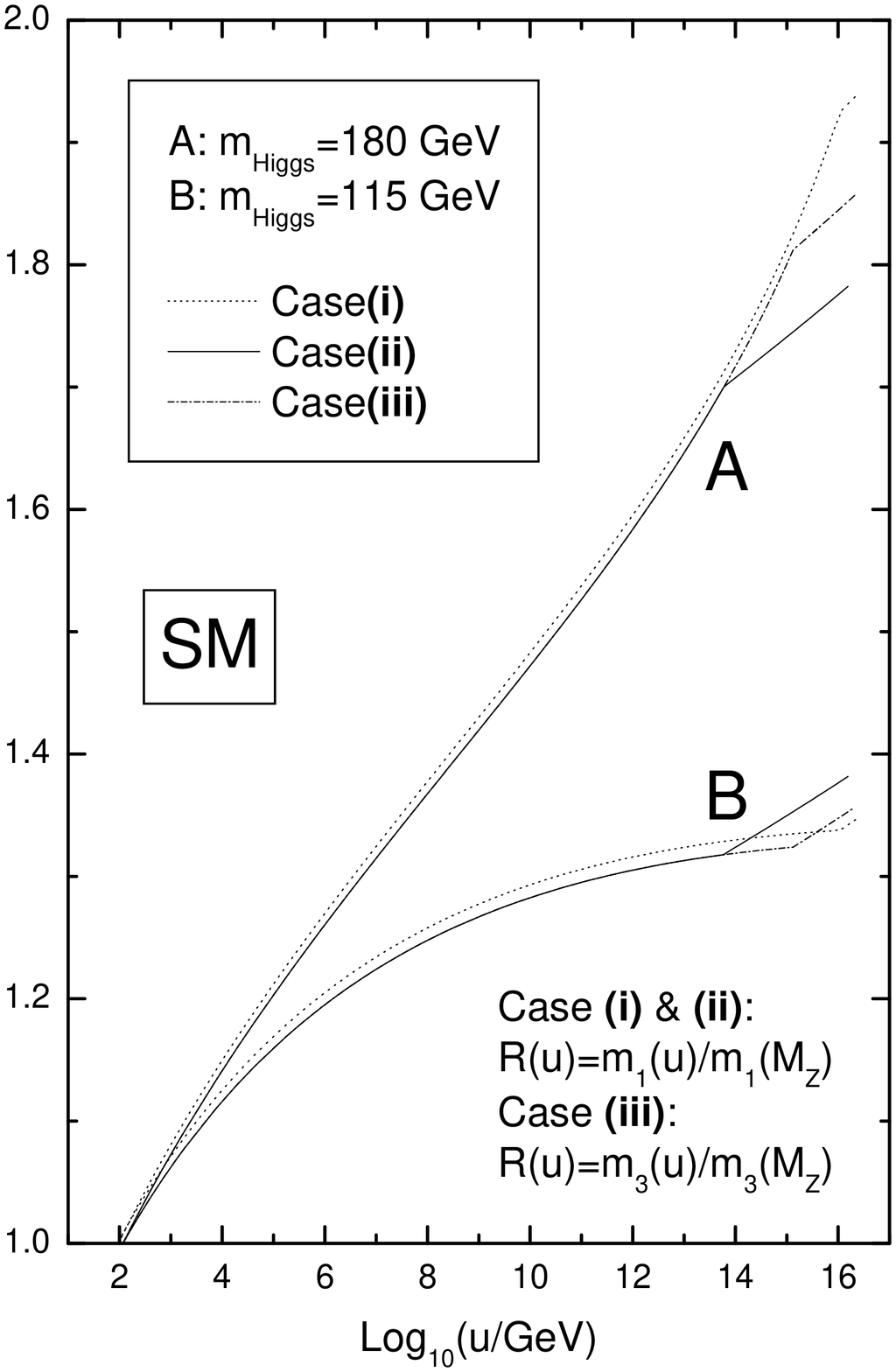}
\includegraphics[width=8cm,height=8cm,angle=0]{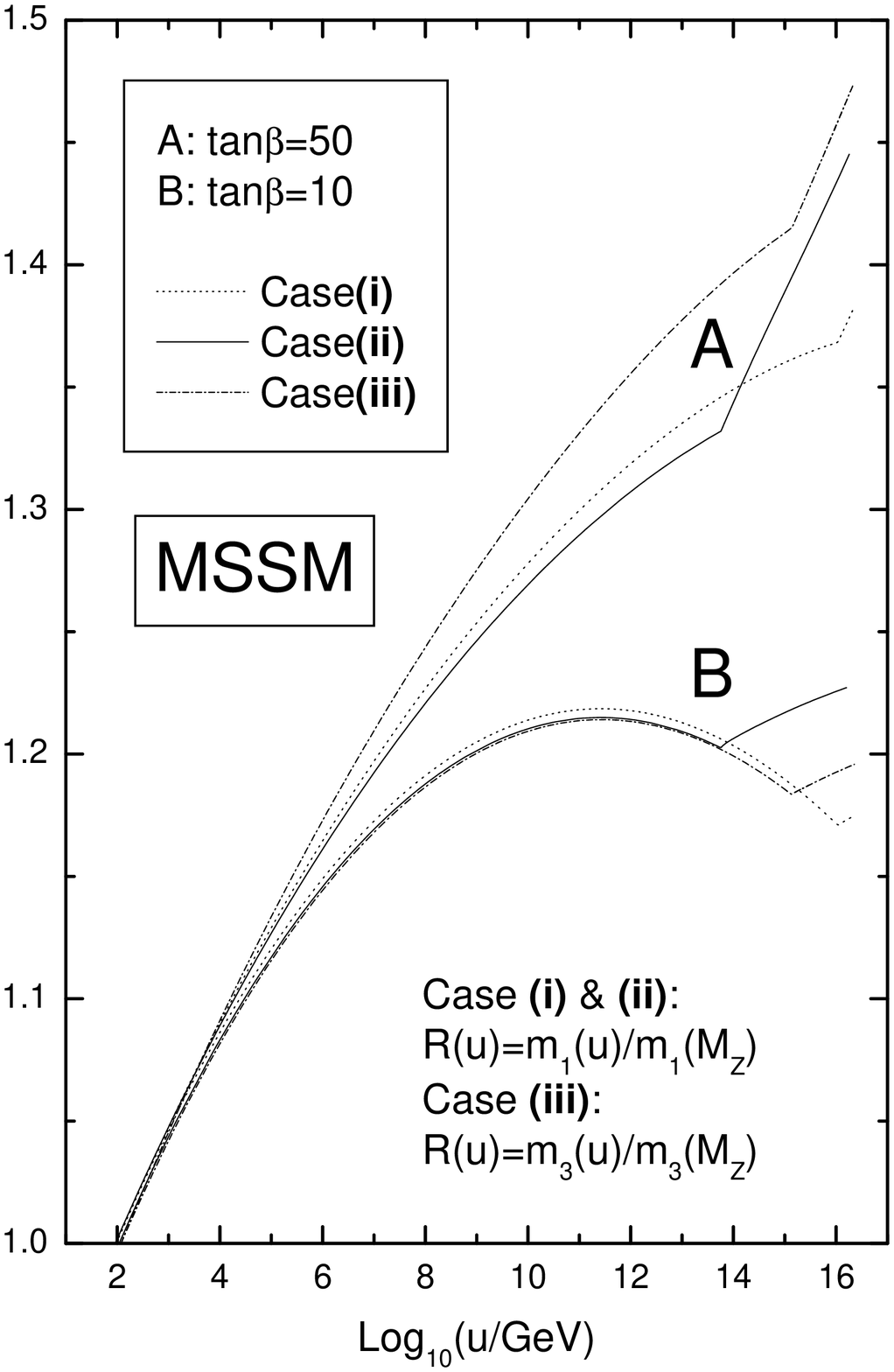}
\end{center}
\caption{The RG evolution of a left-handed neutrino mass ($m_{1}$
of cases (i) and (ii), and $m_{3}$ of case (iii)) from $M_{\rm Z}$
to $\Lambda_{GUT}$, both in the SM and in the MSSM. At $M_{\rm
Z}$, $ \theta_{x}=33.2^{\circ}$, $\theta_{y}=45^{\circ}$, $
\theta_{z}=5^{\circ}$, $\delta =45^{\circ}$ and $\rho =\sigma
=5^{\circ}$. At $M_{3}$, $y_\nu=0.8$, $
\theta_{1}=\theta_{2}=45^{\circ}$ and
$\theta_{3}=\delta_\nu=\phi_{1}=\phi_{2}=5^{\circ}$.}
\label{FigMassCommonScaling}
\end{figure}

%%%%%%%%%%%%%%%%%%%%%%%%%%%%%%%%%%%%%%%%%%%%%%%%%%%%%%%%%%%%%%%%%%%%%%%%%%%%%%%%%%%
\begin{figure}[tbp]
\begin{center}
\includegraphics[width=8cm,height=8cm,angle=0]{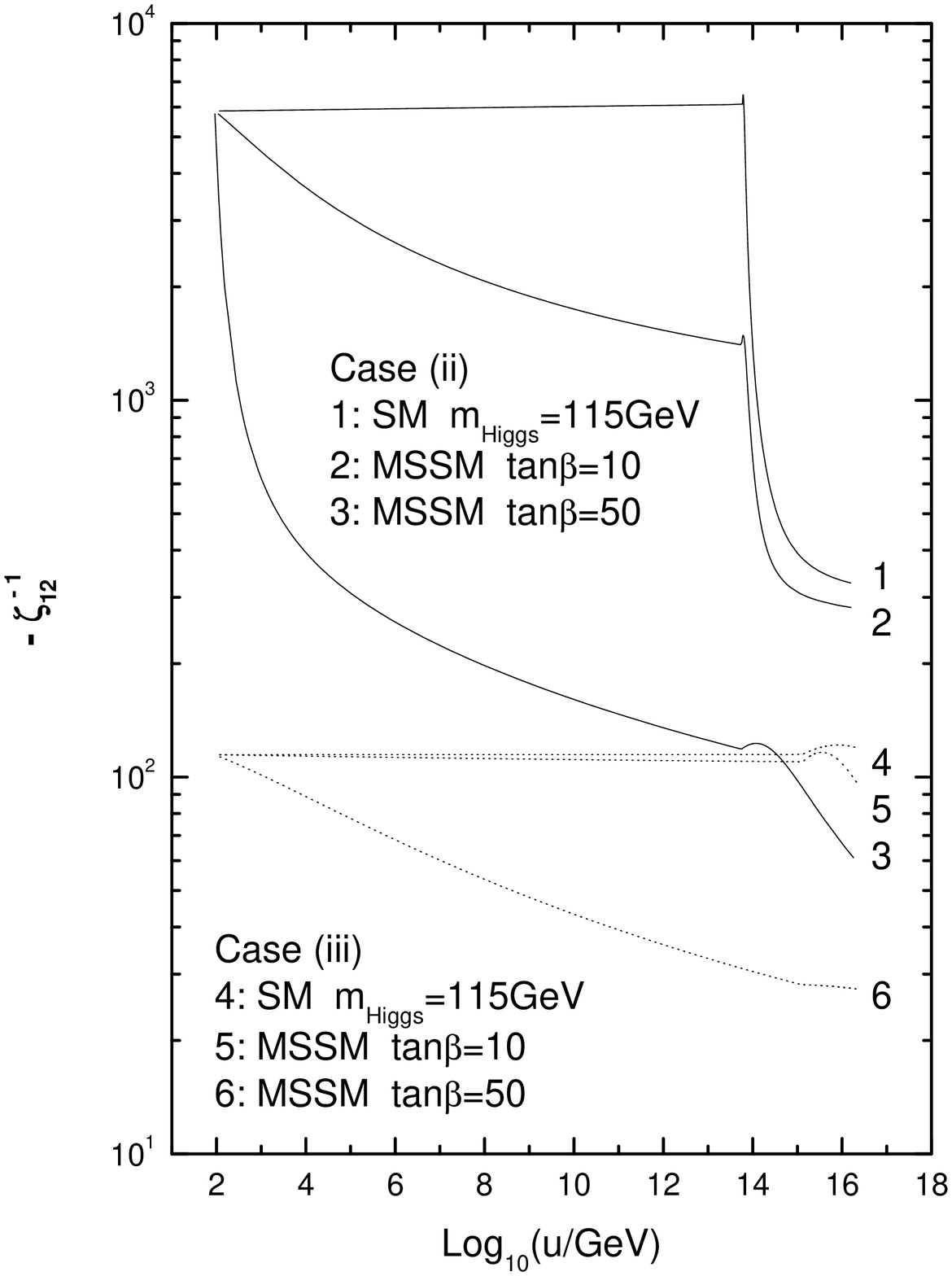} %
\includegraphics[width=8cm,height=8cm,angle=0]{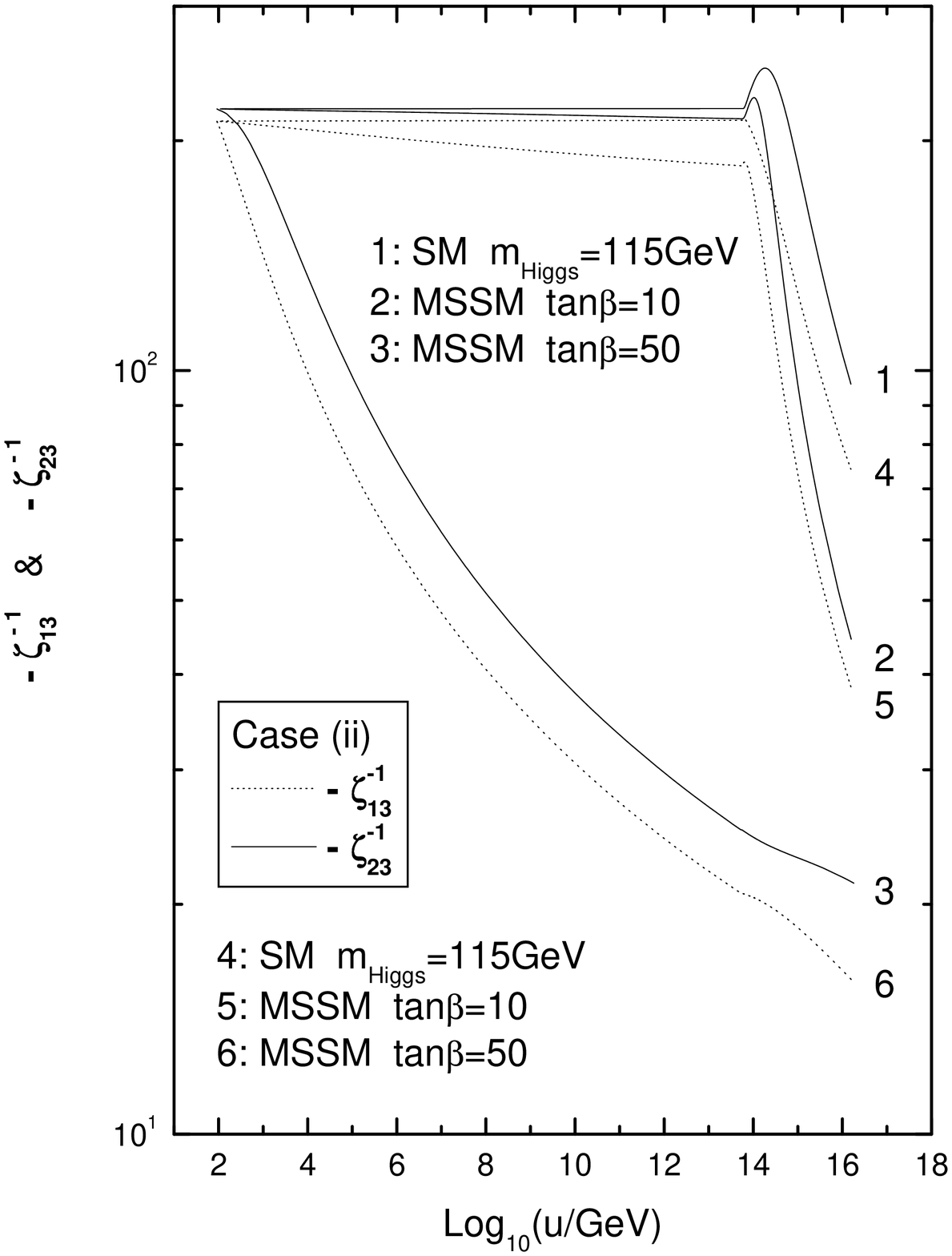}
\end{center}
\caption{The typical evolution behavior of enhancing factors
$\left(-\zeta_{ij}^{-1}\right)$ defined in Eq.(\ref{ratio}), both
in the SM and in the MSSM. At $M_{\rm Z}$, $
\theta_{x}=33.2^{\circ}$, $\theta_{y}=45^{\circ},
~\theta_{z}=5^{\circ}, ~\delta =45^{\circ}$ and $\rho = \sigma
=5^{\circ}$. At $M_{3}$, $y_\nu=0.8$, $\theta_{1}=
\theta_{2}=45^{\circ}$ and $\theta_{3}=
\delta_\nu=\phi_{1}=\phi_{2}=5^{\circ} $.} \label{FigMassZeta}
\end{figure}

%%%%%%%%%%%%%%%%%%%%%%%%%%%%%%%%%%%%%%%%%%%%%%%%%%%%%%%%%%%%%%%%%%%%%%%%%%%%%%%%%%%
\begin{figure}[tbp]
\begin{center}
\includegraphics[width=8cm,height=8cm,angle=0]{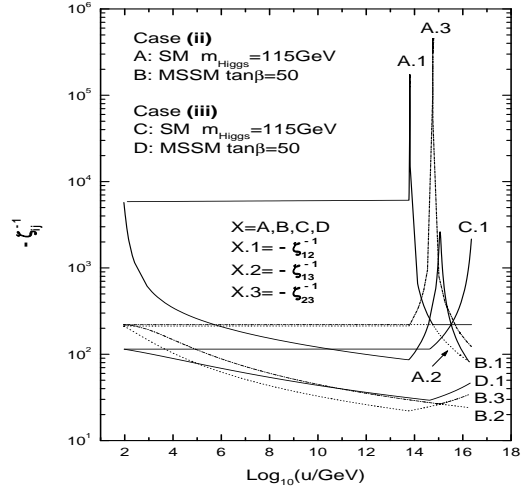}
\end{center}
\caption{Typical peaks of enhancing factors
$\left(-\zeta_{ij}^{-1}\right)$ defined in Eq.(\ref{ratio}) in
nearly singular situations. At $M_{\rm Z}$,
$\theta_{x}=33.2^{\circ}$, $\theta _{y}=45^{\circ}$,
$\theta_{z}=5^{\circ}$, $\delta =55^{\circ}$ and $\rho =\sigma
=5^{\circ}$. At $M_{3}$, $y_\nu=0.8$, $\theta_{1}=$
$\theta_{2}=89^{\circ}$ and $\theta_{3}=\delta_\nu=\phi_{1}=
\phi_{2}=5^{\circ}$.} \label{FigMassPeaks}
\end{figure}

%%%%%%%%%%%%%%%%%%%%%%%%%%%%%%%%%%%%%%%%%%%%%%%%%%%%%%%%%%%%%%%%%%%%%%%%%%%%%%%%%%%
\begin{figure}[tbp]
\begin{center}
\includegraphics[width=8cm,height=8cm,angle=0]{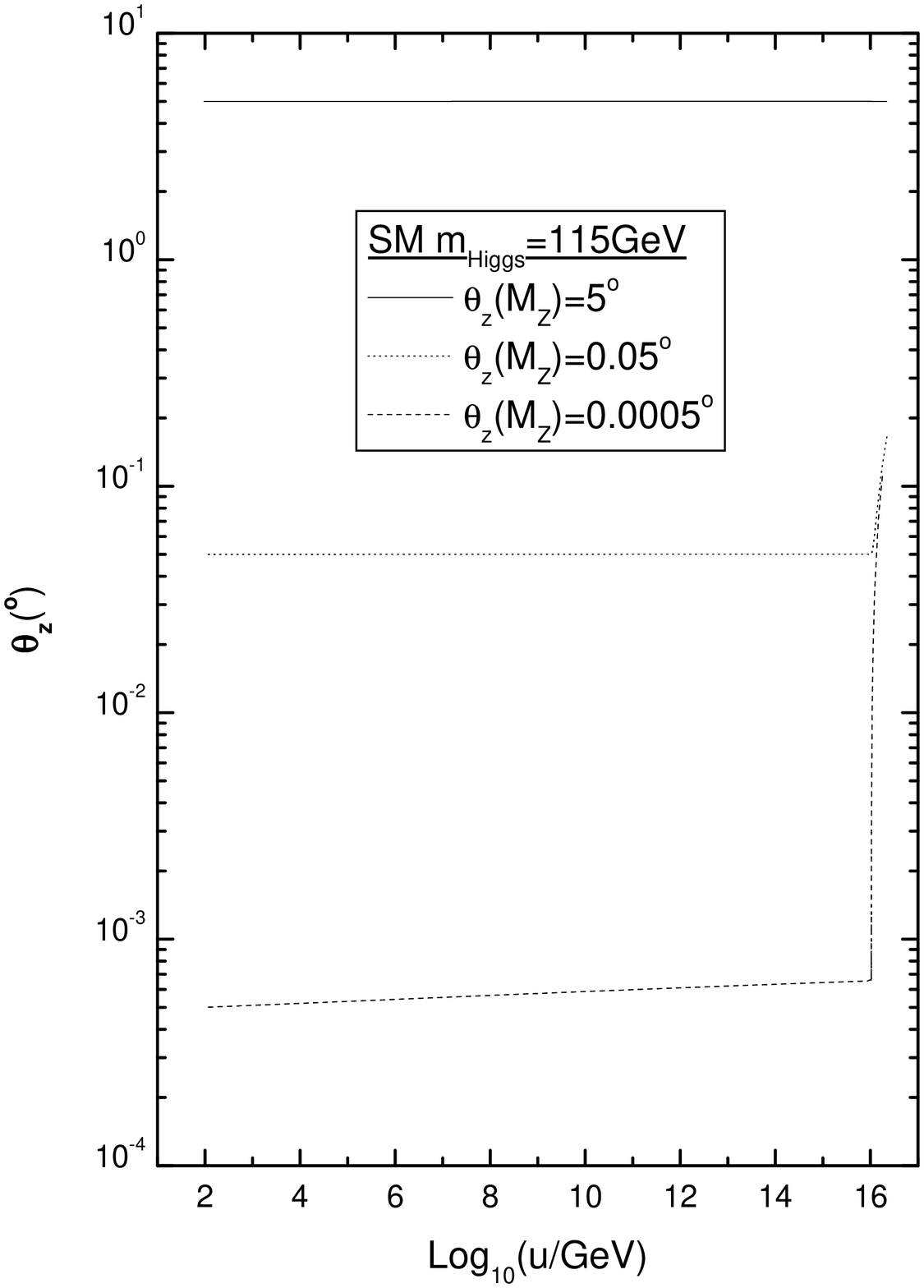}
\includegraphics[width=8cm,height=8cm,angle=0]{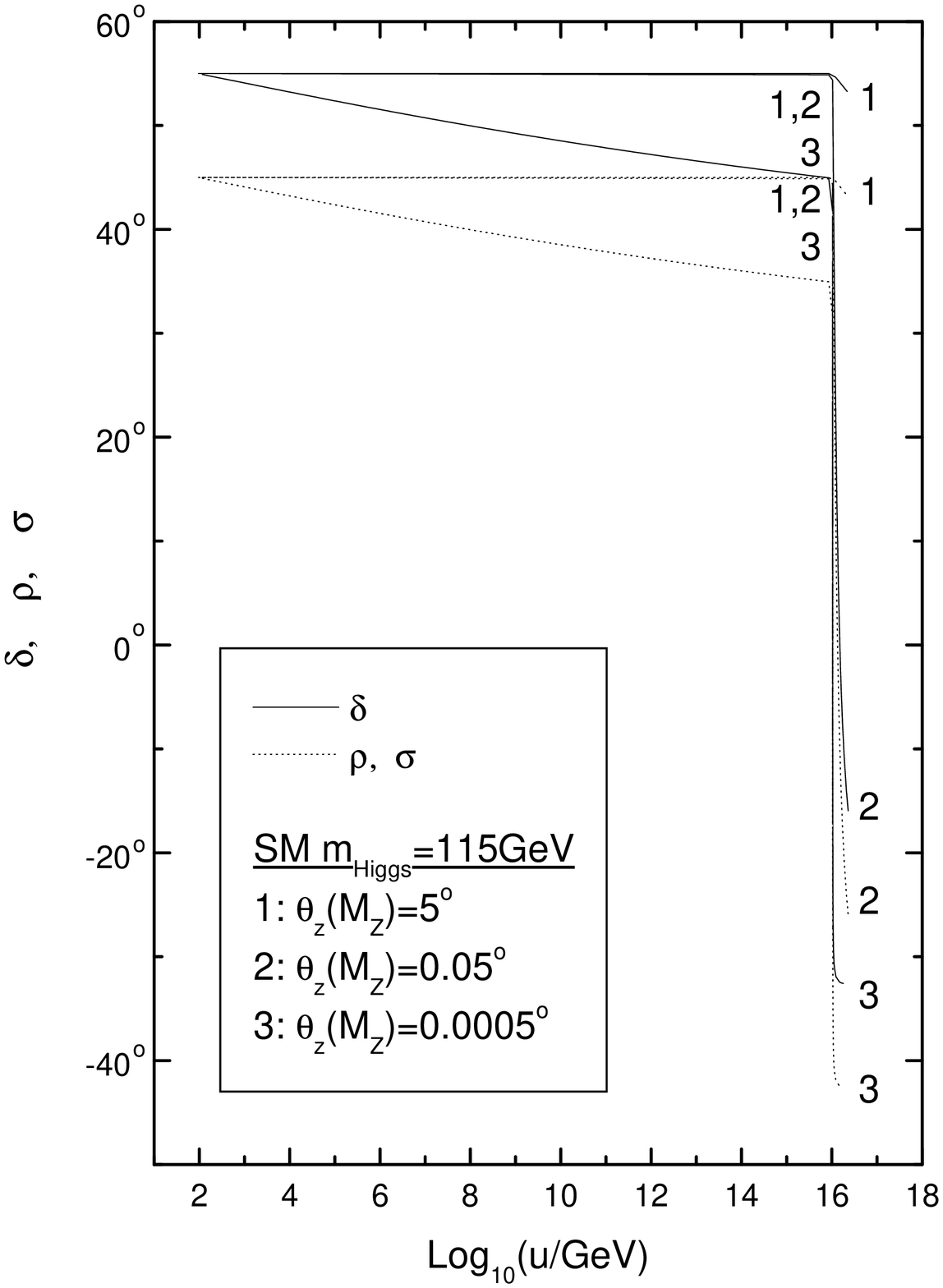}
\end{center}
\caption{The typical evolution behavior of the mixing angle
$\theta_{z}$ and CP-violating phases ($\delta, \rho, \sigma$) in
the SM, in the normal hierarchy case. At $M_{\rm Z}$, $\theta
_{x}=33.2^{\circ}$, $\theta_{y}=45^{\circ}$, $\delta =55^{\circ}$
and $\rho =\sigma =45^{\circ}$. At $M_{3}$, $y_\nu=0.8$,
$\theta_{1}=\theta_{2}=\phi_{1}=\phi_{2}=45^{\circ}$ and
$\theta_{3}=\delta_\nu=5^{\circ}$.} \label{FigCase1SM}
\end{figure}

%%%%%%%%%%%%%%%%%%%%%%%%%%%%%%%%%%%%%%%%%%%%%%%%%%%%%%%%%%%%%%%%%%%%%%%%%%%%%%%%%%%
\begin{figure}[tbp]
\begin{center}
\includegraphics[width=8cm,height=8cm,angle=0]{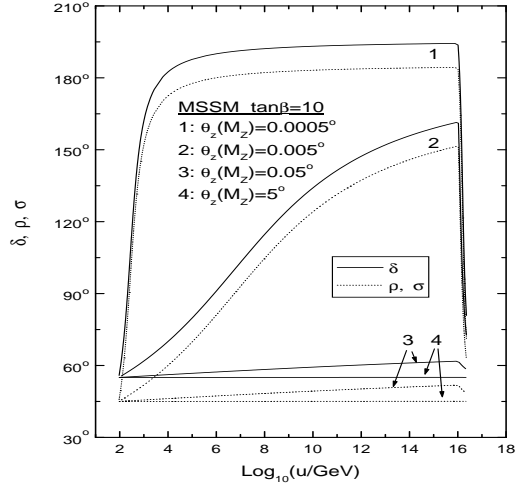}

\includegraphics[width=8cm,height=8cm,angle=0]{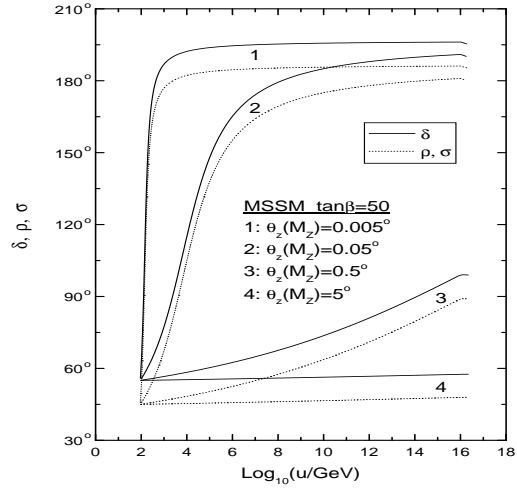}
\end{center}
\caption{The typical evolution behavior of CP-violating phases
($\delta, \rho, \sigma$) in the MSSM, in the normal hierarchy
case. At $M_{\rm Z}$, $\theta_{x}=33.2^{\circ}$,
$\theta_{y}=45^{\circ}$, $\delta =55^{\circ}$ and $\rho = \sigma
=45^{\circ}$. At $M_{3}$, $y_\nu=0.8$, $\theta_{1}=$
$\theta_{2}=\phi_{1}=\phi_{2}=45^{\circ}$ and
$\theta_{3}=\delta_\nu=5^{\circ}$.} \label{FigCase1MSSM}
\end{figure}

%\newpage
%%%%%%%%%%%%%%%%%%%%%%%%%%%%%%%%%%%%%%%%%%%%%%%%%%%%%%%%%%%%%%%%%%%%%%%%%%%%%%%%%%%
\begin{figure}[tbp]
\begin{center}
\includegraphics[width=8cm,height=8cm,angle=0]{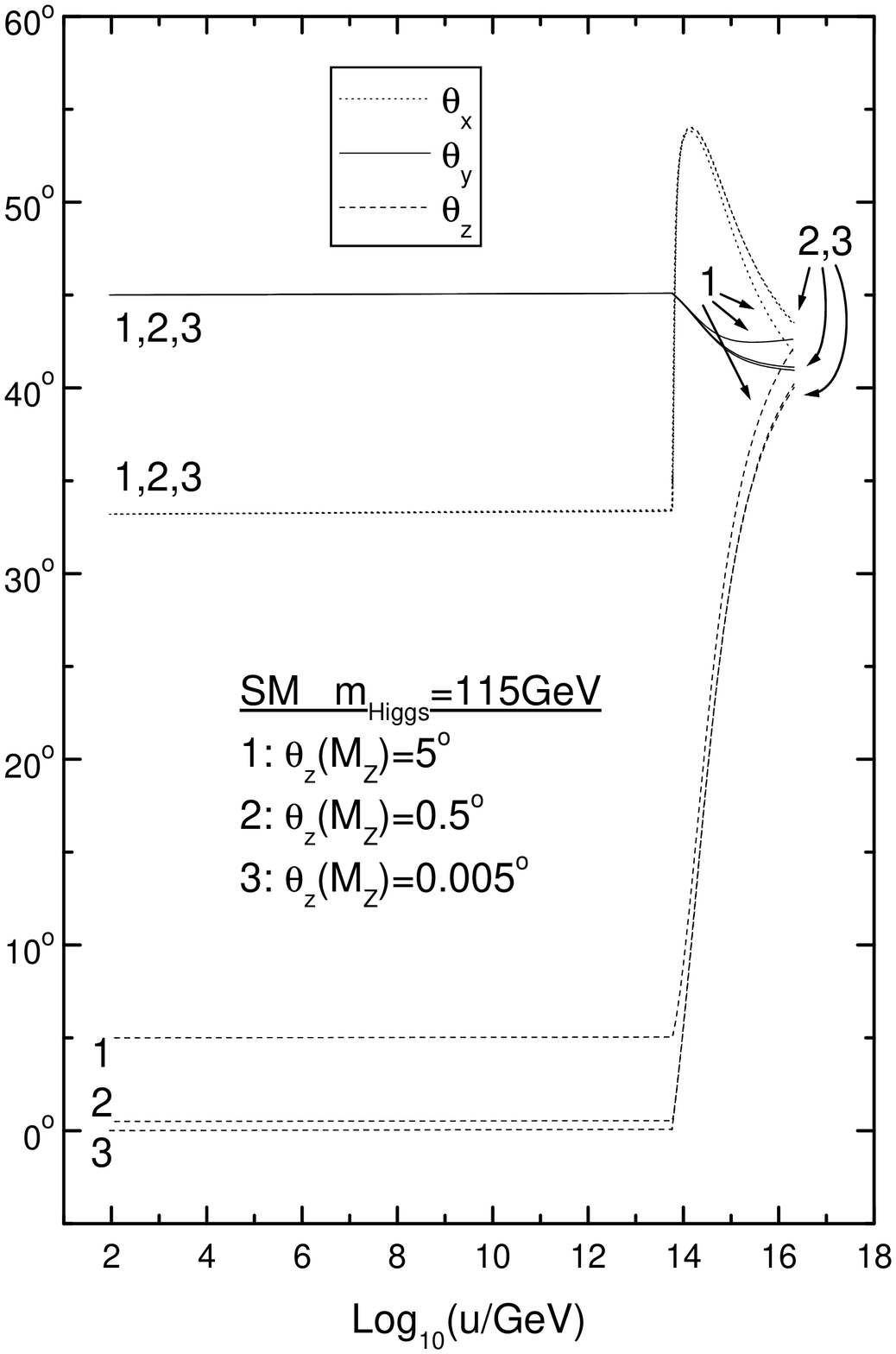}
\includegraphics[width=8cm,height=8cm,angle=0]{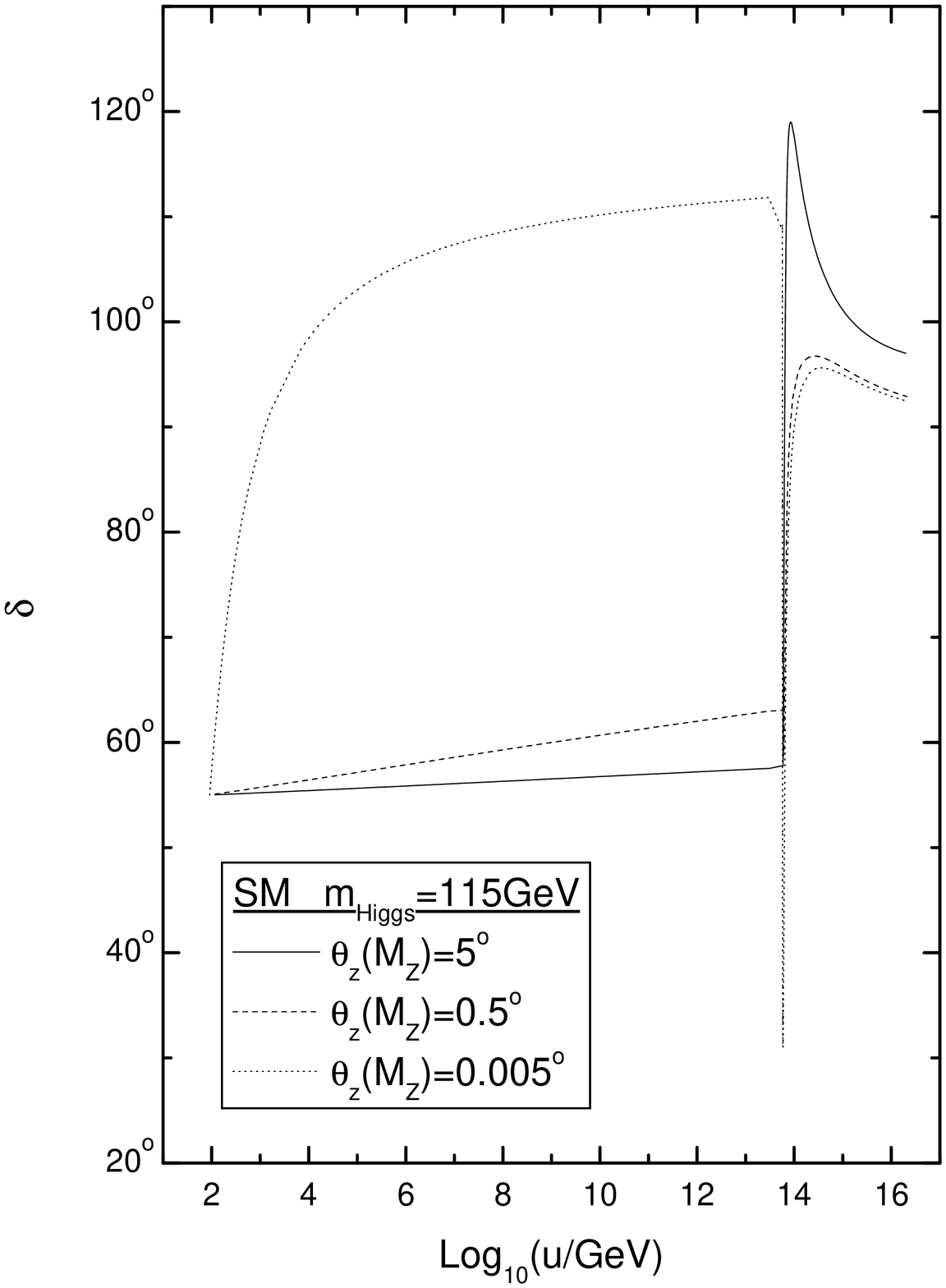}

\includegraphics[width=8cm,height=8cm,angle=0]{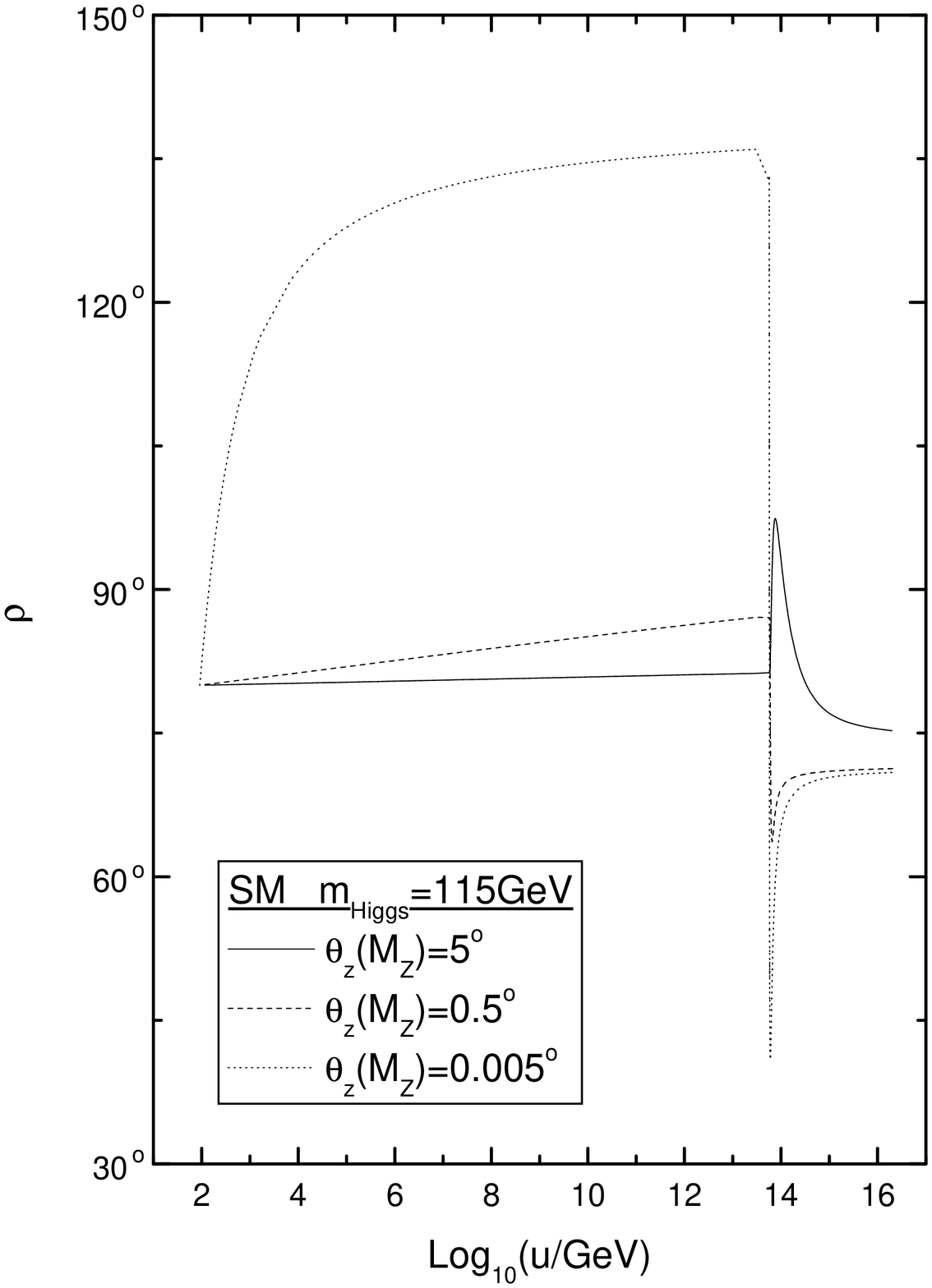}
\includegraphics[width=8cm,height=8cm,angle=0]{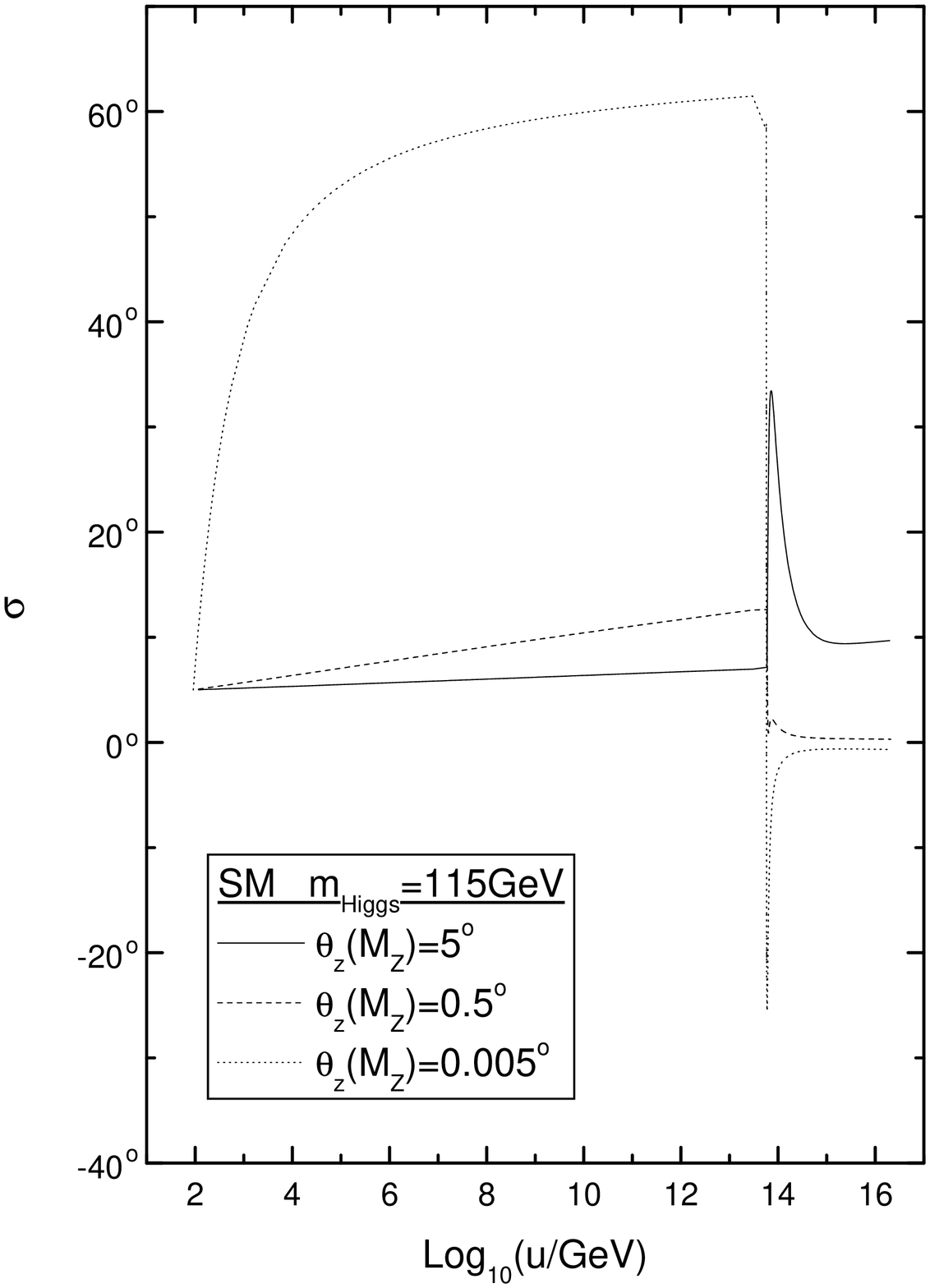}
\end{center}
\caption{The typical evolution behavior of mixing angles
($\theta_{x}, \theta_{y}, \theta_{z}$) and CP-violating phases
($\delta, \rho, \sigma$) in the SM, in the near degeneracy case.
At $M_{\rm Z}$, $\theta_{x}=33.2^{\circ}$, $\theta
_{y}=45^{\circ}$, $\delta =55^{\circ}$, $\rho =80^{\circ}$ and
$\sigma =5^{\circ}$. At $M_{3}$, $y_\nu=0.8$, $\theta_{1}=$
$\theta_{2}=45^{\circ}$ and
$\phi_{1}=\phi_{2}=\theta_{3}=\delta_\nu=5^{\circ}$.}
\label{FigCase2SM}
\end{figure}

%%%%%%%%%%%%%%%%%%%%%%%%%%%%%%%%%%%%%%%%%%%%%%%%%%%%%%%%%%%%%%%%%%%%%%%%%%%%%%%%%%%
\begin{figure}[tbp]
\begin{center}
\includegraphics[width=8cm,height=8cm,angle=0]{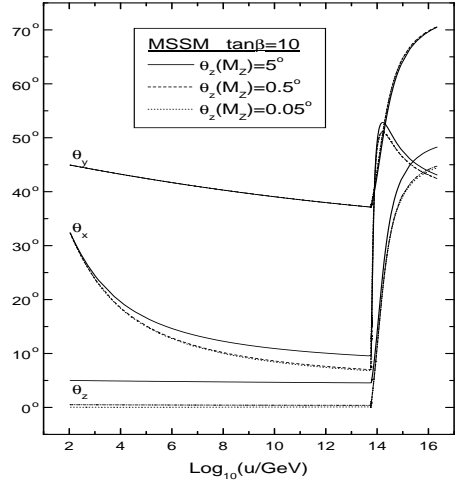}

\includegraphics[width=8cm,height=8cm,angle=0]{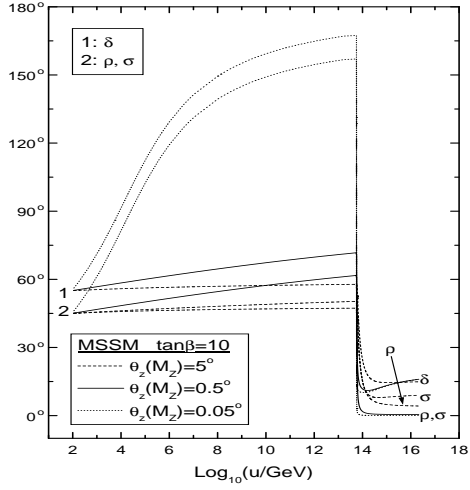}
\end{center}
\caption{The typical evolution behavior of mixing angles
($\theta_{x}, \theta_{y}, \theta_{z}$) and CP-violating phases
($\delta, \rho, \sigma$) in the MSSM when $\tan \beta $ is small,
in the near degeneracy case. At $M_{\rm Z}$,
$\theta_{x}=33.2^{\circ}$, $\theta_{y}=45^{\circ}$, $ \delta
=55^{\circ}$ and $\rho =\sigma =45^{\circ}$. At $M_{3}$,
$y_\nu=0.8$, $\theta_{1}=\theta_{2}=45^{\circ}$ and
$\phi_{1}=\phi_{2}=\theta_{3}= \delta_\nu=5^{\circ}$.}
\label{FigCase2MSSM10}
\end{figure}

%%%%%%%%%%%%%%%%%%%%%%%%%%%%%%%%%%%%%%%%%%%%%%%%%%%%%%%%%%%%%%%%%%%%%%%%%%%%%%%%%%%
\begin{figure}[tbp]
\begin{center}
\includegraphics[width=8cm,height=8cm,angle=0]{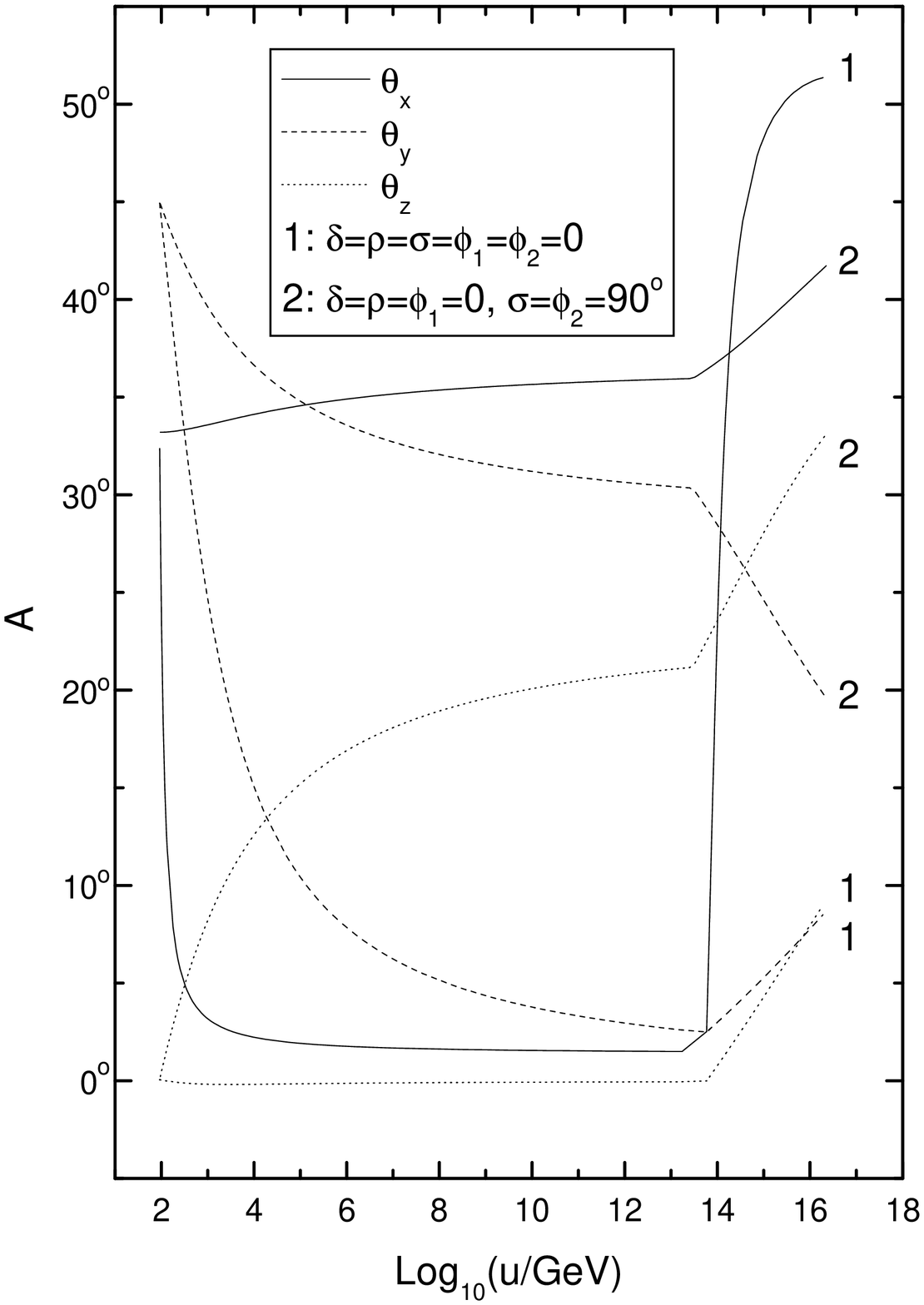}
\includegraphics[width=8cm,height=8cm,angle=0]{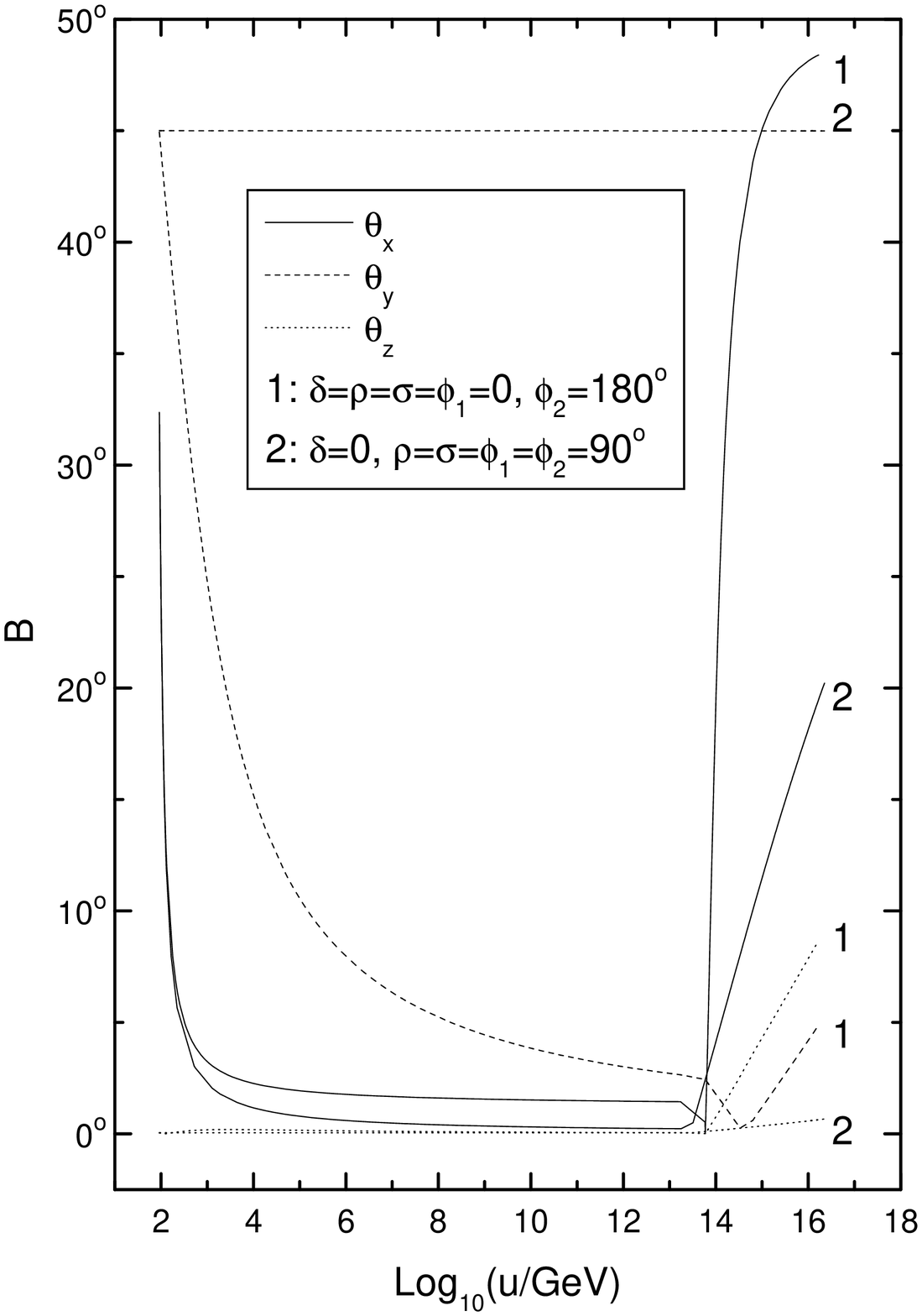}
\par
\includegraphics[width=8cm,height=8cm,angle=0]{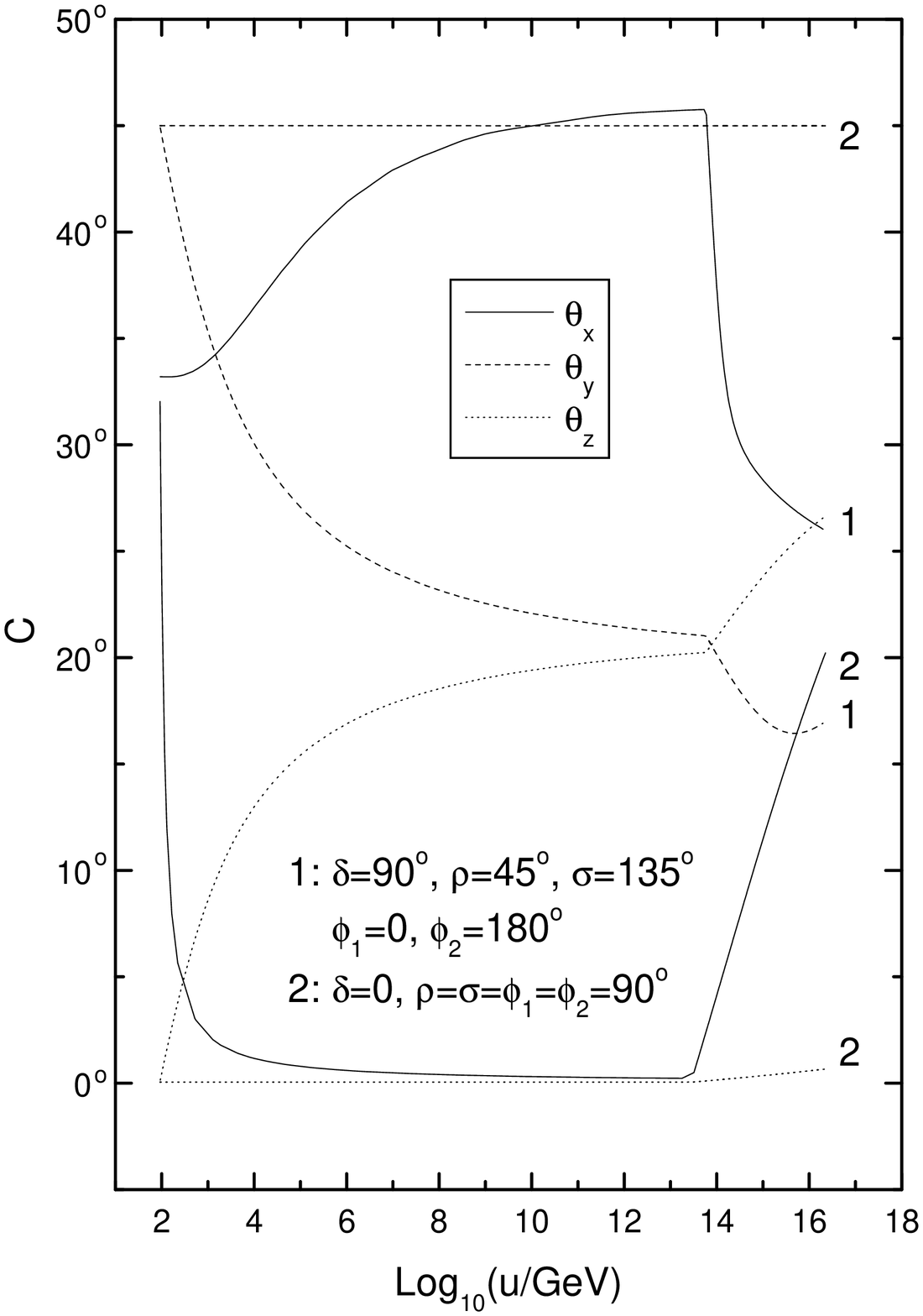}
\includegraphics[width=8cm,height=8cm,angle=0]{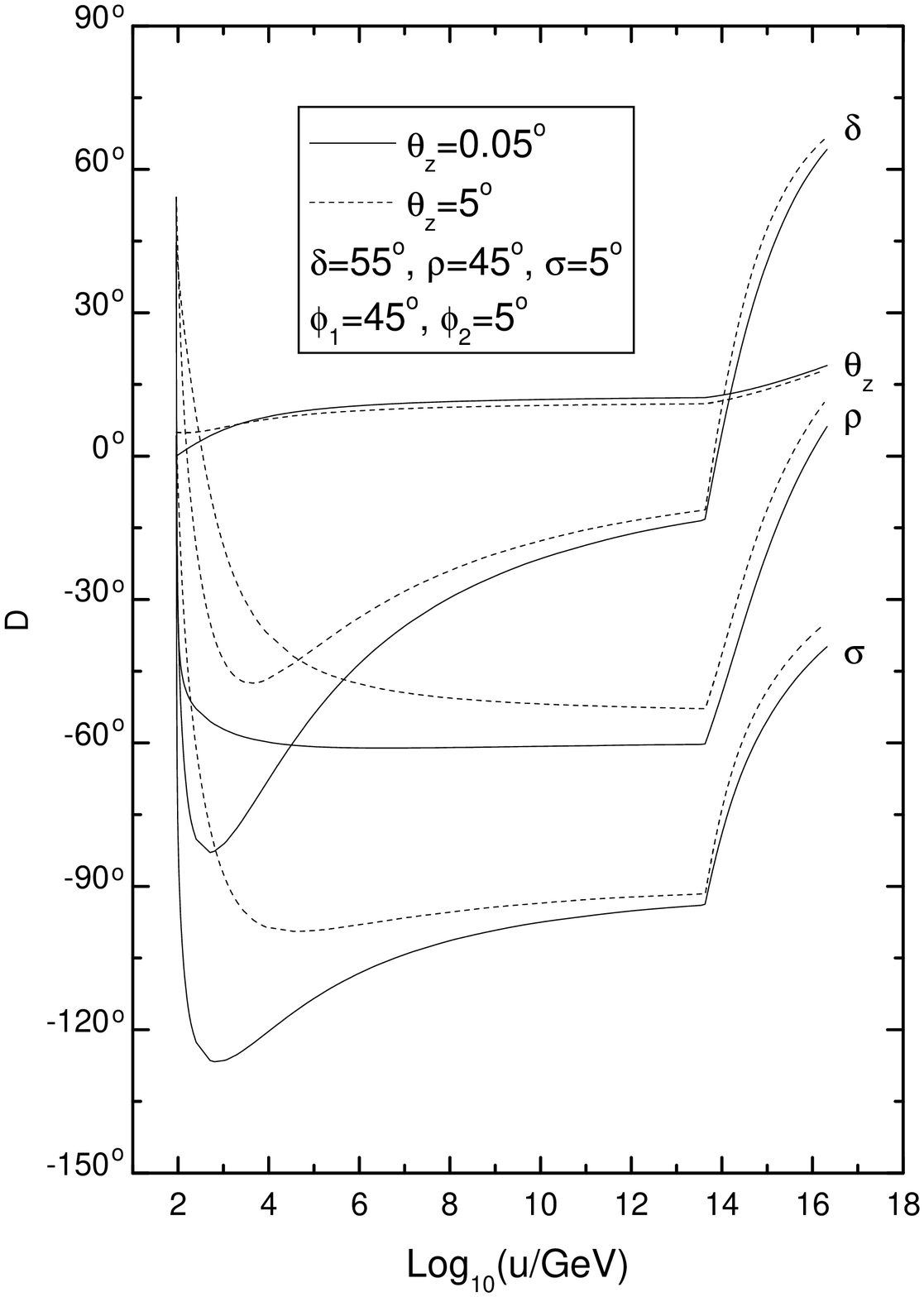}
\end{center}
\caption{The typical evolution behavior of mixing angles
($\theta_{x}, \theta_{y}, \theta_{z}$) and CP-violating phases
($\delta, \rho, \sigma$) in the MSSM when $\tan\beta$ is large, in
the near degeneracy case. Figures A, B and C illustrate cases in
which the correction to a specific mixing angle is mostly enhanced
or damped. Figure A is devoted to $\theta_{x}$, but we have also
included the other two angles in the plot. Similarly, Figure B
corresponds to $\theta_{y}$, and Figure C corresponds to $
\theta_{z}$. Figure D serves to illustrate how $\theta_{z}$ may
affect the RG evolution of CP-violating phases. At $M_{\rm Z}$,
$\theta_{x}=33.2^{\circ}$, $\theta_{y}=45^{\circ}$ and $
\theta_{z}=0.05^{\circ}$ (used in Figures A, B and C); at $M_{3}$,
$y_\nu=0.8$, $ \theta_{1}=$ $\theta_{2}=45^{\circ}$ and $\theta
_{3}=\delta_\nu=5^{\circ}$. Initial values of CP-violating phases
($\delta $, $\rho $ and $\sigma $ at $M_{\rm Z}$, and $\phi_{1}$
and $\phi_{2}$ at $M_{3}$) are marked on the graphs.}
\label{FigCase2MSSM50}
\end{figure}

%%%%%%%%%%%%%%%%%%%%%%%%%%%%%%%%%%%%%%%%%%%%%%%%%%%%%%%%%%%%%%%%%%%%%%%%%%%%%%%%%%%
\begin{figure}[tbp]
\begin{center}
\includegraphics[width=8cm,height=8cm,angle=0]{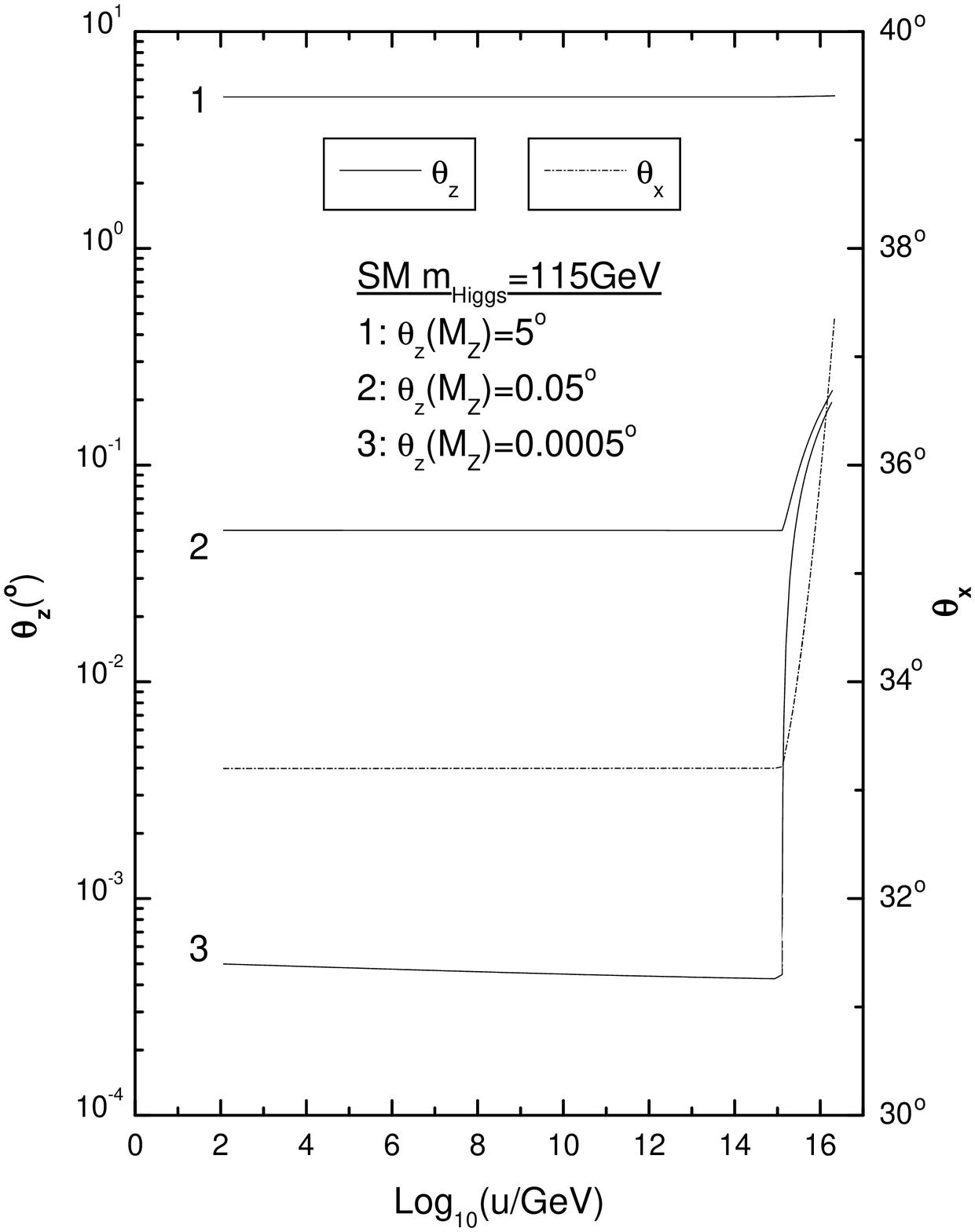}

\includegraphics[width=8cm,height=8cm,angle=0]{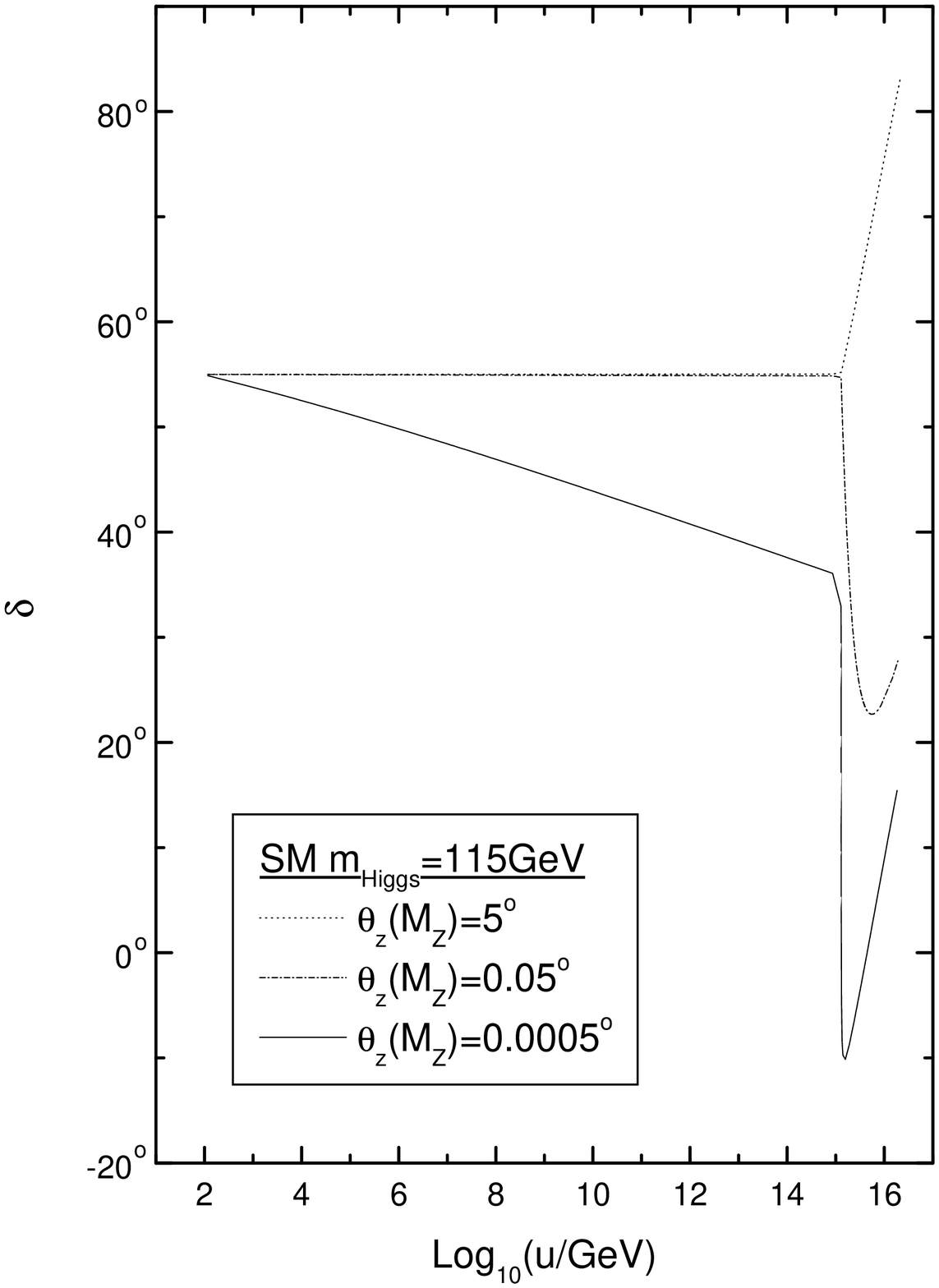}
\includegraphics[width=8cm,height=8cm,angle=0]{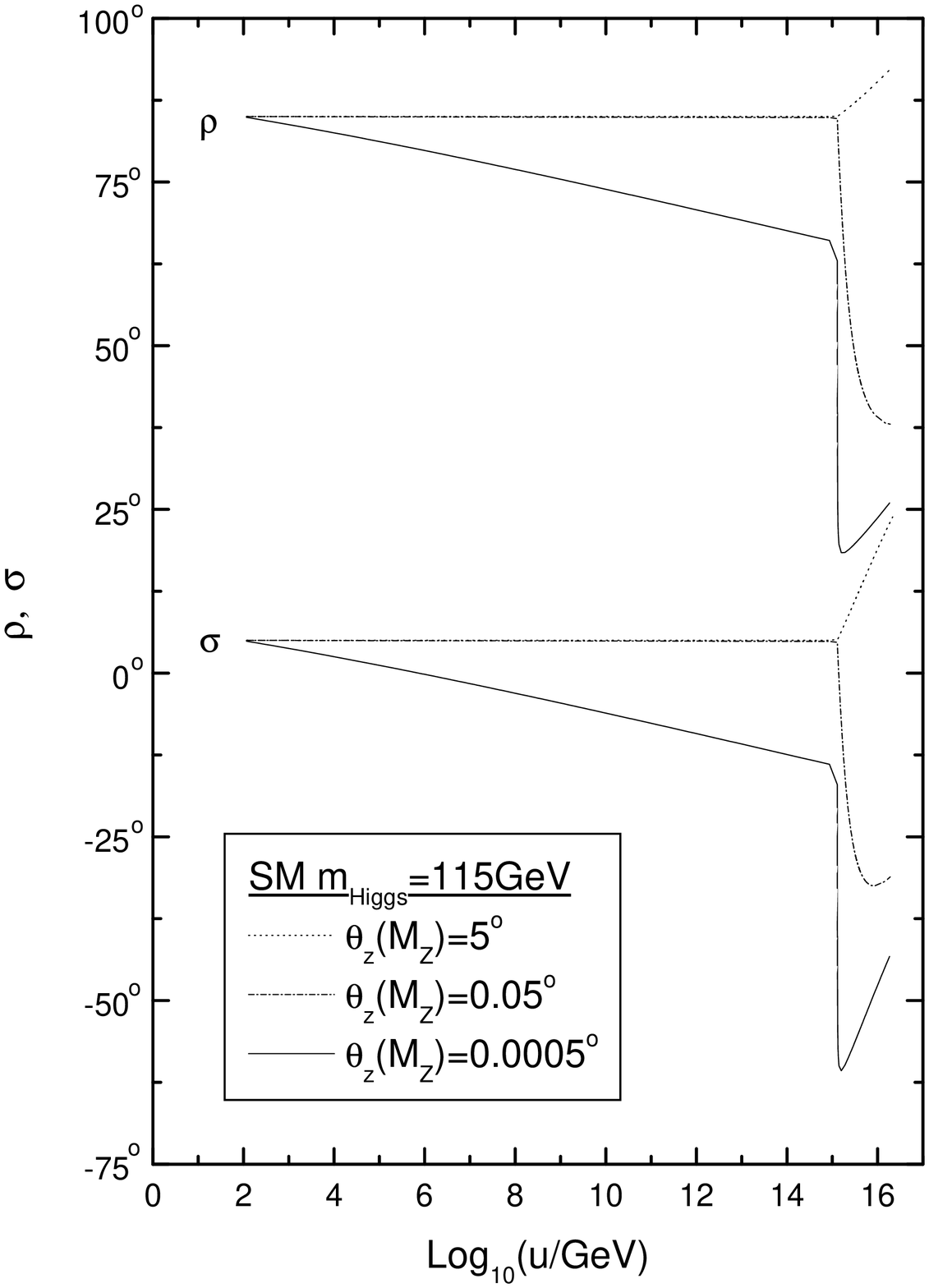}
\end{center}
\caption{The typical evolution behavior of mixing angles
$\theta_{x}$ and $ \theta_{z}$ and CP-violating phases ($\delta,
\rho, \sigma $) in the SM, in the inverted hierarchy case. At
$M_{\rm Z}$, $\theta_{x}=33.2^{\circ}$, $\theta_{y}=45^{\circ}$, $
\delta =55^{\circ}$, $\rho =85^{\circ}$ and $\sigma =5^{\circ}$;
at $M_{3}$, $y_\nu=0.8$, $\theta_{1}=\theta_{2}=45^{\circ}$ and
$\phi_{1}=\phi_{2}= \theta_{3}=\delta_\nu=5^{\circ}$.}
\label{FigCase3SM}
\end{figure}

%%%%%%%%%%%%%%%%%%%%%%%%%%%%%%%%%%%%%%%%%%%%%%%%%%%%%%%%%%%%%%%%%%%%%%%%%%%%%%%%%%%
\begin{figure}[tbp]
\begin{center}
\includegraphics[width=8cm,height=8cm,angle=0]{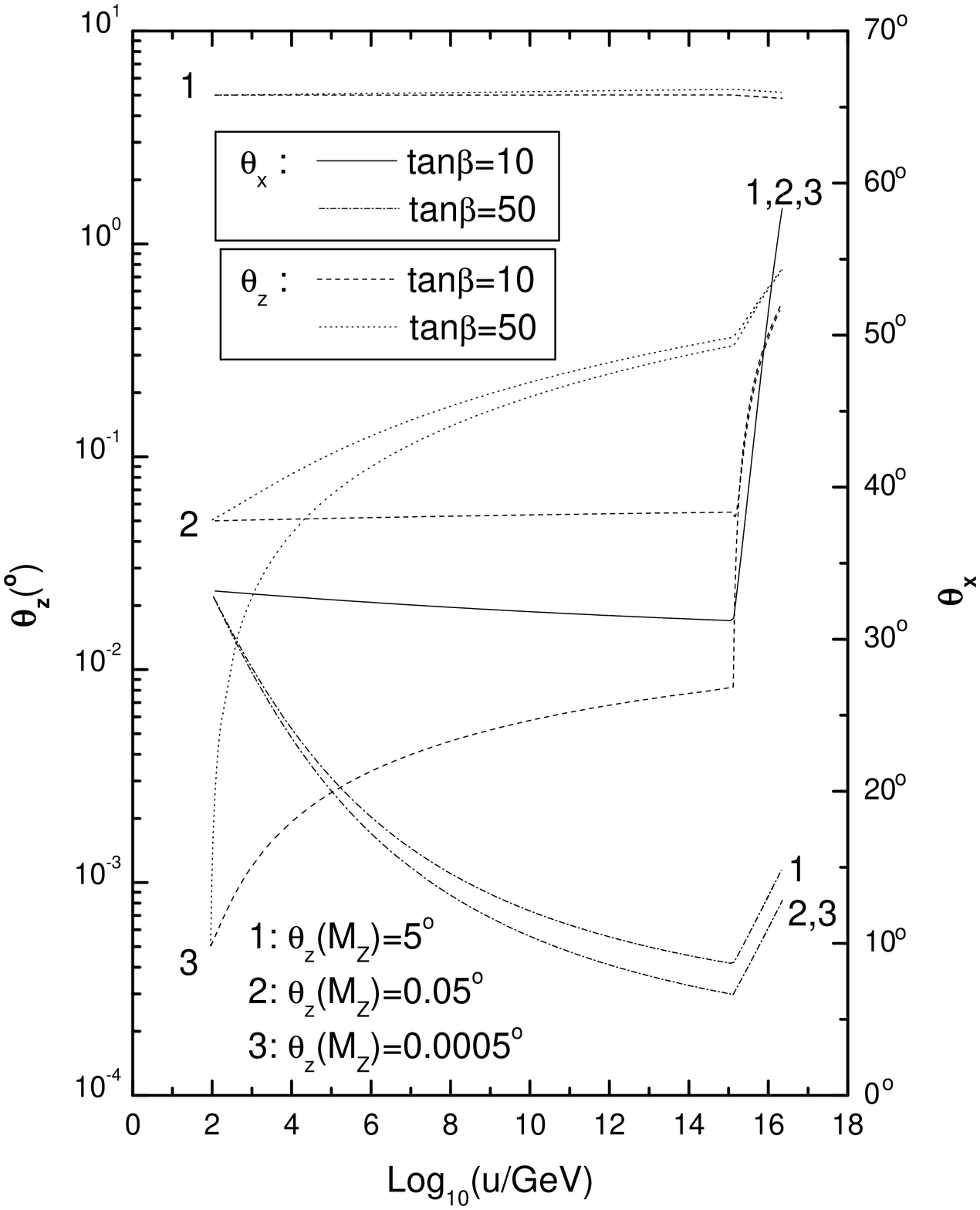}

\includegraphics[width=8cm,height=8cm,angle=0]{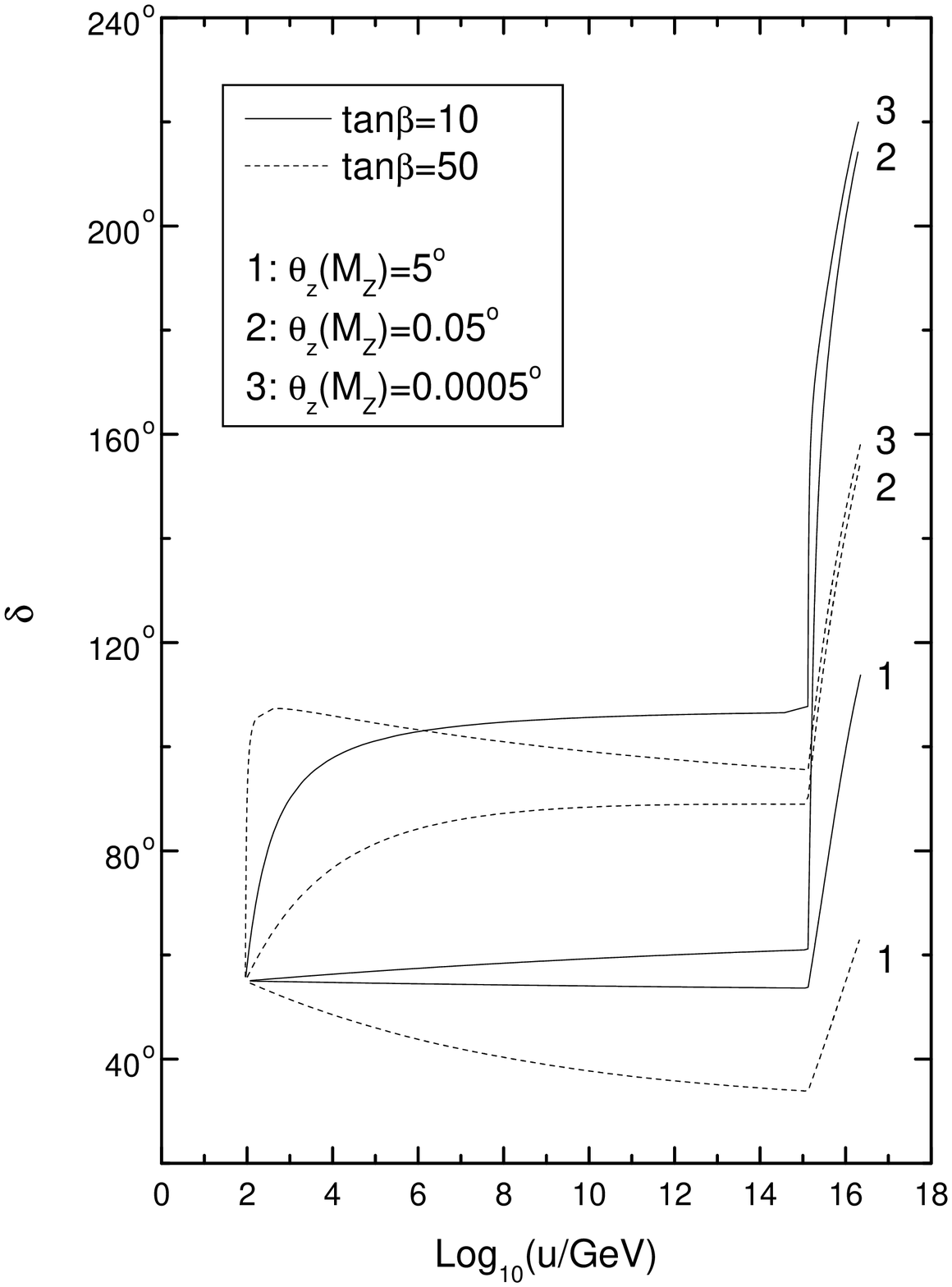}
\includegraphics[width=8cm,height=8cm,angle=0]{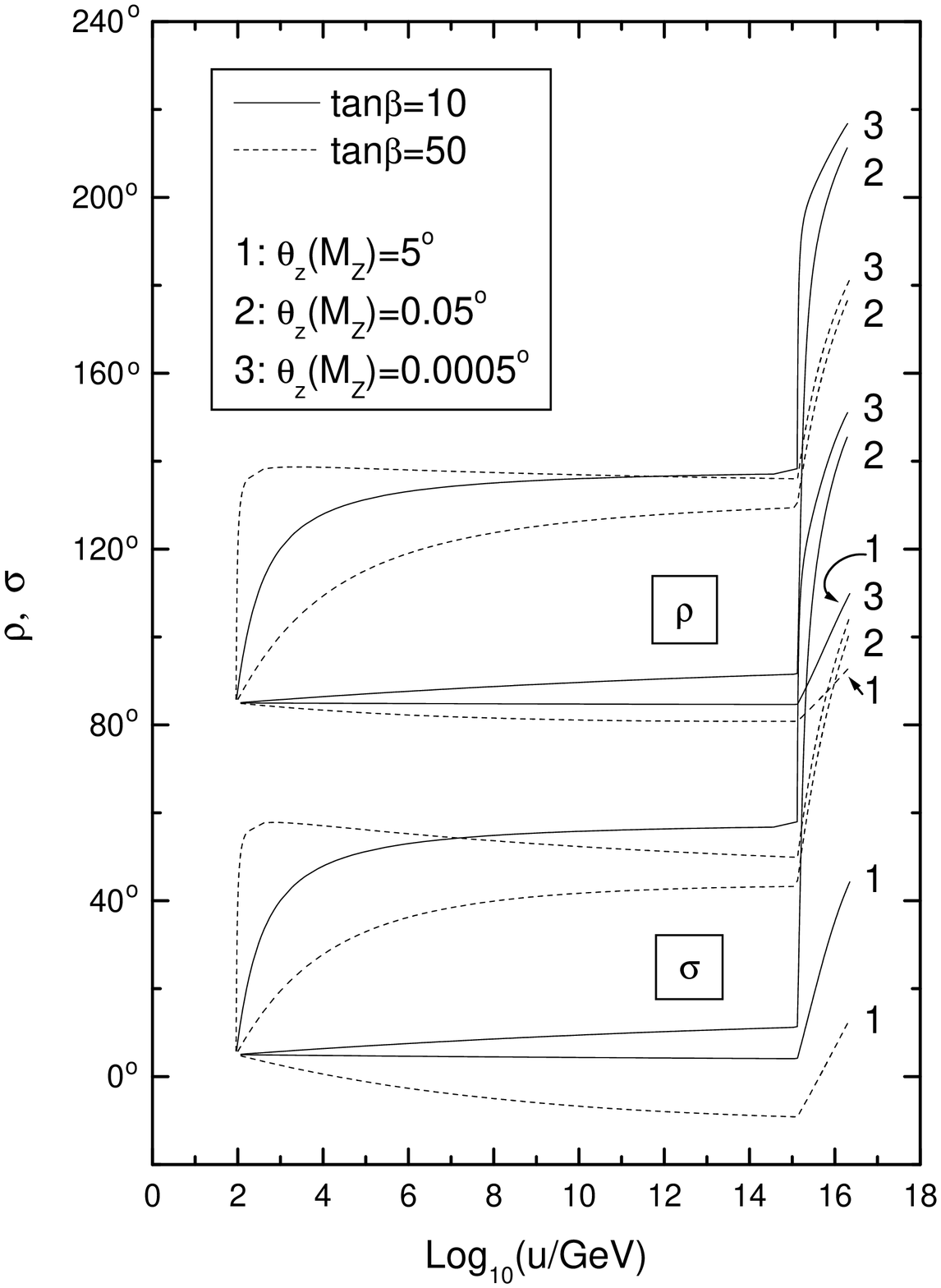}
\end{center}
\caption{The typical evolution behavior of mixing angles
$\theta_{x}$ and $ \theta_{z}$ and CP-violating phases ($\delta,
\rho, \sigma$) in the MSSM, in the inverted hierarchy case. At
$M_{\rm Z}$, $\theta_{x}=33.2^{\circ}$, $\theta_{y}=45^{\circ}$
and $ \delta =55^{\circ}$. Also at $M_{\rm Z}$, $\rho =\sigma
=5^{\circ}$ in the plot of $\theta_x$, but $\rho =85^{\circ}$ and
$\sigma =5^{\circ}$ in the plot of $\theta_z, \delta, \rho $ and
$\sigma $. At $M_{3}$, $y_\nu=0.8$, $\theta_{1}=
\theta_{2}=45^{\circ}$ and $\phi_{1}=\phi_{2}=
\theta_{3}=\delta_\nu=5^{\circ}$.} \label{FigCase3MSSM}
\end{figure}

\end{document}